\documentclass[11pt,english,twoside]{article}

\usepackage[T1]{fontenc}
\usepackage[latin1]{inputenc}
\usepackage[english]{babel}
\usepackage{lmodern}
\usepackage{amsmath,amssymb}
\usepackage{float}
\usepackage{graphicx}
\usepackage[dvips]{epsfig}
\usepackage{psfrag}

\usepackage{hyperref}

\usepackage{lscape}

\catcode`\@=11

\def \be  {\begin{equation}}
\def \ee  {\end{equation}}
\def \ba  {\begin{eqnarray}}
\def \ea  {\end{eqnarray}}
\def \baa {\begin{eqnarray*}}
\def \eaa {\end{eqnarray*}}
\def \bb  {\begin {thebibliography} }
\def \eb  {\end{thebibliography}}
\def \lab #1 {\label{#1}}

\def \qqqquad {\qquad\qquad}
\def \matrix #1 {\left(\begin{array}{cc} #1 \end{array}\right)}

\def \Im {\mathop{\rm Im}\nolimits}
\def \Re {\mathop{\rm Re}\nolimits}

\def \e  {\mathop{\rm e}\nolimits}

\newcommand{\insertfig}[2]{\mbox{\epsfysize=#1cm \epsfbox{#2.eps}}}%
\newcommand{\ft}[2]{{\textstyle\frac{#1}{#2}}}

\def\XXint#1#2#3{{\setbox0=\hbox{$#1{#2#3}{\int}$}
     \vcenter{\hbox{$#2#3$}}\kern-.5\wd0}}

\def\numberbysection{\@addtoreset{equation}{section}
                     \def\theequation{\thesection.\arabic{equation}}}

\numberbysection

\begin{document}

\renewcommand{\thefootnote}{\fnsymbol{footnote}}

\begin{titlepage}
\begin{flushright}
\begin{tabular}{l}
\end{tabular}
\end{flushright}

\vskip3cm

\begin{center}
 {\Large \bf Exciting the GKP string at any coupling}
\end{center}

\vspace{1cm}

\centerline{\sc B.~Basso\footnote{Email: bbasso@princeton.edu}}

\vspace{10mm}

\centerline{\it Princeton Center for Theoretical Science,}
 \centerline{\it Jadwin Hall, Princeton University,}
\centerline{\it NJ 08544, USA}

\vspace{1cm}

\centerline{\bf Abstract}

\vspace{5mm}  
  
We analyze the spectrum of excitations around the Gubser-Klebanov-Polyakov (GKP) rotating string in the long string limit and construct a parametric representation for their dispersion relations at any value of the string tension. On the gauge theory side of the AdS/CFT correspondence, i.e., in the planar $\mathcal{N}=4$ Super-Yang-Mills theory, the problem is equivalent to finding the spectrum of scaling dimensions of large spin, single-trace operators. Their scaling dimensions are obtained from the analysis of the Beisert-Staudacher asymptotic Bethe ansatz equations, which are believed to solve the spectral problem of the planar gauge theory. We examine the resulting dispersion relations in various kinematical regimes, both at weak and strong coupling, and detail the matching with the Frolov-Tseytlin spectrum of transverse fluctuations of the long GKP string. At a more dynamical level, we identify the mechanism for the restoration of the $SO(6)$ symmetry, initially broken by the choice of the Berenstein-Maldacena-Nastase vacuum in the Bethe ansatz solution to the mixing problem.  

\vspace{1cm}

\centerline{\it dedicated to the memory of Jacques Basso}

\end{titlepage}

\setcounter{footnote} 0

{\small \tableofcontents}

\newpage

\renewcommand{\thefootnote}{\arabic{footnote}}

\section{Introduction}\label{Intro}

The AdS/CFT correspondence~\cite{Mal97} has considerably deepened our understanding of the dynamics of strongly coupled gauge theories and shed light on the longstanding problem of constucting their string theory duals. In the most well studied example of the maximally supersymmetric extension of Yang-Mills theory (aka $\mathcal{N}=4$ SYM theory), that we are going to consider here, it offers a geometrical interpretation of the gauge dynamics in terms of strings propagating on $\rm AdS_{5}\times S^{5}$ background. One of the  illustrations of how this dictionary works is well established for the spectrum of scaling dimensions of composite operators in the conformal SYM theory. In the planar limit, the scaling dimensions are complicated functions of the 't Hooft coupling constant $g^2 \equiv \lambda/16\pi^2$, with $\lambda = g^2_{\textrm{{\tiny YM}}}N_{\textrm{c}}$, and can be computed at weak coupling order by order in the perturbative expansion in $g^2$. The prediction from the AdS/CFT correspondence is that this spectrum of scaling dimensions simultaneously describes the spectrum of energies of a single string propagating freely on the background geometry. This interpretation becomes even more manifest at strong coupling, i.e., at large string tension $g \gg 1$, where the quantum fluctuations of the string are suppressed. This conjecture has triggered an enormous flood of theoretical activity over the past decade and has led to dramatic improvement of our understanding of theories on both sides of the duality. For the spectrum of scaling dimensions/energies, it has culminated with a proposal of a set of equations~\cite{BS05,BES06,GKV,AF} which implements hidden integrable structures of the gauge/string system~\cite{L98,BDM98,MZ,BS03,Beisert04,KMMZ} and which is believed to solve exactly the spectral problem for any value of the coupling constant $g$. In this paper, we will apply these equations to unravel the spectrum of excitations around the Gubser-Klebanov-Polyakov (GKP) long rotating string~\cite{GKP02} and construct their dispersion relations at any value of the coupling constant.

There is already a wealth of insights both at weak and strong coupling for the problem that will be the focus of this work. On the string theory side, the perturbative spectrum of fluctuations around the long GKP string, as well as their leading-order dispersion relations, have been known for a long time~\cite{FT02,AM07}. It is also known that this spectrum gets corrected at the non-perturbative level with the massless modes of the string acquiring a mass by dimensional transmutation~\cite{AM07}. At the classical level, but now in the regime of giant excitations, the dispersion relation was recently constructed in~\cite{DL10} using the finite gap machinery and an interesting aspect regarding its large-momentum asymptotics was notably uncovered. On the gauge theory side of the duality, the problem maps into the analysis of the spectrum of scaling dimensions of large spin operators, for which many features have been already elucidated~\cite{BGK03,BGK06,ES06,AM07,FRS07,FRZ09} and for which many results were already extended to all loops. In this paper, we would like to complete the picture and present a comprehensive account of the spectrum of excitations and of their dispersion relations.

This investigation is not purely academic and was greatly motivated by the recent analysis of~\cite{AGMSV10} where it is shown to be directly relevant to the computation of on-shell scattering amplitudes. The bridge between the two comes from their respective relationship with the cusped light-like Wilson loops, which has been known for some time for the large spin spectral problem~\cite{KM,BGK03,Kruczenski,KRTT} and uncovered more recently for scattering amplitudes~\cite{AMg07}. This gives hope that the considerations and results deduced in this paper could be helpful to exploring the relation between the integrable structures observed in the study of scattering amplitudes at strong coupling~\cite{AM09} and the ones that govern the spectrum of scaling dimensions.

Getting back to the problem at hand, it is worth recalling some of the remarkable features of the GKP string. By construction it describes the leading Regge trajectory of string eigenstates carrying single $SO(2,4)$ spin $S$ and minimal global-time energy $\Delta$, and it can be thought as the $\rm AdS_{3}$ analogue of the Berenstein-Maldacena-Nastase (BMN) single $SO(6)$ spin $J$ solution~\cite{BMN,GKP02}, which is the standard reference state for building up the spectrum of scaling dimensions/energies using integrability~\cite{MZ,BS03,Beisert04}. Semiclassically ($g\gg 1$) the GKP string is described by a folded string rotating around its center of mass in $\rm AdS_{3} \subset AdS_{5}\times S^{5}$. At generic values of the spin, it defines a complicated solution to the classical string equations and thus represents itself an intricate background for the semiclassical expansion of the string $\sigma$-model~\cite{FT02}. In the large spin limit however, the string gets long, with proper length $\sim 2\log{S}$,%
\footnote{With the radius of curvature of the warped geometry normalized to one.} ultimately stretching all the way to the boundary of space at $S=\infty$, and becomes homogeneous. Its proper energy%
\footnote{That is the energy measured in the rotating frame of the long GKP string.} $E_{\textrm{vacuum}}$ is then uniformely distributed and scales with the length leading to the logarithmic scaling~\cite{GKP02}
\be\label{LogScalE}
E_{\textrm{vacuum}} = \Delta-S = A_{2}(g)\log{S} + O(\log^0{S})\, ,
\ee
where $A_{2}(g)/2$ stands for the energy per unit of length given classically by the string tension  $A_{2}(g)/2 = 2g+\ldots = \sqrt{\lambda}/2\pi+\ldots\,$. The scaling~(\ref{LogScalE}) persists at higher loops and its string $1/g$ expansion is known explicitely up to two loops~\cite{FT02,RTCusp}. A natural question arises what are the excitations propagating in this background and what are their dispersion relations. As we already alluded above, the answer is known to leading order at strong coupling and can be obtained by considering the spectrum of transverse fluctuations of the string~\cite{FT02,AM07}. It was found that the dispersion relations are all relativitics and that the elementary world-sheet excitations are 5 massless bosons for rotations in $\rm S^{5}$, 2 mass-$\sqrt{2}$ bosons for rotations of $\rm AdS_{3}$ in $\rm AdS_{5}$, 1 mass-2 boson for the transverse fluctuation in $\rm AdS_{3}$, and finally 8 mass-$1$ fermions. The energy of the state made out of $M$ excitations is then given by%
\footnote{Perturbative excitations are free to leading order at strong coupling.}
\be\label{Edec}
E = \Delta - S = E_{\textrm{vacuum}} + \sum_{j=1}^{M}E_{j}(p_{j}) + \ldots\, ,
\ee
with $p_{j} = \pi n_{j}/\log{S}$ for some mode numbers $n_{j}$, and where the ellipsis stands for $1/S$-suppressed corrections. It remains a challenge to compute corrections to this spectrum by conventional $\sigma$-model methods. The AdS/CFT correspondence hints that we can solve this problem by considering the spectrum of scaling dimensions of large spin operators. Integrability suggests that the decomposition~(\ref{Edec}) should persist, in the interacting theory, giving the means to extract individual energies of the excitations. 

The gauge theory dual of the GKP string is a single-trace operator belonging to the (universal) twist-two multiplet~\cite{GKP02}, though it is not entirely clear (to the author at least) what precise gauge-theory operator it corresponds to.%
\footnote{A possible candidate is the superconformal primary which is a $SO(6)$ singlet.} For our purposes, it is not relevant which twist-two operator maps exactly to GKP string, and in practice it is more convenient to work with the following superconformal descendant%
\footnote{The spin of the operator may differ from the spin of the GKP string since we are considering a descendant. This leads to a finite shift that is irrelevant at large spin within the accuracy of our approximation.}
\be\label{GKPvacOp}
\textrm{vacuum} = \textrm{GKP string} = \textrm{tr}\, ZD_{+}^{S}Z + \ldots \, ,
\ee
carrying the Lorentz spin $S$, and where $Z$ is one of the three complex scalars of the gauge theory, $D_{+} = n_{+}^{\mu}D_{\mu} = D_{0}+D_{3}$ is the light-cone covariant derivative, and dots indicate that the form of the operator is renormalized. The operator representation~(\ref{GKPvacOp}) obscures however the fact that the GKP vacuum is a $SO(6)$ singlet. An immediate consequence of the mapping~(\ref{GKPvacOp}) is the recovering of the logarithmic scaling~(\ref{LogScalE}) to all loops~\cite{KM}, with $\Delta$ in~(\ref{LogScalE}) the scaling dimension and with $A_{2}(g) = 2\Gamma_{\textrm{cusp}}(g)= 4g^2 +O(g^4)$ being twice the cusp anomalous dimension~\cite{KR85,P80}.  Since $2\log{S}$ stands for the physical length in the game,%
\footnote{The interpretation of the quantity $2\log{S}$ as defining the physical length in the problem we are considering is not straightforward at weak coupling. It is however possible to construct a geometrical picture for the large spin operator~(\ref{GKPvacOp}) in which this identification becomes transparent~\cite{AM07}. This picture is obtained by conformal mapping of the gauge theory from $\mathbb{R}^{3,1}$ to $\rm AdS_{3}\times S^{1}$. The  twist-two operator then translates into a flux configuration which is stretched along a particular direction with physical size $2\log{S}$. Exciting this configuration is equivalent to exciting the operator~(\ref{GKPvacOp}) and dual to exciting the GKP string.}
the identification~(\ref{LogScalE}) implies that the cusp anomalous dimension has the meaning of the vacuum energy density~\cite{AM07}.

The next step is to add excitations on top of the ground-state operator~(\ref{GKPvacOp}) and extract their energies using the decomposition~(\ref{Edec}). A natural way of doing this is to allow for operator insertions in~(\ref{GKPvacOp}).%
\footnote{It is necessary to increase the length of the operator~(\ref{GKPvacOp}) if we want to put excitation on top of it. This is because all length-two operators fall into the same twist-two multiplet~\cite{KL00,Beisert04} and since taking descendant is a global transformation which `does not propagate'.} For a single excitation, for instance, one expects the following identity to take place,
\be\label{OPstate}
\textrm{one-particle state} = \textrm{tr}\, ZD_{+}^{S_{1}}\Phi D_{+}^{S_{2}}Z + \ldots \, ,
\ee
where $\Phi$ stands for a generic local operator of the gauge theory and with $S \sim S_{1} + S_{2} \gg 1$. Under renormalization, the operator~(\ref{OPstate}) can mix with similar operators that differ in their relative numbers $S_{1}-S_{2}$ of covariant derivatives. The interpretation is that the operator $\Phi$ propagates through the background of covariant derivatives. It carries a momentum $p$ which, roughly, accounts for the inequivalent ways of distributing the covariant derivatives around it. Imposing the condition that the operator~(\ref{OPstate}) carries a definite scaling dimension $\Delta$ should fix its form and translates into a quantization condition for the momentum $p$ of $\Phi$.

Obviously, this picture is directly inspired from the treatment of excitations on top of the BMN vacuum $\textrm{tr}\, Z^J$~\cite{BMN}. The difference here is that since the vacuum operator~(\ref{GKPvacOp}) has a complicated structure, with an intricate mixing, it would be difficult to make use of it outside of the realm of integrability. Keeping this in mind, we can push the analogy with the BMN analysis even further and ask what are the (lightest) elementary excitations. For the BMN case, they are those that make the minimal contribution to the BMN energy $\Delta-J$. For the GKP string, minimizing the energy $E = \Delta-S = \textrm{twist} + O(g^2)$ suggests that the elementary excitations are twist-one partons of the gauge theory. They are also known as light-cone operators~\cite{BDKM03} and are the building blocks of the quasi-partonic operators~\cite{BFLK85}. The pattern of elementary excitations is then composed of eight bosonic and eight fermionic operators that form multiplets under the residual $SO(6)$ symmetry of the GKP vacuum: we get 6 (real) scalar fields in the $\bf{6}$ of $\mathfrak{su}(4)\cong \mathfrak{so}(6)$, $4/4$ twist-one components of left/right Weyl spinors in the $\bf{4}/\bar{\bf{4}}$, and 2 twist-one components of gluon field strength tensor in the $\bf{1}$.%
\footnote{The twist-one component is the one that carries maximal Lorentz spin $S$ in a given $SO(3,1)$ multiplet. Other components possess higher twist.}
Keeping only the $\mathfrak{su}(4)$ highest-weight fields, the twist-one partons are
\be\label{twist-one}
F_{+\perp} \equiv n_{+}^{\mu}n_{\perp}^{\nu}F_{\mu \nu}\, , \qquad \Psi_{+}\, , \qquad Z\, , \qquad \bar{\Psi}_{+}\, , \qquad \bar{F}_{+\perp} = n_{+}^{\mu}\bar{n}_{\perp}^{\nu}F_{\mu \nu}  \, ,
\ee
with $n_{\perp} = (0,1,i, 0), \bar{n}_{\perp} = (0,1, -i, 0)$, the vectors transverse to the lightcone direction, $n_{+}\cdot n_{\perp} = n_{+}\cdot \bar{n}_{\perp} = 0$, and with the $+$ subscripts on the spinor fields indicating their twist-one component.

At the tree level, the twist-one excitations form a supersymmetric pattern and their energy is given by their mass which is equal to their twist. Going beyond this order entails solving the mixing problem for the light-cone operators~(\ref{OPstate}). At one loop, it is equivalent to the diagonalization of the Hamiltonian for the $\mathfrak{sl}(2)$ XXX Heisenberg spin chain, whose spin module depends on the flavor of the parton (see~\cite{BBGK} for the case at hand). It is analyzed systematically by means of a set of Bethe ansatz equations whose solution at large spin $S$ has been given in~\cite{BGK06}. It is immediately found that the energy of a twist-one excitation is~\cite{BGK06}
\footnote{The result of~\cite{BGK06} differs actually from~(\ref{InDRa}) by a vacuum energy shift (namely the energy of the spin chain when no covariant derivatives act on it). This constant depends on the theory in which the conformal spin chain is embedded. For $\mathcal{N}=4$ SYM theory, see~\cite{Beisert04} for instance, we need to add $4g^2(\psi(2s)-\psi(1))$ to the result of~\cite{BGK06} to obtain~(\ref{InDRa}).}
\be\label{InDRa}
E(u) = 1+2g^2\left(\psi\left(s + iu\right) + \psi\left(s - iu\right) -2\psi(1)\right) + O(g^4)\, ,
\ee
where $u$ is a rapidity that parameterizes the motion of the excitation through the ``bath'' of covariant derivatives, and where $\psi(z) = \Gamma'(z)/\Gamma(z)$ is the logarithmic derivative of the Euler Gamma function $\Gamma(z)$. Here $s$ stands for the (light-cone) conformal spin of the excitation and accounts for different flavors, with $2s=1, 2, 3,$ for scalars, fermions and gauge fields, respectively. The analysis of~\cite{BGK06} also contains the effective Bethe ansatz equation that fix the rapidity of the excitation (once confined on the GKP string). From their expression, we read that the momentum $p$, conjugate to the GKP length $2\log{S}$, is
\be\label{InDRb}
p(u) = 2u + O(g^2)\, ,
\ee
for any excitation.%
\footnote{Note that, dispite its appearance, the expression~(\ref{InDRb}) for the momentum is a genuine one-loop computation.} Combining~(\ref{InDRa}) and~(\ref{InDRb}) gives the sought one-loop dispersion relations.

In this paper, we will compute subleading corrections to~(\ref{InDRa}, \ref{InDRb}), and derive the all-loop dispersion relations from the Beisert-Staudacher asymptotic Bethe ansatz equations~\cite{BS05}. The representation for the dispersion relations is parametric and is expressed in terms of particular functions that solve the Beisert-Eden-Staudacher (BES) equation~\cite{BES06} for the ground-state distribution of roots. Our analysis follows the seminal papers~\cite{K95,BGK06,ES06,FRS07} for dealing with the large spin limit of the Bethe ansatz equations, and, as already mentioned above, it overlaps with previous analyses. In the particular case of a scalar, for instance, subleading corrections to the energy~(\ref{InDRa}) were already computed in~\cite{BKP08} up to three loops, by means of the asymptotic Baxter equation for the $\mathfrak{sl}(2)$ sector~\cite{Bel06}. It is also possible to derive the all-loop expression for the energy, given in this paper, directly from the analysis performed in~\cite{BK08} of the Freyhult-Rej-Staudacher (FRS) equation~\cite{FRS07} for the so-called generalized scaling function~\cite{BGK06,FRS07}. Indeed, this function computes the ground-state energy density of a gas of scalars (confined on the GKP string) at finite density~\cite{AM07} and thus encodes the all-loop energy of a scalar. It has to be complemented by a similar expression for the momentum to get the full parameterization of the dispersion relation.

\begin{figure}[h]%
\setlength{\unitlength}{0.14in}
\centering
\begin{picture}(20,18)%
\put(-4,0){\insertfig{6}{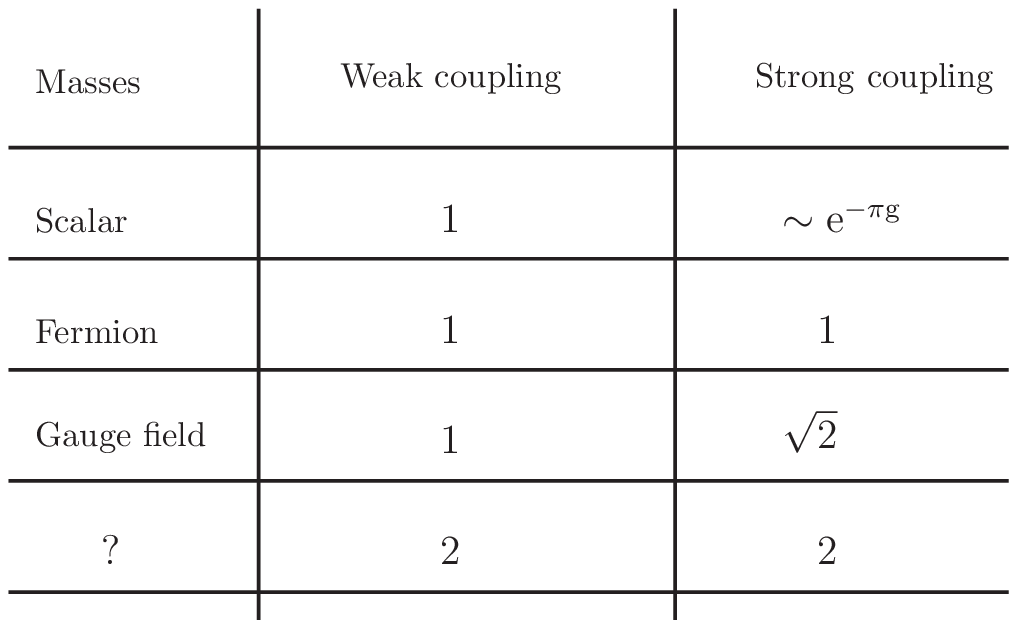}}
\end{picture}%
\caption{Spectrum of masses for gauge and string elementary excitations. The question mark is for the missing mass $2$ boson predicted by the perturbative analysis in the string theory~\cite{FT02,AM07}. Its mass should not receive corrections~\cite{AM07}. There are candidates for it but their stability at non-zero momentum is not guaranteed, since decay channels into two fermions are allowed.}\label{matching}
\end{figure}

As a particular application of our formulae, we can observe the merging of the spectrum of elementary excitations of the gauge and string theories, see Fig.~\ref{matching}. The fate of the scalars has been already elucidated in~\cite{AM07}. Their mass decreases with the coupling and becomes exponentially small at strong coupling. The perturbative string analysis detects only $5$ massless excitations, which are Goldstone bosons for rotations in $\rm S^{5}$. But these are not the genuine asymptotic excitations and the actual spectrum contains $6$ massive scalars with mass $m\sim \e^{-\pi g}$, in agreement with the gauge theory expectations~\cite{AM07}. These features were already tested in the framework of integrability~\cite{FGR08a,BK08,FGR08b} in the context of the FRS equation, and, not surprisingly, are also encoded in our relations. The case of fermions was also covered by the analysis of~\cite{AM07} in which it is explained that they ought to be exactly mass $1$ excitations, as a consequence of the remnant supersymmetries spontaneously broken by the GKP vacuum. Here this property is also a consequence of supersymmetry and is tied to the multiplet joining/splitting~\cite{BS03,BS05}. For the gauge fields, the weak coupling expansion indicates that their mass increases with the coupling and analysis at strong coupling shows that it settles down at $\sqrt{2}$.%
\footnote{Our analysis for the mass of gauge field overlaps with the one of~\cite{FRZ09}, in which the embedding of the gauge-field excitation in the all-loop Bethe ansatz was uncovered (see also~\cite{Beccaria07} for previous discussion of the procedure used in~\cite{FRZ09}). Though the all-loop integral representation is the same here and in~\cite{FRZ09}, the strong coupling expansions disagree.} They find their place in the spectum of perturbative string fluctuations where they are identified with the two transverse mass $\sqrt{2}$ bosons. For illustration of these considerations, we report the one-loop corrections to the masses of twist-one excitations, that immediately follow from~(\ref{InDRa}, \ref{InDRb}) evaluated at $u=0$,%
\footnote{Note that all 8 fermions have the same mass, and idem for the 2 gauge fields, due to the $O(2)$ symmetry that rotates the $\perp$ directions and also exchanges left and right.}
\be
\begin{aligned}
&m_{\textrm{scalar}} \, \, \, \, \, \, \, \, \, \, = 1-8g^2\log{2} + O(g^4) < 1\, , \\
&m_{\textrm{fermion}} \, \, \, \, \, \, = 1 \, , \\
&m_{\textrm{gauge field}} = 1+8g^2(1-\log{2}) + O(g^4) > 1\, .
\end{aligned}
\ee
It is not much difficult to compute corrections to the relativistic laws for the perturbative excitations at strong coupling, which read up to $O(1/g)$,
\be\label{CDRsc}
E_{\star}(p) = \sqrt{m_{\star}^2 + p^2}\bigg[1-c_{\star}\, p^2/ g + O(1/g^2)\bigg]\, ,
\ee
with the masses 
\be
m_{\textrm{scalar}} = O(1/g^\infty)\, , \qquad m_{\textrm{fermion}} = 1\, , \qquad m_{\textrm{gauge field}} = \sqrt{2}-{1\over 8\sqrt{2} g} + O(1/g^2)\, .
\ee
The numerical coefficients entering~(\ref{CDRsc}) at $O(1/g)$ are found to be
\be
c_{\textrm{scalar}} = {\Gamma\left(\ft{1}{4}\right)^4 \over 8(12)^{5/4}\pi^2}\, , \qquad c_{\textrm{fermion}} = {1\over 4}\, , \qquad c_{\textrm{gauge-field}} = {1\over 8}\, .
\ee

Looking further at the table in Fig.\ref{matching} one observes that the twist-one excitations do not exhaust the perturbative spectrum of string fluctuations. The (heavy) mass-2 boson of the string theory is apparently missing. According to~\cite{AM07} its mass is protected and thus should be exactly 2. Twist-two candidates with correct quantum numbers exist for it, like $F_{+-} = n_{+}^{\mu}n^{\nu}_{-}F_{\mu\nu}$, with $n_{-} = (1, 0, 0, -1)$, or possibly $D_{-} = n_{-}\cdot D$. These fields do not find room in our analysis as stable (all-loop) asymptotic excitations. The argument of~\cite{AM07} do not seem actually to guarantee that the boson is stable at non-zero momentum (except at strong coupling due to the boost invariance) and a possible decay into two fermions appears quite realistic, as already advocated in~\cite{AGMSV10}. A similar phenomenon was actually observed in~\cite{Z09} in a relative set up.%
\footnote{I am grateful to Juan Maldacena for pointing this reference out to me.} Whatever the fate of the mass-2 boson can be, its presence raises the question of possible higher-twist excitations. There exist already evidences for them. They come from the spectroscopy of length-three operators~\cite{BBMS04} that shows that higher-conformal-spin $\mathfrak{sl}(2)$ spin chains are embedded into the super spin chain~\cite{BS03} of the planar gauge theory. The conformal spin $s$ is then allowed to take arbitrarily large values $2s = 1, 2, 3, \ldots\, $, the first three ones being realized by the twist-one fields~(\ref{twist-one}) and the others by the (length-one) partons
\be\label{Ibs}
D^{\ell-1}_{\perp}F_{+\perp}\, , \qquad \bar{D}^{\bar{\ell}-1}_{\perp}\bar{F}_{+\perp}\, , \qquad \ell, \bar{\ell} = 2, 3, \ldots\, ,
\ee
where $D_{\perp} = n_{\perp}\cdot D$ and $\bar{D}_{\perp} = \bar{n}_{\perp}\cdot D$. The fields in~(\ref{Ibs}) carry twist $\ell, \bar{\ell}$, and conformal spin $1+\ell/2, 1+\bar{\ell}/2$, respectively, and their one-loop dispersion relations are given by~(\ref{InDRa}, \ref{InDRb}), after plugging the appropriate value of the conformal spin $s$. For instance, the mass of the twist-$\ell$ excitation, equivalently $\bar{\ell}$, is given to one loop by
\be\label{MIbs}
m_{\ell} = \ell+4g^2\left(\psi\left(1+\ell/2\right) -\psi(1)\right) + O(g^4)\, .
\ee
The observation of~\cite{BBMS04} was further analyzed in~\cite{FRZ09} from which one can already extract an all-loop representation for the masses of the fields~(\ref{Ibs}). Extending this analysis, we will construct the all-loop dispersion relations for the fields~(\ref{Ibs}) and argue on top of them that the excitations~(\ref{Ibs}) are bound states of gauge fields. One can already verify at the one-loop level, by plotting~(\ref{MIbs}), that $m_{\ell} < \ell \, m_{\textrm{gauge field}} = \ell \, m_{\ell=1}$, which is obeyed for a bound state. The relation between higher-twist excitations and gauge fields is nevertheless obscured at weak coupling because the $\ell$ gauge-field constituents of the twist-$\ell$ excitation do not all live on the same rapidity sheet (idem for $\bar{\ell}$ by $O(2)$ symmetry). It becomes however manifest as the coupling gets larger and within a suitable kinematical range where the dispersion relations for bound states can be obtained by fusion of the dispersion relation for gauge field, in analogy with bound states of magnons over the BMN vacuum~\cite{Dorey06} though not in the same kinematical domain.%
\footnote{The fusion is performed here by shifting the gauge-fields rapidities in the complex plane assuming $u^2 < (2g)^2$, with $u$ the center-of-mass rapidity (taken real), while it is done for $u^2 > (2g)^2$ for bound states of magnons.}

This completes the pattern of excitations considered in this paper. They are the momentum carriers for the large spin effective Bethe ansatz equations. There are also isotopic roots associated with the $SO(6)$ symmetry of the GKP vacuum. Their embedding into the asymptotic Bethe ansatz equations is asymmetric because the BMN vacuum breaks spontaneously $SO(6)$ down to $SU(2)\times SU(2)$. The root associated with the broken $SU(2)\subset SO(6)$ is embedded in the form of a stack~\cite{BKSZ05} which becomes isotopic at large spin and restores the symmetry. This mechanism is quite similar to what happens for spinons in the thermodynamic limit of the antiferromagnetic Heisenberg spin chain, where the restoration of the $SU(2)$ symmetry is implemented by the formation of $2$-strings of Bethe roots. For the stacks previously mentioned we verify that they do not carry energy nor momentum at any coupling. We show moreover that their equations are identical to the middle-node equations of the inhomogeneous $SO(6)$ Heisenberg spin chain, with length fixed by the total number of scalars and with inhomogeneities given by the scalars' rapidities.

Finally, let us comment on some underlying assumptions. Our analysis relies on the all-loop Bethe ansatz equations~\cite{BS05,BES06}. It is well known that these equations do not capture correctly the interactions that wrap around the spin chain~\cite{AJK05}. For a generic state, the size of these corrections is correlated to the spin-chain length and becomes more important as the length gets short. It seems therefore that our analysis is necessarily bounded to long operators, which is not welcome as we only assume that the spin is large. However, it is widely believed that wrapping issues are circumvented in the large spin limit. This is supported by computations of wrapping corrections of twist $L$ anomalous dimensions in the $\mathfrak{sl}(2)$ sector, using the L\"uscher formalism~\cite{BJ08,BJL08}. These results indicate that wrapping corrections are delayed at large spin, $S\gg1$, affecting contributions of order $O(\log^2{S}/S^2)$. From our point of view, these contributions are effectively of a wrapping type as they are exponentially suppressed in the physical length $R = 2\log{S}$. Hence we expect that the asymptotic Bethe ansatz equations account properly for the anomalous dimensions of large spin operators as long as we restrict ourselves to contributions suppressed by inverse powers of $\log{S}$. It remains, nevertheless, conjectural that wrapping effects are suppressed by powers of the spin for all large spin operators. A more careful analysis would entail the use of the Y system of~\cite{GKV} which is beyond the scope of this paper.

The plan of the paper is the following. In Section~\ref{EO}, we present the embedding of the various excitations into the asymptotic Bethe ansatz equations~\cite{BS05}. We comment on subtleties in the separation between isotopic and momentum-carrying roots at one loop, which are tied to the peculiar embedding of fermions. In Section~\ref{ALA}, we perform the large spin analysis of the all-loop Bethe ansatz equations applying the technology of~\cite{ES06,BES06,FRS07}. In Section~\ref{ALDR}, we derive exact representation for the dispersion relations of the various excitations. Then, we analyze these dispersion relations at weak coupling and derive explicit expressions for them up to three-loop accuracy. We discuss their analytical properties in the complex rapidity plane and derive fusion relations for bound states. In Section~\ref{SCDRs}, we perform the analysis at strong coupling and identify the various kinematical regimes. Section~\ref{Concl} contains concluding remarks and outlooks. Several Appendices contain the technical details of the analysis.

\section{String hypothesis for GKP string}\label{EO}

We start recalling the asymptotic Bethe ansatz equations and then detail the embedding of the various excitations. We deal separately with the momentum-carrying roots and isotopic roots. We further discuss some subtleties in the embedding of fermions, that we relate to an enhancement of symmetry from $SU(4)$ to $SL(2|4)$ at weak coupling.

\subsection{Beisert-Staudacher equations}

The asymptotic Bethe ansatz equations exist in four distinct forms corresponding to different choices for the Dynkin diagram of the $\mathcal{N}=4$ superconformal algebra~\cite{BS05}. They are labelled by the gradings $\eta_{1}, \eta_{2} = \pm 1$ among which we select the one with $\eta_{1} = \eta_{2} =-1$ for our analysis. With the latter choice, the momentum-carrying roots are directly associated with the light-cone covariant derivatives $D_{+}$~\cite{BS05}, with $D_{+} = \mathcal{D}_{1\dot{1}}$ in the bi-spinor notation of~\cite{Beisert04}. The equations in this grading are given by~\cite{BS05}
\be\label{ABAE}
\begin{aligned}
1 =& \prod_{j\neq k}^{K_{2}}{u_{1, k}-u_{2, j}-\ft{i}{2} \over u_{1, k}-u_{2, j}+\ft{i}{2}}\prod_{j=1}^{K_{4}}{1-g^2/x_{1, k}x^{-}_{4, j} \over 1-g^2/x_{1, k}x^{+}_{4, j}}\, , \\
1 =& \prod_{j\neq k}^{K_{2}}{u_{2, k}-u_{2, j}+i \over u_{2, k}-u_{2, j}-i}\prod_{j=1}^{K_{3}}{u_{2, k}-u_{3, j}-\ft{i}{2} \over u_{2, k}-u_{3, j}+\ft{i}{2}}\prod_{j=1}^{K_{1}}{u_{2, k}-u_{1, j}-\ft{i}{2} \over u_{2, k}-u_{1, j}+\ft{i}{2}}\, , \\
1 =& \prod_{j=1}^{K_{2}}{u_{3, k}-u_{2, j}-\ft{i}{2} \over u_{3, k}-u_{2, j}+\ft{i}{2}}\prod_{j=1}^{K_{4}}{x_{3, k}-x^{-}_{4, j} \over x_{3, k}-x^{+}_{4, j}}\, , \\
1 =& \left({x^{-}_{4, k} \over x^{+}_{4, k}}\right)^{L}\prod_{j\neq k}^{K_{4}}{x^{-}_{4, k}-x^{+}_{4, j} \over x^{+}_{4, k}-x^{-}_{4, j}}{1-g^2/x^{+}_{4, k}x^{-}_{4, j} \over 1-g^2/x^{-}_{4, k}x^{+}_{4, j}} \sigma^2(u_{4, k}, u_{4, j}) \\
      &  \times \prod_{j=1}^{K_{3}}{x^{+}_{4, k}-x_{3, j} \over x^{-}_{4, k}-x_{3, j}}\prod_{j=1}^{K_{5}}{x^{+}_{4, k}-x_{5, j} \over x^{-}_{4, k}-x_{5, j}}\prod_{j=1}^{K_{1}}{1-g^2/x^{+}_{4, k}x_{1, j} \over 1-g^2/x^{-}_{4, k}x_{1, j}}\prod_{j=1}^{K_{7}}{1-g^2/x^{+}_{4, k}x_{7, j} \over 1-g^2/x^{-}_{4, k}x_{7, j}}\, ,\\
1 =& \prod_{j=1}^{K_{6}}{u_{5, k}-u_{6, j}-\ft{i}{2} \over u_{5, k}-u_{6, j}+\ft{i}{2}}\prod_{j=1}^{K_{4}}{x_{5, k}-x^{-}_{4, j} \over x_{5, k}-x^{+}_{4, j}}\ , \\
1 =& \prod_{j\neq k}^{K_{6}}{u_{6, k}-u_{6, j}+i \over u_{6, k}-u_{6, j}-i}\prod_{j=1}^{K_{5}}{u_{6, k}-u_{5, j}-\ft{i}{2} \over u_{6, k}-u_{5, j}+\ft{i}{2}}\prod_{j=1}^{K_{7}}{u_{6, k}-u_{7, j}-\ft{i}{2} \over u_{6, k}-u_{7, j}+\ft{i}{2}}\, , \\
1 =& \prod_{j\neq k}^{K_{6}}{u_{7, k}-u_{6, j}-\ft{i}{2} \over u_{7, k}-u_{6, j}+\ft{i}{2}}\prod_{j=1}^{K_{4}}{1-g^2/x_{7, k}x^{-}_{4, j} \over 1-g^2/x_{7, k}x^{+}_{4, j}}\, ,
\end{aligned}
\ee
where $x^{\pm} = x^{\pm}(u) = x(u\pm \ft{i}{2})$ and with $x(u)$ the deformed spectral parameter~\cite{BDS04}
\be
x = {1\over 2}(u+\sqrt{u^2-(2g)^2})\, , \qquad u = x+{g^2 \over x}\, ,
\ee
with asymptotics $x \sim u$ at $u\sim \infty$. The phase $\sigma(u, v)$ entering the middle-node equations for the main (momentum-carrying) roots $u_{4, k}$ is the BES/BHL dressing phase~\cite{BES06,BHL06}. Its expression is recalled in Appendix~\ref{AALE}. At one-loop, $g\rightarrow 0$, we have both $\sigma(u, v)\rightarrow 0$ and $x(u) = u$, and one recovers the Bethe ansatz equations for the $\mathfrak{psu}(2,2|4)$ Heisenberg (super) spin chain~\cite{BS03}. 

Once a solution to Eqs.~(\ref{ABAE}) is known, the anomalous dimension of the associated operator is found as~\cite{BDS04,BS05}
\be\label{Ebs}
\delta \Delta = 2g^2\sum_{j=1}^{K_{4}}\left({i \over x^{+}_{4, k}}-{i \over x^{-}_{4, k}}\right)\, ,
\ee
where only the main roots (the spin-chain momentum-carrying roots) enter. More generally, the total spin-chain higher conserved charges are given by
\be\label{HCbs}
Q_{r+1} = {1 \over r}\sum_{j=1}^{K_{4}}\left({i \over x^{+\, r}_{4}}-{i \over x^{-\, r}_{4}}\right)\, .
\ee

The global charges for a Bethe state with excitations numbers $(K_{1}, \ldots , K_{7})$ are given in~\cite{BS05}. The states we aim at considering are characterized by a large number of main roots $K_{4} \gg 1$, but with $K_{j} = O(1)$ for $j\neq 4$. All these states carry a large Lorentz (or total) spin $S = \ft{1}{2}\left(s_{1}+s_{2}\right) = K_{4} +O(1)$, where $(s_{1}, s_{2})$ are the Dynkin labels (twice the spins) for the $\mathfrak{su}(2)^2 \subset \mathfrak{su}(2,2)$ symmetry algebra~\cite{BS05}. The charges that are kept of order $O(1)$ in the large spin limit include the twist,
\be\label{twist}
\textrm{twist} = \Delta_{0} -S = L + K_{1} + K_{7}\, .
\ee
where $\Delta_{0}$ is the canonical dimension. Another such charge is the transverse $\mathfrak{so}(2)$ spin $= \ft{1}{2}(s_{2}-s_{1}) = \ft{1}{2}\left(K_{1}+K_{3}-K_{7}-K_{5}\right)$. Of course, all the $\mathfrak{su}(4)$ quantum numbers are of order $O(1)$ in this limit, and the GKP vacuum is a singlet under this symmetry. The most important observable is the GKP energy given by
\be\label{GKPenergy}
E = \Delta - S = L + K_{1} + K_{7} + \delta\Delta\, ,
\ee
where $\delta\Delta = O(g^2)$ is the anomalous dimension. Comparing~(\ref{twist}) and~(\ref{GKPenergy}), we see that the twist coincides with the energy of a state at tree level in the gauge theory. However, starting at one loop, the total energy is no longer a finite quantity in the large spin limit. The energy measured above the vacuum is however finite and reads
\be\label{Ephys}
E-E_{\textrm{vacuum}} = \textrm{twist}-2 + \delta\Delta-\delta\Delta_{\textrm{twist-two}}\, ,
\ee
where
\be\label{Evacuum}
E_{\textrm{vacuum}} = 2+\delta\Delta_{\textrm{twist-two}} = A_{2}(g)\left(\log{S}-\psi(1)\right) +2 +B_{2}(g) + O(\log{S}/S)\, ,
\ee
with $A_{2}(g) = 2\Gamma_{\textrm{cusp}}(g)$, $\psi(1) = -\gamma_{\textrm{{\tiny E}}}$ is the Euler-Mascheroni constant, and finally $B_{2}(g)=O(g^4)$ is the so-called (twist-two) virtual scaling function~\cite{FZ09}. The expression~(\ref{Evacuum}) is slightly cumbersome, but it will not be difficult to make use of it. This is because, in the large spin limit, $S\gg1$, the Bethe ansatz equations~(\ref{ABAE}) can be turned into a linear integral equation for the distribution density of roots~\cite{K95,BGK06,ES06,FRS07}. Thanks to the linearity of this equation, the separation of the various contributions to the energy in~(\ref{Ephys}) becomes straightforward.

\subsection{Momentum carriers}

The embedding of the momentum carriers, which were presented in Introduction, can be obtained from the literature~\cite{BGK06,FRS07,Bel07,FRZ09} on the large spin limit of the equations~(\ref{ABAE}).%
\footnote{Except for the fermions, whose embedding turns out to be subtle, as explained in a separate subsection.}
The scalars, for instance, are well-known to be associated with the dual solutions of the main roots equations~\cite{BGK06} and they appear as holes in the distribution of magnons mode numbers~\cite{FRS07}. The total number of holes in a state counts the total number of scalars on top of the GKP vacuum.%
\footnote{Note that we do not take into account the two large holes~\cite{BGK06,FRS07} which cannot change position in the lattice of mode numbers. They form part of the background.} One can easily introduce fermions $\Psi_{+}$ and $\bar{\Psi}_{+}$ by means of the roots $u_{3}$ and $u_{5}$, respectively. Indeed, adding a root $u_{3}$, for instance, performs the change of flavor $D_{+}Z\rightarrow \Psi_{+}$. This is not yet the end of the story for the embedding of fermions and there are subtleties that we discuss below. The embedding of the gauge fields and of the higher-twist excitations~(\ref{Ibs}) was analyzed in~\cite{FRZ09}. It was found that they come out as stacks of strings, that are complex solutions to~(\ref{ABAE}) composed of roots of various flavors~\cite{BKSZ05}. For instance, the gauge field $F_{+\perp}$, with $\mathfrak{so}(2)$ spin $1$, is embedded in the form of a $2$-string of roots $u_{3}$ centered around a real root $u_{2}$. Similarly for $\bar{F}_{+\perp}$ with the replacement $u_{3}\rightarrow u_{5}$ and $u_{2}\rightarrow u_{6}$. Of course the strings that compose the stacks are not `exact strings' at finite spin. But in the large spin limit, they are, as observed in~\cite{FRZ09} and as we would expect from analogy with thermodynamical limit of compact spin chain. The estimate for the departure from the exact string is expected to be exponentially suppressed with the (physical) length, i.e. associated to $\sim 1/S$ corrections. Now the gauge fields are just the first representants of two infinite series of higher-twist excitations~\cite{BBMS04}, associated in~(\ref{Ibs}) to partons of the type $D_{\perp}^{\ell-1}F_{+\perp}$. From the analysis of~\cite{FRZ09} we deduce that the twist-$\ell$ excitations is embedded as a stack of a $(\ell+1)$-string of roots $u_{3}$, a $\ell$-string of roots $u_{2}$, and a $(\ell-1)$-string of roots $u_{1}$. Similarly for $\bar{D}_{\perp}^{\ell-1}\bar{F}_{+\perp}$ after the wing exchange $(u_{1}, u_{2}, u_{3}) \rightarrow (u_{7}, u_{6}, u_{5})$.

As already mentioned in Introduction, we interpret the twist-$\ell$ excitation $\sim D_{\perp}^{\ell-1}F_{+\perp}$ as a bound state of $\ell$ gauge fields. Their form is however a bit puzzling and not immediately evocative of their interpretation. For a bound state of $\ell$ gauge fields, one would have probably expected something like $(F_{+\perp})^\ell$. There seems however to be no room in the ABA equations for accommodating the latter parton as an excitation on the large spin background. Indeed, looking for such an excitation, we would turn to the ($s=3/2$) $\mathfrak{sl}(2)$ spin chain $\textrm{Tr}\, D_{+}^S F_{+\perp}^{2+\ell}$ and ask for a solution describing the bound state $\sim (F_{+\perp})^{\ell}$ on the twist-two background -- here realized by $\sim \textrm{Tr}\, D_{+}^S F_{+\perp}^{2}$. The sought solution should then be given by a distribution of $\ell$ holes%
\footnote{Note that the holes introduced here are distinct, though very similar, to the ones associated to the embedding of the scalar excitations discussed before.} that is expected to form a string in the complex rapidity plane, in analogy with bound states of Bethe roots.%
\footnote{To form a bound state we need to find a way to store less energy in the state than allowed by a configuration of $\ell$ real holes -- hence the need to put some holes in the complex plane.}. But no such complex solutions exist for the non-compact spin chain~\cite{K95}. The operator $D^{\ell-1}_{\perp}F_{+\perp}$, which is not contained in the latter subsector but in a larger $\mathfrak{su}(2,1)$ subsector that supports string solutions, appears as a substitute for the `bound state' $(F_{+\perp})^{\ell}$ in the spin-chain description.

At the same time, how should we understand the claim that the operator $D^{\ell-1}_{\perp}F_{+\perp}$ is a stable excitation over the large spin background. Naively, if we form a spin chain out of such operators, one would certainly expect that the covariant derivatives $D_{\perp}$ would hop from one site $F_{+\perp}$ to another, suggesting that the numbers of bound states are not separately conserved. This is not very welcome as, if the large spin limit is associated to an integrable model of sort, the number of excitations with a given mass (and quantum numbers) should be well defined. The mixing alluded to before certainly occurs, but the question is whether it persists at large spin and for the asymptotic part of the wave function, when the excitations are separated from one another by a large number of background derivatives $D_{+}$. Without the explicit form of the large spin wave function for such states, it is difficult to rigorously address this issue. We will give nevertheless a somewhat dynamical argument in favor of this interpretation of the twist-$\ell$ excitation as a bound state of gauge fields, later on. It will be based on the fusion identities of the type
\be\label{FIintro}
E_{\ell}(u) =  \sum_{j=0}^{\ell-1}E_{\textrm{gf}}(u-i(\ell-1)/2+ij)\, ,
\ee 
and similarly for the momentum, which are natural for bound states. These identities will be observed to hold only when $u^2< (2g)^2$ and thus will not reveal themselves at the level of the weak coupling expansion, which always requires $u^2 > (2g)^2$. The obstructions met in extending these relations at weak coupling are tied to the presence of branch points in the complex $u$-plane, that makes the shifts along the imaginary $u$-axis in~(\ref{FIintro}) inequivalent if performed inside or outside the interval $u^2 < g^2$. This feature might explain the complications mentioned above in the realization of the bound states of gauge fields in the spin-chain picture, since the gauge-field rapidities in a given bound state do not all live on the same rapidity sheet for $u^2>(2g)^2$. 

\subsection{Isotopic roots: one loop versus all loops}

So far we have been considering the roots $u_{1}, u_{2}, u_{6}, u_{7}$, as elements in the construction of stacks. Now we would like to examine them separately. We naively expect that large-spin solutions to the Bethe ansatz equations~(\ref{ABAE}) with solitary (real) roots $u_{1}, u_{2}, u_{6}, u_{7}$, do exist, and so we need to interpret them. We start with the roots $u_{2}, u_{6}$, whose meaning is quite clear.

The roots $u_{2}, u_{6},$ play exactly  the same role for the GKP operators as they do for the BMN ones. Namely, they are associated with the $SU(2)\times SU(2) \subset SU(4)$ (residual) symmetry  group that is common to both backgrounds. They are isotopic roots that do not carry energy nor momentum and implement the latter two $SU(2)$ symmetries. But the residual symmetry of the GKP vacuum is $SU(4)$ not $SU(2)\times SU(2)$. We expect therefore a third type of isotopic root. The missing root is embedded in the form of a stack of roots $u_{3}, u_{4}, u_{5}$. More precisely, the stack is made out of one $2$-string of main roots, $u_{4, \pm} = u_{\textrm{b}}\pm i/2$, with one root $u_{3}$ and one root $u_{5}$ on top of each other, $ u_{3} =  u_{5} = u_{\textrm{b}}$. It is not obvious that this stack does not carry energy nor momentum. This is true however to all loops, as shown in Appendix~\ref{AISR}. What happens is that the bare energy, carried by the $2$-string, is exactly compensated by the induced energy, which measures the shift in the energy stored in the large-spin continuum of real main roots due to the presence of the stack. Moreover, the large-spin equations for these stacks take exactly the same form to all loops 
\be\label{eqir}
\prod_{j=1}^{K_{\textrm{h}}}{u_{\textrm{b}, k}-u_{\textrm{h}, j}+\ft{i}{2} \over u_{\textrm{b}, k}-u_{\textrm{h}, j}-\ft{i}{2}} = \prod_{k\neq j}^{K_{\textrm{b}}}{u_{\textrm{b}, k}-u_{\textrm{b}, j}+i \over u_{\textrm{b}, k}-u_{\textrm{b}, j}-i}\prod_{j=1}^{K'_{2}}{u_{\textrm{b}, k}-u_{2, j}-\ft{i}{2} \over u_{\textrm{b}, k}-u_{2, j}+\ft{i}{2}}\prod_{j=1}^{K'_{6}}{u_{\textrm{b}, k}-u_{6, j}-\ft{i}{2} \over u_{\textrm{b}, k}-u_{6, j}+\ft{i}{2}}\, .
\ee
as shown in Appendix~\ref{AISR}. Here $u_{\textrm{h}, j}$ is a hole (scalar) rapidity, and $K'_{2}, K'_{6},$ are the numbers of solitary roots $u_{2}, u_{6}$. There is no explicit dependence on the spin in these equation, and so there is no momentum for a root $u_{\textrm{b}}$. We recognize in~(\ref{eqir}) the central-node equations for the inhomogeneous $SO(6)$ Heisenberg spin chain, with length equal to the number of scalars and with the scalars' rapidities entering as inhomogeneities.%
\footnote{Note that the scalars' rapidities are the only inhomogeneities in the central-node equations~(\ref{eqir}), in agreement with the fact that the scalars are the only excitations in the $\bf{6}$ of $SO(6)$. Fermions, which are in the $\bf{4}$ or $\bar{\bf{4}}$, couple to the equations for roots $u_{2}$ and $u_{6}$, and gauge fields and bound states, which are neutral, do not couple to any of the $SO(6)$ isotopic roots.} We conclude that the stack $u_{\textrm{b}}$ effectively restores the $SU(2)$ symmetry, initially broken by the BMN vacuum.%
\footnote{Note that at one loop, we could as well work over the pseudo-vacuum $\textrm{Tr}\, F_{+\perp}^L$, associated to the choice of the distinguished Dynkin diagram (the Beast of~\cite{BS05}). The $SU(4)$ symmetry would be manifest from the very beginning. It is not known however how to do it at higher loops. Similarly the choice of the pseudo-vacuum $\textrm{Tr}\, \Psi_{+}^L$ would keep a $SU(3)\subset SU(4)$ symmetry subgroup manifest.}
In essence, the mechanism that restores the $SU(2)$ symmetry here is identical to the one that operates in the thermodynamical limit of the anti-ferromagnetic $SU(2)$ spin chain. The only difference is the necessity here to have fermionic roots to support the 2-strings.

The $SU(4)$ roots $u_{2}, u_{6}$, and $u_{\textrm{b}}$ are isotopic to all loops.  But they do not form the complete set of isotopic roots at one loop, however. Indeed, at one loop, the fermionic roots $u_{1}, u_{7},$ are also isotopic. They do not carry momentum and their energy is canonical and equal to $1$, with no one-loop correction. The reason being that they do not couple to the middle-node equations, at this order, and so cannot induce contributions to the energy. These extra isotopic roots enhance the symmetry from $SU(4)$ to $SL(2|4)$. Under this symmetry, each momentum carriers give rise to an infinite dimensional multiplet, with the `infinite direction' coming from the possibility to extract an arbitrary number of covariant derivatives $D_{-}$. At higher loops, however, roots $u_{1}, u_{7}$, couple to the main roots, see Eqs.~(\ref{ABAE}), and thus acquire energy and momentum.  It raises the question of their role in the spectum of excitations. As we shall see they are in fact fermions, which are trapped in a small momentum domain $p=O(g^2)$ at weak coupling.

\subsection{Fermionic subtleties}

The fact that the roots $u_{1}, u_{7},$ are fermions is supported by the observation that these roots can actually act on a covariant derivative $D_{+}$ at higher loops, due to the supersymmetry transformations~\cite{Beisert04}
\be
\delta_{1} D_{+} \sim g \Psi_{+}\, , \qquad \delta_{7} D_{+} \sim g \bar{\Psi}_{+}\, .
\ee
Accordingly, the one-loop $SL(2|4)$ descendants, obtained by acting on the momentum carriers with the roots $u_{1}, u_{7}$, presumably decay into fermions at higher loops.

\begin{figure}[h]%
\setlength{\unitlength}{0.14in}
\centering
\begin{picture}(28,20)%
\put(-4,0){\insertfig{7}{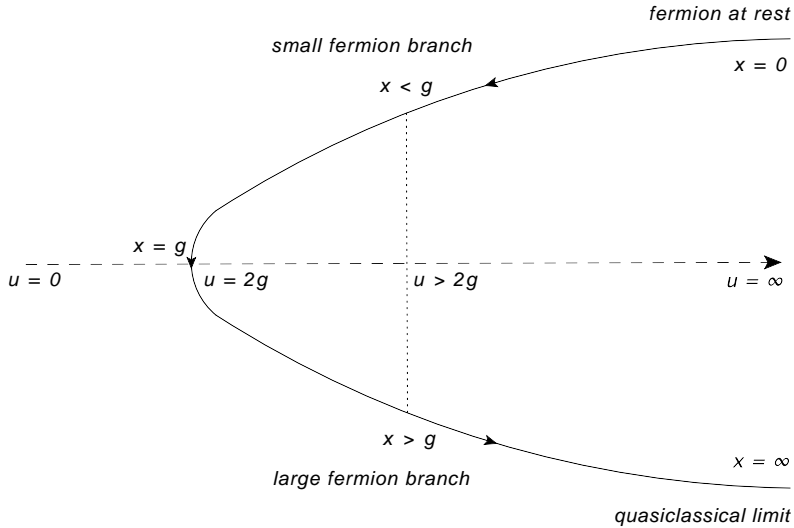}}
\end{picture}%
\caption{The two fermion branches.}\label{fbranch}
\end{figure}

At the kinematical level, it happens that the physical rapidity is not the Bethe rapidity but its Zhukowsky map $x$. The $x$-plane requires to glue together two $u$-planes, and one finds two expressions for energy and momentum of a fermion with rapidity $u$, one coming from $u_{1}$ and the other one from $u_{3}$, idem with $u_{7}$ and $u_{5}$. For illustration, we find
\be
E_{\textrm{sf}}(u) = 1+ {4\zeta_{3}g^4\over u^2} + O(g^6)\, , \qquad p_{\textrm{sf}}(u) = {2g^2\over u} + O(g^4)\, ,
\ee
for a small fermion with $x^2 < g^2$ and 
\be
E_{\textrm{lf}}(u) = 1+ 2g^2\left(\psi(1+iu)+\psi(1-iu)-2\psi(1)\right) + O(g^4)\, , \qquad p_{\textrm{lf}}(u) = 2u + O(g^2)\, ,
\ee
for a large fermion with $x^2>g^2$. These two cases are realized at weak coupling by the roots $u_1$ and $u_3$, respectively, with $x \sim g^2/u_1$ in the first case and $x \sim u_3$ in the second one. The two are however connected with each other, see Fig.~\ref{fbranch}. For instance, in both cases the dispersion relation reads
\be
E(p) = 1+ 2g^4\left(\psi(1+ip/2)+\psi(1-ip/2)-2\psi(1)\right) + O(g^4)\, ,
\ee
but while $p=O(1)$ at $u$ fixed and $g\sim 0$ for a large fermion, it is small $p=O(g^2)$ for a small fermion under the same assumption. In particular, a fermion can be found at rest if it lies at $x=0$. Then its mass is exactly one
\be
m_{\textrm{f}} = E_{\textrm{f}}(p=0) = E_{\textrm{f}}(x=0)\, ,
\ee
due to the special property of the point $x=0$ in the all loop asymptotic equations~\cite{BS05}. At one loop this point can be identified with $u=0$, but at one-loop only. 

For some dynamical reason, tied to the coupling with the $SU(2)\times SU(2)$ roots~\cite{BS05}, some fermions get trapped at weak coupling in the small momentum domain and then behave essentially like if they were isotopic. 
The separation between roots $u_{3}, u_{5},$ and $u_{1}, u_{7},$ seems therefore to be a dynamical necessity at weak coupling. However, for our purpose of constructing dispersion relation at finite value of the coupling, it is quite artificial and it is better to restore the unity between fermions. This can be done by using the dynamic transformations of~\cite{BS05}. The price to pay is that we need to change the length $L$ in~(\ref{ABAE}) as
\be\label{EffLength}
L \rightarrow L+K'_{1}+K'_{7}\, ,
\ee
where $K'_{1}$ and $K'_{7}$ are the numbers of (solitary) roots $u_{1}, u_{7},$ transformed into roots $u_{3}, u_{5}$. The quantity above counts the number of length-one partons in a given states and it is equal to $2+M$ where $M$ is the total number of momentum carriers and $2$ comes from the two vacuum-building partons.

These remarks about the fermions complete our discussion of excitations over the GKP vacuum. We are still lacking a proof of the completeness of the basis presented before. However, it fits remarkably well with the analysis of~\cite{Volin10}, which deals with the string hypothesis for generic super spin chain, and with the independent analysis of the spectrum around the GKP string performed in~\cite{AGMSV10}.

\section{All-loop integral equation}\label{ALA}

Assuming the pattern of excitations presented in Introduction, and the embedding discussed in Section~\ref{EO}, the next step is to derive the dispersion relations. This is done by analyzing the solution to the middle-node equations in~(\ref{ABAE}) for a large number $K_{4}$ of main roots and for a generic set of excitations. This task is greatly simplified by the large spin limit~\cite{K95,BGK06} in which the latter equations can be turned into a linear integral equation for the distribution density of main roots~\cite{ES06,FRS07}. Our analysis, in this section, follows the methodology of~\cite{ES06,BES06,FRS07} to constructing this equation to all loops. We perform a few generalizations, as we need to treat both the symmetric and antisymmetric part of the distribution density, and simultaneously include all the excitations encountered previously. Incidentally, we observe that the symmetric and antisymmetric part totally decouple from one another, leading to integral equations of the type already considered in the literature~\cite{ES06,BES06,FRS07,FZ09,FRZ09}. Applying to them the tools developed in~\cite{Benna06,AABEK07,BKK07,KSV08,BK08,BK09} for dealing with these equations, we derive dispersion relations for the various excitations. Our construction does not provide explicit analytical expressions for the all-loop dispersion relations. But it gives compact expressions for them, parameterized in terms of the solution to the BES equation, which controls the vacuum distribution of roots to leading order at large spin. A sample of the literature which is relevant for the analysis presented below includes~\cite{K95,BGK06,ES06,FRS07,FRZ09} and the reviews~\cite{R09}. 

Some preliminary remarks are in order. In the following, we assume that the dynamic transformations~\cite{BS05} have been performed on the solitary roots $u_{1}, u_{7},$ to unit the fermionic roots. These transformations lead to the replacement of the spin-chain length by
\be
L\rightarrow L+K'_{1}+K'_{7} = 2+M\, , \qquad M = K_{\textrm{h}} + K_{\textrm{f}} + K_{\bar{\textrm{f}}} + \sum_{\ell=1}^{\infty}K_{\ell} + \sum_{\bar{\ell}=1}^{\infty}K_{\bar{\ell}}\, ,  
\ee
where $K'_{1}, K'_{7},$ are the total number of solitary roots $u_{1}, u_{7},$ respectively, $M$ is the total number of momentum carriers, and $K_{\star}$ is the total number of scalars ($\star = \textrm{h}$), fermions ($\star = \textrm{f}$), ..... There is actually no need to separate between excitations carrying positive or negative $\mathfrak{so}(2)$ spin. Since the main roots do not distinguish between them, they are degenerate, and it is enough to consider $\star = \textrm{h}, \textrm{f}, \ell$. Finally, the particular case of the $SU(2)$ isotopic root will not be treated below. Its analysis is deferred to Appendix~\ref{AISR}, where it is shown that it does not carry energy/momentum, nor any of the spin-chain conserved charges, to all loops.%
\footnote{There is a slight contribution to the total spin-chain momentum $\e^{iP}$ in the form of the factor $(-1)^{K_{\textrm{b}}}$, where $K_{\textrm{b}}$ is the total number of $SU(2)$ isotopic roots.}

\subsection{Middle-node equations}

In the presence of a generic set of excitations, the middle-node equations in~(\ref{ABAE}) can be written
\be\label{MNeq}
1 = \left({x^{-}_{4,k} \over x^{+}_{4,k}}\right)^{2+M}\prod_{j\neq k}^{K_{4}}S^{\circ}_{44}(u_{4, k}, u_{4, j})\sigma^{2}(u_{4, k}, u_{4, j})\prod_{\star \neq \textrm{h}}^{M}S_{4\star}(u_{4, k}, u_{\star})\, ,
\ee
where all the $K_{4}$ main roots are real.%
\footnote{In our analysis, complex roots $u_{4}$ appear in the form of 2-strings, which are part of the $SU(2)$ isotopic roots, considered separately in Appendix~\ref{AISR}.} For later convenience, we have factorized the scattering phase for the self-interaction of main roots into its undressed component, denoted $S^{\circ}_{44}(u_{4, k}, u_{4, j})$, and the dressing phase, $\sigma^{2}(u_{4, k}, u_{4, j})$. The symbol $\star$ denotes a generic excitation, which can be a scalar (hole) $\star = \textrm{h}$, a fermion $\star = \textrm{f}$, a gauge field $\star = \ell = 1$ or a bound state thereof $\star = \ell > 1$. The total number of excitations within the state is fixed equal to $M$ and the phase $S_{4\star}(u_{4}, u_{\star})$ stands for the scattering phase between a main root $u_{4}$ and an excitation $\star$ with rapidity $u_{\star}$. The product over the excitations is restricted to $\star \neq \textrm{h}$ because there is no scattering phase for scalars at this time. They are indeed dual to the main roots~\cite{BGK06} and appear as holes in their distribution of mode numbers~\cite{FRS07}. A scattering phase for them will be induced later, during the process of taking the large spin (continuum) limit. For any of the other excitations, the scattering phase with a main root is obtained directly from the middle-node equations in~(\ref{ABAE}), taking into account the embedding of each excitation into the Bethe ansatz equations. For gauge field ($\ell=1$) and bound states ($\ell>1$), which appear as stacks of various strings, they are obtained by fusion of the relevant elementary scattering phases (see Appendix~\ref{AALS} for details).

Now we would like to analyze the equations~(\ref{MNeq}) at large spin, $S = K_{4} + O(1) \gg 1$. Following the lines of~\cite{ES06,FRS07}, we cast them into the form
\be\label{nZeq}
2\pi n_{k} = Z(u_{k})\, , 
\ee
which follows from taking the logarithm of~(\ref{MNeq}), after multiplying both sides by the phase $(-1)^{K_{4}+K_{\textrm{h}}-1}$ leading to a more convenient choice for the branch of the logarithm. Then the (fermionic) mode numbers $n_{k}$, entering~(\ref{nZeq}), belong to a finite lattice $\mathfrak{S}$ of integers or half-integers -- depending on the parity of $K_{4}+K_{\textrm{h}}-1$ -- and the so-called counting function $Z(u)$ is defined by 
\be\label{CFhbis}
Z(u) \equiv Z(0) +2\pi \int_{0}^{u}dv\, \rho(v) = -i\sum_{j=1}^{K_{4}}\log{(-S^{\circ}_{44}(u, u_{4,j}))} + \ldots\, ,
\ee
where the dots includes contributions from dressing, ...., that we will restore later. Meanwhile we have introduced the density distribution $\rho(u)$ as the derivative, with respect to the rapidity $u$, of the counting function. The lattice of mode numbers $\mathfrak{S}$ is almost filled up by the magnons up to a few holes~\cite{FRS07}. These holes are associated to dual solutions~\cite{BGK06} of the equations~(\ref{nZeq}) and they can be put in correspondence with a set of rapidities by using~(\ref{nZeq}). Namely, a hole occupying the mode number $n_{\textrm{h}}$ in $\mathfrak{S}$ is given the rapidity $u_{\textrm{h}} = Z^{-1}(2\pi n_{\textrm{h}})$. We are assuming here that the counting function $Z(u)$ can be inverted, at least for real rapidity $u$, and thus that the mapping between rapidities and mode numbers is one-to-one. This is definitely the case in the $\mathfrak{sl}(2)$ sector, which is obtained by imposing the restriction that the only momentum carriers are the holes (scalars), $M = K_{\textrm{h}}$ with $K_{\textrm{h}}$ the total number of holes. This is not obviously the case here, but, as we shall see below, this is a quite reasonable assumption at large spin. We will therefore push further the analogy with the $\mathfrak{sl}(2)$ sector and now take the large spin limit.

The main simplification at large spin limit is that the equations~(\ref{nZeq}) can be turned into an integral equation for the distribution density $\rho(u)$. The first step in the derivation of this equation is to approximate the sum over the main roots in~(\ref{CFhbis}) by an integral weighted by the density. To this end, we extend the latter sum to the full lattice of mode numbers, subtracting simultaneously the holes' contribution,
\be\label{EMsa}
-i\sum_{j=1}^{K_{4}}\log{(-S^{\circ}_{44}(u, u_{4,j}))} = -i\sum_{n\in \mathfrak{S}}\log{(-S^{\circ}_{44}(u, u(n)))}+i\sum_{j=1}^{K_{\textrm{h}}}\log{(-S^{\circ}_{44}(u, u_{\textrm{h}, j}))}\, ,
\ee
where $u(n) = Z^{-1}(2\pi n)$ is the rapidity, either of a magnon or a hole, corresponding to the mode number $n$ in $\mathfrak{S}$. Now in the large spin limit, the distribution of roots $u(n)$ is dense~\cite{K95,BGK06}, the typical distance between the roots of order $O(S^0)$ being $\sim 1/\log{S}$. It allows us to apply the Euler-MacLaurin summation formula to evaluate the first sum in the right-hand side of~(\ref{EMsa}), giving  
\be\label{EMsb}
-i\sum_{j=1}^{K_{4}}\log{(-S^{\circ}_{44}(u, u_{4,j}))} =  -i\int dv \, \rho(v)\log{(-S^{\circ}_{44}(u, v))}+i\sum_{j=1}^{K_{\textrm{h}}}\log{(-S^{\circ}_{44}(u, u_{\textrm{h}, j}))}+\ldots \, ,
\ee
where dots stand for $1/S$-suppressed corrections at large spin. Note that we have assumed that the distribution density $\rho(u)$, as defined in~(\ref{CFhbis}) is positive over the full domain of integration. This property, which is related to the invertibility of the counting function alluded before, is not guaranteed, but seems quite realistic, in the large spin limit. The reason being that the distribution density $\rho(u)$ receives a dominant contribution from the vacuum (twist-two) distribution of roots. The latter scales like~\cite{K95,BGK06,ES06}
\be\label{ScalingVac}
\rho_{\textrm{vacuum}}(u) = {2\over \pi}\log{S} + O(\log^0{S})\, ,
\ee
at large spin, for a rapidity $u=O(S^0)$, which is the regime of interest here. When holes, or other excitations, are introduced with rapidities of order $O(S^{0})$, they generate corrections to the distribution density $\rho(u)$ which are of order $O(\log^{0}{S})$, and thus subleading as compared to~(\ref{ScalingVac}). The density, as defined in~(\ref{CFhbis}), should therefore be positive, at least for $u=O(S^0)$.

It remains to specify the domain of integration in~(\ref{EMsb}). For our purpose, it can be considered to be the full real line, but it requires some comments. Indeed, it is known~\cite{K95} that the twist-two distribution is supported on the symmetric interval $u^2 < (S/2)^2$, at large spin. It does not immediately implies, however, that we can extend the domain of integration in~(\ref{EMsb}) to $\pm \infty$, even if we are interested in constructing the density $\rho(u)$ for $u=O(S^0)$, which is our ambition here. The reason is that the vacuum density scales like
\be\label{BCtt}
\rho_{\textrm{vacuum}}(u) = {2\over \pi} \log{\left(S/u\right)} + O(1/u)\, ,
\ee
at large rapidity. It is so to all loops~\cite{ES06}, as long as $S\gg 1$ and $S\gg g$, which are both always implicitely assumed here. Then one needs to proceed with care concerning the domain of integration. But, fortunately, the situation is different for corrections to the density sourced by the excitations. Indeed, the latters are suppressed at large rapidity, if the excitations carry rapidities of order $O(S^0)$. It follows that in our case, we can effectively extend the domain of integration to the full real line. We note finally that this implies that the large rapidity regime of the density $\rho(u)$ is controlled by the vacuum contribution and scales like~(\ref{BCtt}). One can think therefore of the behavior~(\ref{BCtt}) as a boundary condition for our problem. One can change this boundary condition by sending a few excitations (holes, ...) to infinity, or more precisely by giving them rapidities of order $O(S)$. The boundary conditon~(\ref{BCtt}) then gets replaced by~\cite{K95}
\be
\rho(u) = {(2+n)\over \pi} \log{\left(S/u\right)} + C_{n} + O(\bar{u})\, ,
\ee 
where $n$ is the number of large rapidity excitations and where $C_{n}$ is some constant. This corresponds to working around higher-excited trajectories in the spectrum of large spin scaling dimensions~\cite{BGK03,BGK06}. A recent literature on this subject with analysis performed at both weak and strong coupling, and with relation to spiky strings~\cite{Kruczenski04}, includes~\cite{BKP08,D8Spiky,FKT09}.

The next step in the derivation of the integral equation for the density $\rho(u)$ is to differentiate with respect to $u$ in~(\ref{CFhbis}), using~(\ref{EMsb}) to evaluate the right-hand side of the last equalities in~(\ref{CFhbis}). This step is performed in the next subsection after restoring contributions associated to dressing phase, .... in~(\ref{CFhbis}).

\subsection{Large spin integral equation}

The complete counting function, deduced from the equations~(\ref{MNeq}), can be decomposed as
\be\label{AlZ}
Z(u) = Z^{\circ}(u) + (2+M) Z_{\textrm{length}}(u) + Z_{\textrm{dressing}}(u) + Z_{\textrm{excitations}}(u)\, ,
\ee
where the first term, $Z^{\circ}(u)$, coincides with the one previously considered in~(\ref{CFhbis}). It is the term that generates the main kernel of the (large spin) integral equation~\cite{ES06,BES06}. Its expression follows from the relation~\cite{ES06}
\be\label{Zmb}
-i\log{\left(-S^{\circ}_{44}(u, v)\right)} = -i\log{\left({1+iu-iv \over 1-iu+iv}\right)} -2i\log{\left({1-g^2/x^{+}y^{-} \over 1-g^2/x^{-}y^{+}}\right)}\, ,
\ee
where $x(u) = (u+\sqrt{u^2-(2g)^2})/2, y(v) = (v+\sqrt{v^2-(2g)^2})/2$ and $x^{\pm} = x(u\pm i/2)$ (similarly for $y^{\pm}$). The second term in~(\ref{AlZ}) takes care of the contribution proportional to the spin-chain length,
\be\label{Zl}
Z_{\textrm{length}}(u) = -i\log{\left(-{x^{-}(u) \over x^{+}(u)}\right)}\, .
\ee
Concerning the contribution originating from the dressing phase, it is better not to expand it for the time being. So we simply write
\be\label{Zd}
Z_{\textrm{dressing}}(u) =   2\sum_{j=1}^{K_{4}}\theta(u, u_{4,j})\, ,
\ee
where $\theta(u,v)$ is given in terms of the BES/BHL dressing phase $\sigma(u,v)$~\cite{BES06,BHL06} as $\theta(u,v) = -i\log{\sigma(u,v)}\, .$ Finally, we have the contributions from the various excitations
\be\label{Zi}
Z_{\textrm{excitations}}(u) = - i\sum_{\star \neq \textrm{h}}^{M}\log{\left(-S_{4\star}(u, u_{\star})\right)}\, .
\ee

The integral equation for the density is linear and, as we have said before, we are only interested in the contributions sourced by the excitations (holes, ...). We therefore decompose $\rho(u)$ into two parts%
\footnote{The minus signs are conventional~\cite{ES06,FRS07}.}
\be\label{Dec}
\rho(u) = \rho_{\textrm{vacuum}}(u) -\sigma(u)-\tilde{\sigma}(u)\, ,
\ee
where the first term stands for the twist-two (vacuum) distribution of roots. It is controlled to leading order at large $S$ by the BES equation~\cite{ES06,BES06} and to next-to-leading order by a similar equation~\cite{FZ09}. The other terms in~(\ref{Dec}) account for the deformation of the vacuum distribution induced by the excitations. The two densities $\sigma(u)$ and $\tilde{\sigma}(u)$ are univocally distinguished by their parity in $u$. The former is even, $\sigma(-u) = \sigma(u)$, while the latter is odd, $\tilde{\sigma}(-u) = -\tilde{\sigma}(u)$. This separation anticipates on the fact that the equation we will have to consider is not of difference form. Now, implementing the decomposition~(\ref{Dec}) consistently requires to remove the terms in the counting function~(\ref{AlZ}) that are captured by the vacuum distribution density. This is done by performing the replacement $2+M \rightarrow M$ in~(\ref{AlZ}) and by subtracting the vacuum contribution to the dressing-phase correction $Z_{\textrm{dressing}}(u)$. The latter quantity admits an expansion over the spin-chain higher conserved charges~\cite{BS05,BHL06,BES06}. Substracting the vacuum contribution amounts to evaluating these charges above the vacuum,
\be\label{DecChar}
Q_{r} = q^{\textrm{vacuum}}_{r} + q_{r}\, ,
\ee
where the charges $q_{r}$ are the total charges induced by the excitations. In the large spin limit, they are found as
\be\label{CharAV}
q_{r} = -\int du\, (\sigma(u)+\tilde{\sigma}(u))q_{r}(u) -\sum_{j=1}^{K_{\textrm{h}}}q_{r}(u_{\textrm{h}, j})\, , 
\ee
where $q_{r}(u)$ is the charge carried by a magnon with rapidities $u$.

The equation for the distribution densities $\sigma(u), \tilde{\sigma}(u)$, follows from differentiating~(\ref{CFhbis}) with respect to $u$, applying~(\ref{AlZ}) and~(\ref{EMsb}). Keeping only the relevant contributions for the analysis of the excitations (see paragraph above), we get
\be\label{PFIE}
0 = 2\pi \sigma(u) + 2\pi \tilde{\sigma}(u) + \partial_{u}Z(u)\, ,
\ee
where $Z(u)$ is given by~(\ref{AlZ}) with the modifications explained before. The contribution to~(\ref{PFIE}) from $Z^{\circ}(u)$ splits into two pieces, after using~(\ref{EMsb}),
\be\label{ZmEx}
\partial_{u}Z^{\circ}(u)\big|_{\textrm{above vacuum}} = i\int dv\, \partial_{u}\log{\left(-S^{\circ}_{44}(u, v)\right)}\left(\sigma(v)+\tilde{\sigma}(v)\right) + i\sum_{j=1}^{K_{\textrm{h}}}\partial_{u}\log{\left(-S^{\circ}_{44}(u, u_{\textrm{h}, j})\right)}\, .
\ee
The first term generates the so-called main kernel~\cite{BES06,ES06} for the integral equation while the second one is the source term associated with the holes rapidities. Now, remarkably enough, the parity odd and even densities decouple from each other at the level of the kernel. Namely, we have
\be
\begin{aligned}\label{KenE}
i\int dv\, \partial_{u}\log{\left(-S^{\circ}_{44}(u, v)\right)}\left(\sigma(v)+\tilde{\sigma}(v)\right)\big|_{\textrm{even in $u$}} &= \mathcal{K}^{\circ}\star \sigma (u)\, , \\
i\int dv\, \partial_{u}\log{\left(-S^{\circ}_{44}(u, v)\right)}\left(\sigma(v)+\tilde{\sigma}(v)\right)\big|_{\textrm{odd in $u$}} &= \tilde{\mathcal{K}}^{\circ}\star \tilde{\sigma} (u)\, ,
\end{aligned}
\ee
where explicit expressions for the two kernels $\mathcal{K}^{\circ}, \tilde{\mathcal{K}}^{\circ},$ will be given later. It is not clear to us whether this property should have been expected on general grounds. One could still expect a coupling at the level of the dressing-phase correction $\partial_{u}Z_{\textrm{dressing}}(u)$, but here again there is none. The even part depends on the even charges and the odd part on the odd ones. So the equation~(\ref{PFIE}) splits into two integral equations that read
\be\label{LSeqf}
\begin{aligned}
0 = 2\pi \sigma(u) + \mathcal{K}^{\circ}\star \sigma (u) +\mathcal{I}(u)\, , \\
0 = 2\pi \tilde{\sigma}(u) + \tilde{\mathcal{K}}^{\circ}\star \tilde{\sigma} (u) +\tilde{\mathcal{I}}(u)\, ,
\end{aligned}
\ee
where the inhomogeneous contributions $\mathcal{I}(u), \tilde{\mathcal{I}}(u)$, are respectively even and odd in $u$. Note that our terminology here is slightly abusive since the inhomogeneous terms contain the contributions from the dressing phase, that will correct the kernel of the integral equation later on. It is convenient nevertheless to keep them enclosed in $\mathcal{I}(u), \tilde{\mathcal{I}}(u)$, together with the genuine inhomogeneous contributions, at this moment.

We are now in position to detail each term in the equation~(\ref{LSeqf}). We start with the kernels. From their definitions~(\ref{KenE}), and taking into account the expression~(\ref{Zmb}) and formulae in Appendix~\ref{AALE}, we get
\be
\begin{aligned}
\mathcal{K}^{\circ}\star \sigma (u) &= \,  -2\int_{0}^{\infty}dt\, \cos{(ut)}\e^{-t}\int dv\, \sigma(v)\cos{(vt)} \\
&+2(2g)^2\int_{0}^{\infty}dt\, \cos{(ut)}\e^{-t/2}\int_{0}^{\infty}ds\, t K(2gt, 2gs)\e^{-s/2}\int dv\, \sigma(v)\cos{(vs)}\, , \\
\tilde{\mathcal{K}}^{\circ}\star \tilde{\sigma} (u) &= \,  -2\int_{0}^{\infty}dt\, \sin{(ut)}\e^{-t}\int dv\, \tilde{\sigma}(v)\sin{(vt)} \\
&+2(2g)^2\int_{0}^{\infty}dt\, \sin{(ut)}\e^{-t/2}\int_{0}^{\infty}ds\, t K(2gt, 2gs)\e^{-s/2}\int dv\, \tilde{\sigma}(v)\sin{(vs)}\, ,
\end{aligned}
\ee
where $K(t,s)$ is the symmetric, $K(t,s) = K(s,t)$, kernel density of~\cite{ES06,BES06}. It can be decomposed into even and odd parts in $t$ as
\be\label{Kern}
K(t, s) = K_{+}(t, s) + K_{-}(t, s)\, ,
\ee
and the two components $K_{\pm}(t, s)$ can be found to admit the representations~\cite{ES06,BES06}
\be\label{Kernpm}
\begin{aligned}
K_{+}(t, s) &= {2 \over ts}\sum_{n\ge 1}(2n-1)J_{2n-1}(t)J_{2n-1}(s)\, , \\
K_{-}(t, s) &= {2 \over ts}\sum_{n\ge 1}(2n)J_{2n}(t)J_{2n}(s)\, ,
\end{aligned}
\ee
where $J_{2n-1}(t), J_{2n}(t),$ stand for the parity odd, even, Bessel's functions, respectively.

The inhomogeneous terms in~(\ref{LSeqf}) can be written as
\be
\begin{aligned}
\mathcal{I}(u) &= \sum_{\star}^{M}\mathcal{I}_{\star}(u, u_{\star}) +  \mathcal{I}_{\textrm{dressing}}(u) \, ,\\
\tilde{\mathcal{I}}(u) &= \sum_{\star}^{M}\tilde{\mathcal{I}}_{\star}(u, u_{\star}) +  \tilde{\mathcal{I}}_{\textrm{dressing}}(u) \, ,
\end{aligned}
\ee
where the sums run over all the $M$ excitations above the vacuum, including the holes. For any excitation with rapidity $u_{\star}$ we have
\be\label{ITE}
\begin{aligned}
\mathcal{I}_{\star}(u, u_{\star}) &= - i\partial_{u}\log{\left(-S_{4\star}(u, u_{\star})\right)}\big|_{\textrm{even in $u$}} -i\partial_{u}\log{\left(-{x^{-}(u) \over x^{+}(u)}\right)}\, , \\
\tilde{\mathcal{I}}_{\star}(u, u_{\star}) &= - i\partial_{u}\log{\left(-S_{4\star}(u, u_{\star})\right)}\big|_{\textrm{odd in $u$}}\, ,
\end{aligned}
\ee
where for a hole, $\star = \textrm{h}$, the scattering phase has been induced in Eq.~(\ref{ZmEx}),
\be\label{ITEbis}
- i\partial_{u}\log{\left(-S_{4\textrm{h}}(u, u_{\textrm{h}})\right)} = + i \partial_{u}\log{\left(-S^{\circ}_{44}(u, u_{\textrm{h}})\right)}\, .
\ee
Note that we have distributed the contribution coming from $\partial_{u}Z_{\textrm{length}}(u)$ among the $M$ excitations. To proceed further with the inhomogeneous terms~(\ref{ITE}) we need explicit expressions for the scattering phases between main root and excitations. Unfortunately, we did not find a closed formula that would encompass all these contributions, and each type of excitation has to be treated more or less separately. However, they have in common that they can all be written as
\be
\begin{aligned}
\mathcal{I}_{\star}(u, u_{\star}) &= -2\int_{0}^{\infty}dt\, \cos{(ut)}\e^{-t/2}\pi_{\star}(t, u_{\star})\, , \\
\tilde{\mathcal{I}}_{\star}(u, u_{\star}) &= -2\int_{0}^{\infty}dt\, \sin{(ut)}\e^{-t/2}\tilde{\pi}_{\star}(t, u_{\star}) \, .
\end{aligned}
\ee
The expressions for the functions $\pi_{\star}(t, u_{\star}), \tilde{\pi}_{\star}(t, u_{\star}),$ are given in Appendix~\ref{AALS}. The contribution coming from the dressing phase can be written as (see Appendix~\ref{AALS})
\be\label{Idress}
\begin{aligned}
\mathcal{I}_{\textrm{dressing}}(u) &= 2(2g)^2\int_{0}^{\infty}dt\, \cos{(ut)}\e^{-t/2}\int_{0}^{\infty}ds\, t K_{-}(2gt, 2gs){q_{-}(2gs) \over \e^{s}-1}\, , \\
\tilde{\mathcal{I}}_{\textrm{dressing}}(u) &= 2(2g)^2\int_{0}^{\infty}dt\, \sin{(ut)}\e^{-t/2}\int_{0}^{\infty}ds\, t K_{+}(2gt, 2gs){\tilde{q}_{+}(2gs) \over \e^{s}-1} \, ,
\end{aligned}
\ee
where we have introduced generating functions for the higher conserved charges as
\be\label{HCC}
\begin{aligned}
q_{-}(t) &= 2\sum_{n\ge 1}(2n-1)(-1)^n g^{2n-1}q_{2n}J_{2n-1}(t)\, ,\\
\tilde{q}_{+}(t) &= 2\sum_{n\ge 1}(2n)(-1)^n g^{2n}q_{2n+1}J_{2n}(t)\, .
\end{aligned}
\ee
Note that the subscripts `$-/+$' refer to the parity in $t$, and not to the parity of the charges that are generated. We recall that the charges $q_{r} (r\ge 2)$, entering~(\ref{HCC}), are evaluated above the vacuum. We can now verify the claim concerning the coupling between $\sigma(u)$ and $\tilde{\sigma}(u)$ at the level of the dressing-phase corrections. Indeed, the charges $q_{2n}, q_{2n+1}$, only depend on $\sigma(u), \tilde{\sigma}(u)$, respectively.

Looking at the various contributions to the integral equations~(\ref{LSeqf}) suggests to switch to a Fourier representation. Following~\cite{ES06,BES06}, we introduce it as
\be\label{FLtra}
\Omega(t) = \e^{-t/2}\int du\,  \sigma(u) \cos{(ut)}\, , \qquad \tilde{\Omega}(t) = \e^{-t/2}\int du\, \tilde{\sigma}(u) \sin{(ut)}\, ,
\ee
for $t>0$, and reciprocally
\be\label{FLtrb}
\sigma(u) = {1\over \pi}\int_{0}^{\infty}dt \cos{(ut)}\e^{t/2}\Omega(t)\, , \qquad \tilde{\sigma}(u) = {1\over \pi}\int_{0}^{\infty}dt \sin{(ut)}\e^{t/2}\tilde{\Omega}(t)\, .
\ee
Note that the functions $\Omega(t), \tilde{\Omega}(t),$ do not have a definite parity in $t$. Taking the representations above into account, the integral equations~(\ref{LSeqf}) can be brought into the form
\be\label{EqT}
0 = \int_{0}^{\infty}dt\, \cos{(ut)} f(t)\, , \qquad 0 = \int_{0}^{\infty}dt\, \sin{(ut)} \tilde{f}(t)\, ,
\ee
with some functions $f(t), \tilde{f}(t),$ that collect all the terms given before. The equations~(\ref{EqT}) are valid for any real value of $u$ by construction. Under some minimal assumptions of regularity for the functions $f(t), \tilde{f}(t)$, the unique solutions to~(\ref{EqT}) are $f(t) = \tilde{f}(t) = 0$ for any $t\ge0$. Expanding now the expressions for $f(t)$ and $\tilde{f}(t)$, we get the equations for $\Omega(t), \tilde{\Omega}(t)$, as
\be
\begin{aligned}\label{EqTF}
(\e^{t}-1)\Omega(t) &+ (2g)^2\int_{0}^{\infty}ds\, t K(2gt, 2gs)\Omega(s) \\
&+ (2g)^2\int_{0}^{\infty}ds\, t K_{-}(2gt, 2gs){q_{-}(2gs) \over \e^{s}-1} = \sum_{\star=1}^{M}\pi_{\star}(t, u_{\star})\, , \\
(\e^{t}-1)\tilde{\Omega}(t) &+ (2g)^2\int_{0}^{\infty}ds\, t K(2gt, 2gs)\tilde{\Omega}(s) \\
&+(2g)^2\int_{0}^{\infty}ds\, t K_{+}(2gt, 2gs){\tilde{q}_{+}(2gs) \over \e^{s}-1} = \sum_{\star=1}^{M}\tilde{\pi}_{\star}(t, u_{\star})\, .
\end{aligned}
\ee
To make the picture complete, we rewrite the generating functions for the conserved charges~(\ref{HCC}) in terms of $\Omega(t), \tilde{\Omega}(t)$~(\ref{FLtra}). Starting from~(\ref{CharAV}) and using formulae in Appendix~\ref{AALE}, we get
\be\label{HCCF}
\begin{aligned}
q_{-}(2gt) &= 2(2g)^2\int_{0}^{\infty}ds\, t K_{+}(2gt, 2gs)\Omega(s)& \\
&+2(2g)^2\sum_{j=1}^{K_{\textrm{h}}}\int_{0}^{\infty}ds\, t K_{+}(2gt, 2gs)\cos{(u_{\textrm{h}, j} s)}\e^{-s/2}\, , \\
\tilde{q}_{+}(2gt) &= 2(2g)^2\int_{0}^{\infty}ds\, t K_{-}(2gt, 2gs)\tilde{\Omega}(s) \\
&+2(2g)^2\sum_{j=1}^{K_{\textrm{h}}}\int_{0}^{\infty}ds\, t K_{-}(2gt, 2gs)\sin{(u_{\textrm{h}, j} s)}\e^{-s/2}\, .
\end{aligned}
\ee

The equations~(\ref{EqTF}) are of type previously encountered in the literature~\cite{ES06,BES06,FRS07,FZ09}. In the following subsection, we apply the methodology that has been developed to deal more conveniently with them in~\cite{Benna06,AABEK07, BKK07,KSV08,BK08,BK09}. 

\subsection{Simplifying the equations}

The equations~(\ref{EqTF}) being linear both in the unknows $\Omega(t), \tilde{\Omega}(t),$ and in the sources $\sum_{\star} \pi, \sum_{\star} \tilde{\pi}_{\star}$, we can treat each contribution separately. Thus we look for a solution in the form
\be
\Omega(t) = \sum_{\star=1}^{M}\Omega_{\star}(t)\, , \qquad \tilde{\Omega}(t) = \sum_{\star=1}^{M}\tilde{\Omega}_{\star}(t)\, , 
\ee
where $\Omega_{\star}(t),\tilde{\Omega}_{\star}(t),$ are sourced by a single excitation carrying a rapidity $u_{\star}$. In parallel, we have a similar decomposition for the conserved charges, or equivalently their generating functions~(\ref{HCC}),
\be\label{HCCb}
q_{-}(2gt) = \sum_{\star} q^{\star}_{-}(2gt)\,, \qquad \tilde{q}_{+}(2gt) = \sum_{\star} \tilde{q}^{\star}_{-}(2gt)\,,
\ee
where in particular case of the holes, $\star = \textrm{h}$, we need not to forget the extra contribution in the right-hand side of~(\ref{HCCF}).

Now, to turn the equations for $\Omega_{\star}(t),\tilde{\Omega}_{\star}(t),$ into a form which is more appropiate for our purposes, we introduce auxiliary functions $\gamma^{\star}_{\pm}(t), \tilde{\gamma}^{\star}_{\pm}(t)$, with definite parity in $t$, $\gamma^{\star}_{\pm}(t) = \pm \gamma^{\star}_{\pm}(-t)$ and similarly for $\tilde{\gamma}^{\star}_{\pm}(t)$. This is done schematically as
\be\label{DecBa}
\begin{aligned}
&(\e^{t}-1)\Omega_{\star}(t) = \textrm{one-loop value} + \gamma^{\star}_{-}(2gt)+\gamma^{\star}_{+}(2gt)\, , \\
&(\e^{t}-1)\tilde{\Omega}_{\star}(t) = \textrm{one-loop value} + \tilde{\gamma}^{\star}_{-}(2gt)+\tilde{\gamma}^{\star}_{+}(2gt)\, .
\end{aligned}
\ee
This is sketchy because, actually, it is convenient to add a bit of higher-loop corrections into the `one-loop values' above. The precise decompositions can be found in Appendix~\ref{AALS}, for each type of excitations~(\ref{SGH}, \ref{SGGF}, \ref{SGLF}, \ref{SGSF}). Plugging the ans\"atze~(\ref{DecBa}) into the integral equations~(\ref{EqTF}), one finds that the functions $\gamma^{\star}_{\pm}(t), \tilde{\gamma}^{\star}_{\pm}(t)$, are naturally given in an expansion over the Bessel's functions,
\be\label{SGg}
\begin{aligned}
\gamma^{\star}_{-}(t) &= 2\sum_{n\ge 1}(2n-1)\gamma^{\star}_{2n-1}J_{2n-1}(t)\, , \\
\gamma^{\star}_{+}(t) &= 2\sum_{n\ge 1}(2n)\gamma^{\star}_{2n}J_{2n-1}(t)\, ,
\end{aligned}
\ee
and similarly for $\tilde{\gamma}^{\star}_{\pm}(t)$. The expansion coefficients $\gamma^{\star}_{n}, \tilde{\gamma}^{\star}_{n}$, are functions of the coupling constant $g$, and they implicitely depend on the rapidity $u_{\star}$. Note also that there is no contribution $\sim J_{0}(t)$ for $\gamma^{\star}_{+}(t), \tilde{\gamma}^{\star}_{+}(t)$. It implies that $\gamma^{\star}_{+}(t) \sim \tilde{\gamma}^{\star}_{+}(t) \sim t^2$ at small $t$. For $\gamma^{\star}_{-}(t),\tilde{\gamma}^{\star}_{-}(t)$, we have the small $t$ asymptotics $\gamma^{\star}_{-}(t) = \gamma^{\star}_{1}(g) t + O(t^3)$, and similarly for $\tilde{\gamma}^{\star}_{-}(t)$.

Now, adopting the notations~(\ref{DecBa}, \ref{SGg}), the integral equations~(\ref{EqTF}) can be casted into the form of infinite systems of equations for $\gamma^{\star}_{n}, \tilde{\gamma}^{\star}_{n}$,
\be\label{EQN}
\begin{aligned}
&\gamma^{\star}_{n} + \int_{0}^{\infty}{dt\over t}J_{n}(2gt){\gamma^{\star}_{+}(2gt)-(-1)^n\gamma^{\star}_{-}(2gt) \over \e^{t}-1} = \kappa^{\star}_{n}(u_{\star})\, , \\
&\tilde{\gamma}^{\star}_{n} + \int_{0}^{\infty}{dt\over t}J_{n}(2gt){\tilde{\gamma}^{\star}_{-}(2gt)+(-1)^n\tilde{\gamma}^{\star}_{+}(2gt) \over \e^{t}-1} = \tilde{\kappa}^{\star}_{n}(u_{\star})\, ,
\end{aligned}
\ee
with $n\ge 1$. The two series of coefficients $\kappa^{\star}_{n}(u_{\star}), \tilde{\kappa}^{\star}_{n}(u_{\star}),$ are given in Appendix~\ref{AALS}, for each type of excitations~(\ref{KH}, \ref{KGFa}, \ref{KGFb}, \ref{KLFa}, \ref{KLFb}, \ref{KSF}). In this form, the equations are amenable to a numerical analysis along the lines of~\cite{Benna06}. For our purposes, it is the relationship of~(\ref{EQN}) with the anologue system of equations for the vacuum solution that is important. As we have said before, the vacuum distribution density is controlled, to leading order at large $S$, by the solution to the BES equation~\cite{BES06}. This equation is known to support the form~(\ref{EQN}), see~\cite{Benna06,AABEK07,BKK07}, with
\be\label{STBES}
\kappa_{n}^{\o} = 2g\, \delta_{n, 1}\, ,\qquad \tilde{\kappa}_{n}^{\o} = 0\, .
\ee
The two functions $\gamma^{\o}_{\pm}(t)$ that solve~(\ref{EQN}) with the source terms~(\ref{STBES}), are fundamental within our analysis.%
\footnote{Note that since the vacuum distribution density is even, there is no need for $\tilde{\gamma}^{\o}_{\pm}(t)$, which are indeed zero due to $\tilde{\kappa}_{n}^{\o} = 0$.} Indeed, as we shall see, they control the dispersion relations of all the excitations. These functions can be constructed at both weak and strong coupling~\cite{BES06,AABEK07,BKK07,KSV08,BK09} and given their knowledge in these regimes we will derive explicit expressions for the dispersion relations. Finally, note that the subleading, $O(\log^{0}{S})$, correction to the vacuum distribution density can also be casted into the form~(\ref{EQN}), as shown in~\cite{FZ09}. The source terms in that case can be written as
\be
\delta\kappa_{n}^{\o} = 2\int_{0}^{\infty}{dt\over t}{J_{n}(2gt)J_{0}(2gt)-gt\delta_{n, 1} \over \e^{t}-1}\, , \qquad \delta\tilde{\kappa}_{n}^{\o} = 0\, . 
\ee
They differ slightly from the ones given in~\cite{FZ09}, due to a different way of introducing $\delta\gamma^{\o}_{+}(t)$. The relation between $\gamma^{\o}_{\pm}(t), \delta\gamma^{\o}_{\pm}(t)$, and the vacuum distribution density is given in Appendix~\ref{AALS} for completeness.

To conclude this subsection, let us point out that the coefficients $\gamma^{\star}_{2n-1}$ and $\tilde{\gamma}^{\star}_{2n}$ have a simple spin-chain interpretation. They are directly related to the spin-chain conserved charges. Indeed, it is not difficult to prove that
\be\label{CGR}
q^{\star}_{2n} = 2{(-1)^{n-1} \over g^{2n-1}}\gamma^{\star}_{2n-1}\, , \qquad q^{\star}_{2n+1} = 2{(-1)^{n-1} \over g^{2n}}\tilde{\gamma}^{\star}_{2n}\, ,
\ee
where the notation $q^{\star}_{r}$ stands for the individual contribution of the excitation to the total charge (above the vacuum) $q_{r}$, with $q_{r} = \sum_{\star} q^{\star}_{r}$. The relations~(\ref{CGR}) are also valid for the vacuum solution. Namely, we have ($q^{\o}_{2n+1} =0$)
\be\label{CGRvac}
q^{\o}_{2n} = 2{(-1)^{n-1} \over g^{2n-1}}\gamma^{\o}_{2n-1}\, ,
\ee
and similarly for the vacuum subleading correction $\delta q^{\o}_{2n}$. We recall that the spin-chain total charge $Q_{r}$ admits the decomposition
\be
Q_{r} = q^{\o}_{r}\left(\log{S}-\psi(1)\right) +\delta q^{\o}_{r} + \sum_{\star} q^{\star}_{r}\, ,
\ee
where the first two terms stand for the leading and subleading vacuum charge, previously denoted $q^{\textrm{vacuum}}_{r}$ in~(\ref{DecChar}). Finally, generating functions for $q^{\star}_{r}$ can be introduced as in~(\ref{HCC}), leading to the identities
\be\label{HCCbis}
q^{\star}_{-}(2gt) = -2\gamma^{\star}_{-}(2gt)\, , \qquad \tilde{q}^{\star}_{+}(2gt) = -2\tilde{\gamma}^{\star}_{+}(2gt)\, ,
\ee
and similarly for $q^{\o}_{r}$ and $\delta q^{\o}_{r}$. The physical meaning, if any,  of the functions $\gamma_{+}^{\star}(t)$ and $\tilde{\gamma}_{-}^{\star}(t)$ remains obscure, however.

\subsection{Fermionic subtleties}

The analysis for the fermions turns out to be subtle, as mentioned before, and requires additional comments. Fixing a fermion at a rapidity $u_{\textrm{f}}$, one could ask what are the coefficients $\kappa^{\textrm{f}}_{n},\tilde{\kappa}^{\textrm{f}}_{n}$, that source the equations~(\ref{EQN}). Looking for them in Appendix~(\ref{AALS}), one would find two possible expressions, denoted $\kappa^{\textrm{lf}}_{n},\tilde{\kappa}^{\textrm{lf}}_{n},$ and $\kappa^{\textrm{sf}}_{n},\tilde{\kappa}^{\textrm{sf}}_{n},$ for `large fermion' and `small fermion', respectively. The reason for this doubling is that the `physical' rapidity for a fermion is the string spectral parameter $x_{\textrm{f}}$, not the spin-chain one $u_{\textrm{f}}$. Fixing $u_{\textrm{f}}$ leaves two possible values for $x_{\textrm{f}}$, as it is well known. These are the two solutions to the quadratic equation
\be\label{ZhuM}
u_{\textrm{f}} = x_{\textrm{f}} + g^2/x_{\textrm{f}} \, .
\ee  
The two kinematical domains to which we refer as large and small fermion in this paper correspond to $x_{\textrm{f}}^2 >  g^2$ and $x_{\textrm{f}}^2 < g^2$, respectively. In each of these domains, we can invert~(\ref{ZhuM}) leading to
\begin{align}
x_{\textrm{f}} = \ft{1}{2}(u_{\textrm{f}} + \sqrt{u_{\textrm{f}}-(2g)^2}) \, , \notag \\
x_{\textrm{f}} = \ft{1}{2}(u_{\textrm{f}} - \sqrt{u_{\textrm{f}}-(2g)^2}) \, , 
\end{align}
for large and small fermion, respectively. Of course, one would like to work directly in terms of $x_{\textrm{f}}$ and forget about the separation between large and small rapidity domain. However, it turns out to be simpler to parameterize the dispersion relation in terms of the Bethe rapidity $u_{\textrm{f}}$;  hence the necessity to deal with two set of coefficients, $\kappa^{\textrm{lf}}_{n}, \tilde{\kappa}^{\textrm{lf}}_{n}$ and $\kappa^{\textrm{sf}}_{n}, \tilde{\kappa}^{\textrm{sf}}_{n}$.

What happens at the level of the system of equations~(\ref{EQN}) when $x_{\textrm{f}}$ crosses the boarder between the two domains is that the coefficients $\kappa^{\cdot \textrm{f}}_{n}, \tilde{\kappa}^{\cdot \textrm{f}}_{n}$ (with `$\cdot \textrm{f}$' = `$\textrm{lf}$'  or  `$\textrm{sf}$'), are no longer suppressed at large $n$. More precisely, they contain contributions of the type $(g/x_{\textrm{f}})^n, (x_{\textrm{f}}/g)^n$, for large, small, fermion, respectively. To get ride of these terms, one can redefine $\gamma^{\cdot \textrm{f}}_{+}(2gt), \tilde{\gamma}^{\cdot \textrm{f}}_{-}(2gt)$, yielding a new system of equations, still of the type~(\ref{EQN}), but with a different set of coefficients $\kappa^{\cdot \textrm{f}}_{n}, \tilde{\kappa}^{\cdot \textrm{f}}_{n}$. This is the way one can switch from one domain to the other. In more sensible terms, the equations for large and small fermion can be obtained from two different manners of looking at the scattering phase $S_{4\textrm{f}}$. From the weak-coupling point of view, the domain $x_{\textrm{f}}^2 \ge  g^2$ is the most natural one. In that case, one factors out the one-loop scattering phase and write
\be
S_{4\textrm{f}}(u, x_{\textrm{f}}) = {u-u_{\textrm{f}}+\ft{i}{2} \over u-u_{\textrm{f}}-\ft{i}{2}}{1-g^2/x^{-}x_{\textrm{f}} \over 1-g^2/x^{+}x_{\textrm{f}}}\, .
\ee
Then proceeding along the lines presented before, one computes the inhomogeneous terms $\mathcal{I}_{\textrm{f}}(u, u_{\textrm{f}})$, $\tilde{\mathcal{I}}_{\textrm{f}}(u, u_{\textrm{f}})$, by means of~(\ref{ITE}), and after a few of algebra get to the system of equations~(\ref{EQN}) with source terms $\kappa^{\textrm{lf}}_{n}, \tilde{\kappa}^{\textrm{lf}}_{n}$, as quoted in~(\ref{KLFa}, \ref{KLFb}). But, obviously, the expansion around the one-loop scattering phase is not a good starting point for a fermion with $x_{\textrm{f}}^2 \le g^2$, as this kinematical window shrinks to zero size at weak coupling. Instead of this, one would look for an expansion around $x_{\textrm{f}}=0$, which means expanding around a fermion at rest. At exactly $x_{\textrm{f}}=0$, the only effect of introducing a fermion is to reduce the spin-chain length by one unit~\cite{BS05}, meaning that
\be
S_{4, \textrm{f}}(u, x_{\textrm{f}}=0) = {x^{+} \over x^{-}}\, .
\ee
As it is the leading order contribution at small $x_{\textrm{f}}$, let us factorize it and rewrite
\be
S_{4, \textrm{f}}(u, x_{\textrm{f}}) = {x^{+} \over x^{-}}\left({1-x_{\textrm{f}}/x^{+}\over 1-x_{\textrm{f}}/x^{-}}\right)\, .
\ee
Then, applying~(\ref{ITE}), one finds that the inhomogeneous terms $\mathcal{I}_{\textrm{sf}}(u, u_{\textrm{f}}),$ $\tilde{\mathcal{I}}_{\textrm{sf}}(u, u_{\textrm{f}})$, are simply given by
\be\label{ITsfM}
\mathcal{I}_{\textrm{sf}}(u, u_{\textrm{f}}) + \tilde{\mathcal{I}}_{\textrm{sf}}(u, u_{\textrm{f}}) = -i\partial_{u}\log{\left({1-x_{\textrm{f}}/x^{+}\over 1-x_{\textrm{f}}/x^{-}}\right)} \, .
\ee
In that form it is clear that at $x_{\textrm{f}}=0$ there is no deformation of the vacuum solution induced by the fermion, since $\mathcal{I}, \tilde{\mathcal{I}} =0$, in agreement with the special role played by this point from the symmetry point of view~\cite{BS05}. Performing the steps explained before, starting with the expression~(\ref{ITsfM}), one would easily derive the set of equations~(\ref{EQN}) with coefficients $\kappa^{\textrm{sf}}_{n},\tilde{\kappa}^{\textrm{sf}}_{n}$, as quoted in~(\ref{KSF}). For consistency, one should observe that the two descriptions given above match when $x_{\textrm{f}}^2 \sim g^2$. We shall see that this is indeed the case for the two expressions we will obtain in the large and small fermion domain for both the energy and momentum.

\section{Dispersion relations}\label{ALDR}

In the previous section, we derived a set of equations that control the contribution to the distribution density brought by the addition of an arbitrary excitation on top of the GKP string. In this section, we will explain how to construct a representation for the energy and momentum of the latter impurity.

\subsection{General expressions}

The GKP energy for an excitation with rapidity $u_{\star}$ is given by the formula
\be\label{BEi}
E_{\star} = \textrm{twist}_{\star} + 2g^2q_{2}^{\star}\, ,
\ee
where $q_{2}^{\star}$ stands for the contribution of the excitation to the spin-chain energy, and $\textrm{twist}_{\star} =1$ for $\star =$ scalar, fermion, gauge field, while $\textrm{twist}_{\star} =\ell$ for $\star =$ bound states of gauge fields  ($\ell>1$). The identity~(\ref{BEi}) can be converted into
\be\label{BEii}
E_{\star} = \textrm{twist}_{\star} + 4g\gamma_{1}^{\star}\, ,
\ee
by using~(\ref{CGR}). So solving the system of equations~(\ref{EQN}) for $\gamma_{n}^{\star}$ would give us the energy, as expected. What about the GKP momentum? In principle, to obtain the momentum, we need to look at the effective Bethe ansatz equations for the type of excitation we consider and extract the quantity that is conjugated to the length $2\log{S}$. We did this analysis and derived the expressions given below for the momenta of the various excitations. Incidentally, we observed that the GKP momentum can also be obtained from the large $u$ asymptotics of the odd part of the distribution density $\tilde{\sigma}(u)$. Namely, we found that
\be\label{SPLU}
\tilde{\sigma}_{\star}(u) \sim {p_{\star} \over 2\pi u}\, ,
\ee
at large $u$, with $\tilde{\sigma}_{\star}(u)$ the correction to the total density sourced by the excitation. The identity~(\ref{SPLU}) can be rephrased in terms of the Fourier-like transform of $\sigma_{\star}(u)$, see Eq.~(\ref{FLtrb}), as $p_{\star} = 2\tilde{\Omega}(0)$. Now, using the relation between $\tilde{\Omega}_{\star}(t)$ and the auxiliary functions $\tilde{\gamma}^{\star}_{\pm}(t)$ (see Appendix~\ref{AALS}), one easily deduces that
\be\label{BPi}
p_{\star} = p_{\star}^{\textrm{one-loop}}+4g\tilde{\gamma}^{\star}_{1}\, .
\ee
For any excitation, we have $p_{\star}^{\textrm{one-loop}} = 2u_{\star}$, except for a small fermion, $p_{\star}^{\textrm{one-loop}}(u_{\star})=0$. We conclude that to find the dispersion relation, we need to evaluate $\gamma^{\star}_{1}, \tilde{\gamma}^{\star}_{1},$ and then apply~(\ref{BEii}, \ref{BPi}).

At this point we are still facing the difficulty to solve the system of equations~(\ref{EQN}). Fortunately we are only interested in extracting the components $\gamma^{\star}_{1}, \tilde{\gamma}^{\star}_{1},$ of the corresponding solution. For them it is possible to derive a representation in terms of $\gamma^{\o}_{\pm}(t)$. The specificity of the latter functions is that they solve the system of equations~(\ref{EQN}) with the inhomogeneous terms $\kappa^{\o}_{n} = 2g\, \delta_{n, 1}$. As explained in~\cite{BK08}, this observation allows us to use the functions $\gamma^{\o}_{\pm}(t)$ to project out the components $\gamma^{\star}_{1}, \tilde{\gamma}^{\star}_{1},$ for functions $\gamma^{\star}_{\pm}(t), \tilde{\gamma}^{\star}_{\pm}(t)$, satisfying the same system of equations~(\ref{EQN}) but with different inhomogeneous terms $\kappa^{\star}_{n}, \tilde{\kappa}^{\star}_{n}$. Going along these lines~\cite{BK08}, we find that
\be\label{WR}
\begin{aligned}
4g\gamma_{1}^{\star} &= \sum_{n\ge 1}2(2n-1)\gamma^{\o}_{2n-1}\kappa^{\star}_{2n-1}-\sum_{n\ge 1}2(2n)\gamma^{\o}_{2n}\kappa^{\star}_{2n} \, , \\
4g\tilde{\gamma}_{1}^{\star} &= \sum_{n\ge 1}2(2n-1)\gamma^{\o}_{2n-1}\tilde{\kappa}^{\star}_{2n-1}+\sum_{n\ge 1}2(2n)\gamma^{\o}_{2n}\tilde{\kappa}^{\star}_{2n} \, .
\end{aligned}
\ee
As we know explicitely the coefficients $\kappa^{\star}_{n}, \tilde{\kappa}^{\star}_{n}$, these identities provide a parameterization of the dispersion relations in terms of the solution to the BES equation for the large spin vacuum distribution of roots.

For the cases we will consider, it is possible to find more tractable integral representations, starting from~(\ref{WR}). Indeed, the coefficients $\kappa^{\star}_{n}, \tilde{\kappa}^{\star}_{n},$ are typically given by some integrals over the Bessel functions $J_{n}(t)$. It suggests to commute summation and integration and then make use of~(\ref{SGg}) to form the functions $\gamma^{\o}_{\pm}(t)$.%
\footnote{Commuting summation and integration is permitted by the fact that the Bessel's functions Neumann series~(\ref{SGg}) for $\gamma^{\o}_{\pm}(t)$ are uniformely convergent over the real $t-$axis, because the coefficients $\gamma^{\o}_{n}$ are suppressed at large $n$.}
Simplifying a bit the obtained expressions, with help of identities for the Bessel's functions and using the equations for $\gamma^{\o}_{\pm}(t)$ when necessary~\cite{BK08}, one arrives at the dispersion relations given below.

\subsubsection*{Scalar}

The dispersion relation for a hole carrying the rapidity $u$ reads
\be\label{DRH}
\begin{aligned}
E(u) &= 1+\int_{0}^{\infty}{dt \over t}{\gamma^{\o}_{+}(2gt)-\gamma^{\o}_{-}(2gt) \over \e^{t}-1}\left(\e^{t/2}\cos{(ut)}-1\right)-\int_{0}^{\infty}{dt \over t}\gamma^{\o}_{+}(2gt)\, , \\
p(u) &= 2u-\int_{0}^{\infty}{dt \over t}{\gamma^{\o}_{+}(2gt)+\gamma^{\o}_{-}(2gt) \over \e^{t}-1}\e^{t/2}\sin{(ut)}\, .
\end{aligned}
\ee
Integrating the energy in~(\ref{DRH}) with an even distribution density for the hole rapidities would lead to the representation of the generalized scaling function proposed in~\cite{BK08}. Taking the derivative of the momentum in~(\ref{DRH}) with respect to the rapidity gives the leading-order contribution to the vacuum distribution density at large spin,
\be\label{MHbes}
\rho_{\textrm{vacuum}}(u) = {\log{S} \over \pi}p'(u) + O(\log^0{S})\, .
\ee
This equality is due to the fact that the effective equations for the holes rapidities are obtained via the counting function $Z(u) \sim \int^{u}\rho$ for the main roots,
\be
2\pi n_{\textrm{h}} = Z(0) + 2\pi \int_{0}^{u_{\textrm{h}}}dv \, \rho(v) = Z(0) + 2p(u_{\textrm{h}})\log{S} + O(\log^{0}{S})\, ,
\ee
where we used that $\rho(v) = \rho_{\textrm{vacuum}}(v) + O(\log^0{S})$ and where $n_{\textrm{h}}$ is a hole mode number. The identity~(\ref{MHbes}) implies that constructing the momentum of a scalar for arbitrary rapidity $u$ is equivalent to solving the BES equation.

\subsubsection*{Gauge field and bound states}

The dispersion relation for a gauge field ($\ell=1$), and its bound states ($\ell > 1$), carrying a rapidity $u$ is given by
\be\label{DRGF}
\begin{aligned}
E(u) &= \ell+\int_{0}^{\infty}{dt \over t}\bigg[{\gamma^{\o}_{+}(2gt) \over 1-\e^{-t}} -{\gamma^{\o}_{-}(2gt) \over \e^{t}-1}\bigg]\left(\cos{(ut)}\e^{-\ell t/2}-1\right)\, , \\
p(u) &= 2u-\int_{0}^{\infty}{dt \over t}\bigg[{\gamma^{\o}_{-}(2gt) \over 1-\e^{-t}} +{\gamma^{\o}_{+}(2gt) \over \e^{t}-1}\bigg]\sin{(ut)}\e^{-\ell t/2}\, .
\end{aligned}
\ee
When $u=0$, we have $p(u=0)=0$, and the expression for the mass $E(u=0)$ is in agreement with the findings of~\cite{FRZ09}.%
\footnote{In the notations of~\cite{FRZ09}, the index $\ell$ is related to $m$, which counts the number of roots $u_{3}$ in the stack $\ell$, by the equality $\ell = m-1$, }

\subsubsection*{Large fermion}

The dispersion relation for a large fermion with rapidity $u$ is
\be\label{DRLF}
\begin{aligned}
E(u) = 1\, \, +&\, \int_{0}^{\infty}{dt \over t}{\gamma^{\o}_{+}(2gt)-\gamma^{\o}_{-}(2gt) \over \e^{t}-1}\left(\cos{(ut)}-1\right) \\
&+{1 \over 2}\int_{0}^{\infty}{dt \over t}\cos{(ut)}\gamma^{\o}_{+}(2gt) -\int_{0}^{\infty}{dt \over t}\gamma^{\o}_{+}(2gt) \, , \\
p(u) = 2u\, \, -&\, \int_{0}^{\infty}{dt \over t}{\gamma^{\o}_{+}(2gt)+\gamma^{\o}_{-}(2gt) \over \e^{t}-1}\sin{(ut)} -{1\over 2}\int_{0}^{\infty}{dt \over t}\sin{(ut)}\gamma^{\o}_{-}(2gt)\, .
\end{aligned}
\ee
Note that this representation assumes that $u^2 >(2g)^2$. At $u^2 = (2g)^2$ the two integrals
\be
\int_{0}^{\infty}{dt \over t}\cos{(ut)}\gamma^{\o}_{+}(2gt)\, , \qquad \int_{0}^{\infty}{dt \over t}\sin{(ut)}\gamma^{\o}_{-}(2gt)\, ,
\ee
have square-root branch points, pointing toward the fact that the physical rapidity is $x = (u+\sqrt{u^2-(2g)^2})/2$, not $u$, as discussed already. The analytic continuation of~(\ref{DRLF}) in the complex $u$-plane will be explained later.

\subsubsection*{Small fermion}

The dispersion relation for a small fermion with rapidity $u$ is given by
\be\label{DRSF}
\begin{aligned}
E(u) &= 1-{1 \over 2}\int_{0}^{\infty}{dt \over t}\cos{(ut)}\gamma^{\o}_{+}(2gt) \, , \\
p(u) &= {1 \over 2}\int_{0}^{\infty}{dt \over t}\sin{(ut)}\gamma^{\o}_{-}(2gt) \, ,
\end{aligned}
\ee
assuming that $u^2 >(2g)^2$. It can be written as a Taylor series around $x = 0$, which for $x = {1 \over 2}(u-\sqrt{u^2-(2g)^2})$ maps to $u=\infty$.
It reads
\be\label{TSSF}
\begin{aligned}
E(x) &=  1-\sum_{n\ge 1}(-1)^n\gamma^{\o}_{2n}\left({x \over g}\right)^{2n}\, , \\
p(x) &=  -\sum_{n\ge 1}(-1)^n\gamma^{\o}_{2n-1}\left({x \over g}\right)^{2n-1}\, ,
\end{aligned}
\ee
with the coefficients $\gamma^{\o}_{n}$ defined by the expansion of $\gamma^{\o}_{\pm}(t)$ over the Bessel's functions~(\ref{SGg}). In particular we have
\be\label{SXA}
\begin{aligned}
E(x) = 1 + O(x^2)\, , \qquad p(x) = {\Gamma_{\textrm{cusp}}(g) \over 2g^2} x  + O(x^3)\, ,
\end{aligned}
\ee
where we used that $\Gamma_{{\textrm{cusp}}}(g) = 2g\gamma^{\o}_{1}$. The asymptotics~(\ref{SXA}) shows that the fermion has zero momentum at $x=0$ and that it is mass $1$ exactly to all loops, $E(x=0)=1$, in agreement with the discussion of~\cite{AM07}. The fact that the cusp anomalous dimension appears in the small $x$ behavior of the fermion momentum~(\ref{SXA}) should not surprise us too much. Indeed, this is a direct consequence of the coupling between a fermion and the main roots, which comes through the spin-chain conserved charges. At small $x$, we have
\be\label{EFsk}
\prod_{j=1}^{K_{4}}{x-x^{-}_{4, j} \over x-x^{+}_{4, j}} = \exp{\left(-iP-iQ_{2}x +O(x^2)\right)}\, ,
\ee
where $P, Q_{2},$ are the total spin-chain momentum and energy. At large spin $S\gg 1$, the energy $Q_{2}$ is dominated by the contribution of the vacuum distribution of main roots, $2g^2Q_{2} = 2\Gamma_{\textrm{cusp}}(g)\log{S} + O(\log^{0}{S})$, such that the GKP momentum (which is conjugated to the length $2\log{S}$) reads $p(x) = \Gamma_{\textrm{cusp}}(g)x/(2g^2)+O(x^3)$. Expanding further in powers of $x$ and using the relations~(\ref{CGRvac}) between vacuum conserved charges and coefficients $\gamma^{\o}_{2n-1}$, one would easily reproduce the Taylor representation~(\ref{TSSF}) for the momentum. 

\subsection{Weak coupling expansion}

At weak coupling, the dispersion relations~(\ref{DRH}, \ref{DRGF}, \ref{DRLF}, \ref{DRSF}) can be evaluated by expanding the functions $\gamma^{\o}_{+}(t)$ and $\gamma^{\o}_{-}(t)$ over the Bessel's functions, see Eq.~(\ref{SGg}). We just need the expressions for the expansion coefficients $\gamma_{n}^{\o}$, see~(\ref{SGg}), at weak coupling. They are easily obtained in the form of an expansion in $g^2$ by solving iteratively the system of equations~(\ref{EQN}) for $\gamma_{n}^{\o}$, i.e. with source terms~(\ref{STBES}). Doing so, we find that the coefficients $\gamma_{n}^{\o}$ start as
\be\label{GLOa}
\begin{aligned}
&\gamma^{\o}_{1} = 2g\left(1-{\pi^2 \over 3}g^2 +{11\pi^4\over 45}g^4\right) +O(g^7)\, , \qquad \, \, \, \, \, \gamma^{\o}_{3} = -{2\pi^4 \over 45}g^5 + O(g^7)\, , \\
&\gamma^{\o}_{2} = 4\zeta_{3}g^4-4\left({\pi^2 \over 3}\zeta_{3}+10\zeta_{5}\right)g^6 + O(g^8) \, , \qquad \gamma^{\o}_{4} = 4\zeta_{5}g^6 + O(g^8)\, ,
\end{aligned}
\ee
and otherwise satisfy $\gamma^{\o}_{n} = O(g^{n+2})$ for $n>1$. Here $\zeta_{n}$ denotes the Riemann Zeta function $\zeta(z)$ evaluated at $z=n$. With the help of~(\ref{GLOa}), one can compute the dispersion relations parametrically in terms of the rapidity $u$ up to $O(g^6)$ (i.e. three loops for energy and four loops for momentum).%
\footnote{Actually, the expression given in~(\ref{GLOa}) for the coefficient $\gamma^{\o}_{3}$ is not necessary for the evaluation of the dispersion relations at $O(g^6)$.} The results are given below, where, in order to save space, we have introduced the notations
\be
\begin{aligned}
\psi^{(+)}_{n}\left(k, u\right) &\equiv {1\over 2}\left(\psi_{n}\left(k + iu\right)+\psi_{n}\left(k - iu\right)\right) \, , \\
\psi^{(-)}_{n}\left(k, u\right) &\equiv {i\over 2}\left(\psi_{n}\left(k + iu\right)-\psi_{n}\left(k - iu\right)\right) \, ,
\end{aligned}
\ee
with $\psi_{n}(z) = \partial^{n+1}_{z}\log{\Gamma(z)}$. We also use
\be
\Gamma_{\textrm{cusp}}(g) = 2g\gamma^{\o}_{1} = 4g^2\left(1-{\pi^2 \over 3}g^2 +{11\pi^4\over 45}g^4+\ldots\right)\, ,
\ee
for the cusp anomalous dimension.

\subsubsection*{Scalar}

At weak coupling, the dispersion relation for a scalar is given parametrically as
\begin{align}\label{WCH}
E(u) =& \, \, 1 + \Gamma_{\textrm{cusp}}(g)\left[\psi^{(+)}_{0}\left(\ft{1}{2}, u\right)-\psi_{0}(1)\right] -2g^4\left[\psi^{(+)}_{2}\left(\ft{1}{2}, u\right) + 6\zeta_{3}\right] \notag \\
&+ {g^6 \over 3}\bigg[\psi_{4}^{(+)}\left(\ft{1}{2}, u\right) +2\pi^2\psi^{(+)}_{2}\left(\ft{1}{2}, u\right) + 24\zeta_3\psi^{(+)}_{1}\left(\ft{1}{2}, u\right)+8\left(\pi^2\zeta_{3}+30\zeta_{5}\right)\bigg] + O\left(g^8\right)\, ,\notag \\
p(u) =& \, \, 2u + \Gamma_{\textrm{cusp}}(g)\, \psi^{(-)}_{0}\left(\ft{1}{2}, u\right) -2g^4\psi^{(-)}_{2}\left(\ft{1}{2}, u\right) \notag \\
&+{g^6 \over 3}\bigg[\psi^{(-)}_{4}\left(\ft{1}{2}, u\right) +2\pi^2\psi^{(-)}_{2}\left(\ft{1}{2}, u\right)-24\zeta_3\psi^{(-)}_{1}\left(\ft{1}{2}, u\right)\bigg] + O\left(g^8\right)\, .
\end{align}
The expression quoted in~(\ref{WCH}) for the energy is in agreement with the three-loop computation of~\cite{BKP08} using the long-range Baxter equation~\cite{Bel06}. Moroever, the mass of a scalar $E(u=0)$ is directly related to the first contribution to the generalized scaling function $f_{1}(g)$~\cite{BGK06,FRS07}. Evaluating $E(u=0)$ with~(\ref{WCH}) and simplifying the expression, one finds agreement with the weak coupling expansion of $f_{1}(g)$~\cite{FRS07}. This matching is actually guaranteed to all loops, by comparing the all-loop expression we found for $E(u=0)$ with the formula given in~\cite{BK08} for $f_{1}(g)$. Finally, note that it is possible to reexpress the momentum in terms of more familiar hyperbolic functions (at least for the first few terms in~(\ref{WCH})). For instance, for the momentum of a hole at two-loop order, one has
\be
p(u) = 2u -2\pi g^2\tanh{(\pi u)} + O(g^4)\, .
\ee

\subsubsection*{Gauge field and bound states}

The dispersion relations for a gauge field ($\ell=1$) and its bound states ($\ell>1$) are given at weak coupling by
\begin{align}\label{WCGF}
E(u) =&\, \, \ell + \Gamma_{\textrm{cusp}}(g)\left[\psi^{(+)}_{0}\left(s, u\right)-\psi_{0}(1)\right] -2g^4\left[\psi^{(+)}_{2}\left(s, u\right) + 6\zeta_{3}\right] \notag \\
&+ {g^6 \over 3}\bigg[\psi_{4}^{(+)}\left(s, u\right) +2\pi^2\psi^{(+)}_{2}\left(s, u\right) + 24\zeta_3\psi^{(+)}_{1}\left(s-1, u\right)+8\left(\pi^2\zeta_{3}+30\zeta_{5}\right)\bigg] + O\left(g^8\right)\, ,\notag \\
p(u) =&\, \, 2u + \Gamma_{\textrm{cusp}}(g)\, \psi^{(-)}_{0}\left(s-1, u\right) -2g^4\psi^{(-)}_{2}\left(s-1, u\right)  \notag \\
&+{g^6 \over 3}\bigg[\psi^{(-)}_{4}\left(s-1, u\right) +2\pi^2\psi^{(-)}_{2}\left(s-1, u\right)-24\zeta_{3}\psi^{(-)}_{1}\left(s, u\right)\bigg] + O\left(g^8\right)\, ,
\end{align}
where $s\equiv 1+\ell/2$. For $u=0$, we have $p(u)=0$, and the masses $E(u=0)$ are in agreement with the results of~\cite{FRZ09,Beccaria07}.

\subsubsection*{Large fermion}

The dispersion relation for a large fermion is
\begin{align}\label{WCLF}
E(u) =&\, \, 1 + \Gamma_{\textrm{cusp}}(g)\left[\psi^{(+)}_{0}\left(1, u\right)-\psi_{0}(1)\right] -2g^4\left[\psi^{(+)}_{2}\left(1, u\right) + 6\zeta_{3}\right] \notag \\
&+ {g^6 \over 3}\bigg[\psi_{4}^{(+)}\left(1, u\right) +2\pi^2\psi^{(+)}_{2}\left(1, u\right) + 24\zeta_3\psi^{(+)}_{1}\left(1, u\right)-{12\zeta_{3}\over u^2}+8\left(\pi^2\zeta_{3}+30\zeta_{5}\right)\bigg] + O\left(g^8\right)\, ,\notag \\
p(u) =&\, \, 2u + \Gamma_{\textrm{cusp}}(g)\, \psi^{(-)}_{0}\left(1, u\right) -{\Gamma_{\textrm{cusp}}(g)\over 2u}-2g^4\psi^{(-)}_{2}\left(1, u\right)-{2g^4 \over u^3} \notag \\
&+{g^6 \over 3}\bigg[\psi^{(-)}_{4}\left(1, u\right) -{12 \over u^5} +2\pi^2\psi^{(-)}_{2}\left(1, u\right)+{2\pi^2 \over u^3}-24\zeta_{3}\psi^{(-)}_{1}\left(1, u\right)\bigg] + O\left(g^8\right)\, .
\end{align}
The singularities at small $u$ signal the presence of the square-root branch cut associated to the map
\be\label{XUlf}
x = {1\over 2}(u+\sqrt{u^2-(2g)^2}) = u -{g^2 \over u} + O(g^4)\, .
\ee
When expressed in terms of the rapidity $x$, the energy and momentum are regular for any real value of $x$, as verified below.

\subsubsection*{Small fermion}

The dispersion relation for a small fermion is
\be\label{WCSF}
\begin{aligned}
E(u) &= 1+{4\zeta_{3} \over u^2}g^6 + O(g^8)\, , \\
p(u) &= {\Gamma_{\textrm{cusp}}(g) \over 2u} + {2g^4 \over u^3} +{2g^6 \over u^3}\left({2\over u^2}-{\pi^2 \over 3}\right)+ O(g^8)\, .
\end{aligned}
\ee
Note that the weak coupling expansion is taken at fixed $u$. It means that the rapidity $x$ scales as
\be\label{XUsf}
x = {1\over 2}(u-\sqrt{u^2-(2g)^2}) = {g^2 \over u} + O(g^4)\, .
\ee 
Here again the singular behavior at $u\sim 0$ indicates the emergence of the cut with branch points at $u=\pm 2g$. But in distinction with a large fermion, we note that there seems to be no other singularities in the $u$-plane of a small fermion. This is actually correct and both energy and momentum are holomorphic functions of $x$ for $|x| < g$. 

\subsubsection*{Interpolating large/small fermion}

Let us verify here, at the level of the weak coupling expansion~(\ref{WCLF}, \ref{WCSF}), that large and small fermion are analytically related to one another. To this end, we start with the dispersion relation~(\ref{WCLF}) for a large fermion and reexpress it in terms of the physical rapidity $x = u+g^2/u$. We find to two loops that
\begin{align}\label{WCPF}
E(x) &= 1+\Gamma_{\textrm{cusp}}(g)\bigg[\psi_{0}^{(+)}(1, x)-\psi_{0}(1)\bigg] -2g^4\left[\psi^{(+)}_{2}(1,x)+6\zeta_{3}-{2 \over x}\psi_{1}^{(-)}(1, x)\right] + O(g^6)\, , \notag\\
p(x) &= 2x + \Gamma_{\textrm{cusp}}(g)\, \psi^{(-)}_{0}\left(1, x\right) -2g^4\left[\psi_{2}^{(-)}(1, x) +{1\over x}\left(2\psi_{1}^{(+)}(1,x)-{\pi^2 \over 3}\right)\right] + O(g^6)\, .
\end{align}
In spite of their appearance, all the terms in square brackets are regular at small $x$, and keeping only the leading contributions, we get
\be
E(x) = 1 + (4g^2\zeta_{3}+O(g^4))x^2 + O(x^4)\, , \qquad p(x) = {\Gamma_{\textrm{cusp}}(g)\over 2g^2}x + O(x^3)\, ,
\ee
in agreement with~(\ref{WCSF}, \ref{XUsf}). We could have started from the small fermion domain as well. From the Taylor series for energy and momentum given in~(\ref{TSSF}) and with the help of
\be
\gamma^{\o}_{n} = 2g\, \delta_{n, 1} + 4(-1)^n\zeta_{n+1}g^{n+2} + O(g^{n+4})\, ,
\ee
one finds
\begin{align}
E(x) &= 1-4g^2\sum_{n\ge 1}(-1)^n \zeta_{2n+1}x^{2n} + O(g^4) = 1+4g^2(\psi_{0}^{(+)}(1,x)-\psi_{0}(1)) + O(g^4)\, , \notag \\
p(x) &= 2x + 4g^2\sum_{n\ge 1}(-1)^n\zeta_{2n} x^{2n-1} + O(g^4) = 2x + 4g^2\psi^{(-)}_{0}(1,x)+O(g^4)\, ,
\end{align}
which is consistent with~(\ref{WCLF}, \ref{XUlf}), since $\Gamma_{\textrm{cusp}}(g) = 4g^2 + O(g^4)$.

\subsection{Complex rapidity plane}

To go beyond the weak coupling expansion and improve our understanding of the complex rapidity plane, it is helpful to recall few facts about the solution to the BES equation and derive alternative representation for the dispersion relations. After these steps, we will show that fermions in the small and large rapidity domain are analytically related to one another. Then we will discuss what are the genuine singularities in the complex rapidity plane of the various dispersion relations and derive their large rapidity  asymptotics. Finally, we explain why the higher-twist excitations can be thought as bound states of gauge fields by constructing a set of fusion identities for their dispersion relations.

\subsubsection*{Alternative representation}

The prerequisite for the exploration of the complex rapidity plane is a brief reminder of the main properties of the solution to the BES equation~\cite{BKK07,KSV08,BK08,BK09}. A remarkable account of the analytical properties of the functions $\gamma_{\pm}(t)$, which are directly relevant for our investigation, can be obtained from~\cite{KSV08}, where the solution to the BES equation was constructed in terms of resolvants. %
\footnote{It is not difficult to establish the mapping between the resolvants of~\cite{KSV08} and the functions we manipulate in this paper. The dictionary is given in appendix of~\cite{KSV08} and it can be applied here after performing an overall rescaling: $\gamma^{\o}_{\pm}(t) = 4g\gamma_{\pm}(t)$ with $\gamma_{\pm}(t)$ the notations used in the dictionary of~\cite{KSV08}. Similarly for the functions $\Gamma^{\o}_{\pm}(t)$ (introduced below) and $\Gamma_{\pm}(t)$ in~\cite{KSV08}.} Within our approach, most of these properties can be understood as being inherited from the Bessel's functions over which $\gamma_{\pm}^{\o}(t)$ expand in a very controlled way.%
\footnote{The expansion coefficients $\gamma^{\o}_{n}$, solving the BES equation in the form~(\ref{EQN}, \ref{STBES}), are well suppressed at large index $n\gg 1$ (taken at a fixed value of the coupling), which follows from the asymptotics of the kernel of the BES equation~\cite{Benna06}.} The main point is that the Neumann series~(\ref{SGg}) are absolutely convergent and converge uniformly toward the functions $\gamma_{\pm}^{\o}(t)$ in the complex $t$-plane. It follows that the functions $\gamma^{\o}_{\pm}(t)$ are entire functions over the complex $t$-plane, with an essential singularity at $t=\infty$. They are also bounded over the real $t$-axis which allows us to perform the transformations
\be\label{LapTrs}
\lambda_{+}(u) = \int_{0}^{\infty}{dt\over t}\gamma^{\o}_{+}(2gt)\e^{iut}\, , \qquad \lambda_{-}(u) = -i\int_{0}^{\infty}{dt\over t}\gamma^{\o}_{-}(2gt)\e^{iut}\, ,
\ee
converging for $\Im{(u)} \geqslant 0$. Differentiating~(\ref{LapTrs}) with respect to $u$ and performing few rescalings give two of the resolvants of~\cite{KSV08}. 

From the analysis of~\cite{KSV08} or from the expansion over the Bessel's functions, we learn that $\lambda_{\pm}(u)$ have square-root branch points at $u=\pm 2g$, such that locally around these points
\be
\lambda_{\pm}(u) = \alpha_{\pm}(u)\sqrt{u^2-(2g)^2}+\beta_{\pm}(u)\, ,
\ee
with $\alpha_{\pm}(u), \beta_{\pm}(u)$, being analytic at $u=\pm 2g$. The functions $\lambda_{\pm}(u)$ can furthermore be continued to holomorphic functions of $u$ in the complex plane $\mathbb{C}\cup \{\infty\}$ outside the cut $u^2 \leqslant (2g)^2$. Explicit expressions for the functions $\lambda_{\pm}(u)$ in the lower-half plane then follow from~(\ref{LapTrs}) and from the properties $(\lambda_{\pm}(u))^* = \lambda_{\pm}(u^*)$, with $*$ the complex conjugation.%
\footnote{We recall that the functions $\gamma^{\o}_{\pm}(t)$, which are integrated in~(\ref{LapTrs}), are real, if $g$ is real.} In particular,  for $u^2 > (2g)^2$, the functions $\lambda_{\pm}(u)$ are real and can be written as
\be\label{RealForm}
\lambda_{+}(u) = \int_{0}^{\infty}{dt\over t}\gamma^{\o}_{+}(2gt)\cos{(ut)}\, , \qquad \lambda_{-}(u) = \int_{0}^{\infty}{dt\over t}\gamma^{\o}_{-}(2gt)\sin{(ut)}\, ,
\ee
as a consequence of~(\ref{LapTrs}) and of the equalities~(\ref{Beq}) for Bessel's functions. 

It is easily recognized that the functions $\lambda_{\pm}(u)$ permit to analytically continued energy and momentum of a small fermion. Indeed, comparing~(\ref{DRSF}) and~(\ref{RealForm}), we conclude that
\be
E_{\textrm{sf}}(u) = 1-{1\over 2}\lambda_{+}(u)\, , \qquad p_{\textrm{sf}}(u) = {1\over 2}\lambda_{-}(u)\, .
\ee
For later purpose, we also note that
\be\label{EpPsf}
E_{\textrm{sf}}(u)+p_{\textrm{sf}}(u) = 1-{1\over 2}\lambda(u)\, ,
\ee
where
\be\label{DefSmallG}
\lambda(u) \equiv \lambda_{+}(u)-\lambda_{-}(u) = \int_{0}^{\infty}{dt \over t}\gamma^{\o}(2gt)\e^{iut}\, , \qquad \gamma^{\o}(t) \equiv \gamma^{\o}_{+}(t)+i\gamma^{\o}_{-}(t)\, ,
\ee
where, in the first equality, the integral representation is for $\Im{(u)} \geqslant 0$. One could as well consider the quantity $E_{\textrm{sf}}(u)-p_{\textrm{sf}}(u)$, which can be obtained from~(\ref{EpPsf}) through the parity $u\rightarrow -u$.%
\footnote{Note that $\lambda_{\pm}(-u) = \pm \lambda_{\pm}(u)$.} Now, applying what we have learnt before, we conclude that both energy and momentum of a small fermion are holomorphic functions of the rapidity $u$ outside the cut $u^2 \leqslant (2g)^2$. Equivalently, energy and momentum of fermion are holomorphic function of the Zhukowski rapidity $x$ in the disk $|x| \leqslant g$.

Around $u=\infty$ (i.e., $x = 0$) we get from~(\ref{LapTrs}) that $\lambda_{+}(u) \sim 1/u^2$ and $\lambda_{-}(u) \sim 1/u$, since $\gamma^{\o}_{+}(t) \sim t^2$ and $\gamma^{\o}_{-}(t) \sim t$, implying that $E_{\textrm{sf}} = 1 + O(p_{\textrm{sf}}^2)$ when $p_{\textrm{sf}} \sim 0$. When we go away from this point and get closer to the cut $u^2 \geqslant (2g)^2$, the momentum typically increases in modulus, and, if done along  the real $u$-line, the energy will also increase.%
\footnote{We do not know how to prove these naive expectations from general properties of the functions $\gamma^{\o}_{\pm}(t)$, but we observe that they are correct at both weak and strong coupling. Note also that if we approach the cut starting form $u=\infty$ and running along the imaginary $u$-line, then the momentum increases in modulus, but, since it is imaginary, the energy is decreased.}
Note finally that energy and momentum are generically complex around the cut, with values above and below the cut being complex conjugate of one another. So in order to increase energy and momentum further, one needs to enter the cut and change sheet. This is the connection with the large fermion domain.

In order to verify that small and large fermion domain are correctly glued together, we need to analytically continue the functions $\lambda_{\pm}(u)$ through the cut. This will be done later. For the time being, we simply report on the additional information we need to perform this step. This additional information comes from the fact that the functions $\gamma^{\o}_{\pm}(t)$ solve the BES equation. A more useful statement is that they solve the equations
\be\label{SystII}
\begin{aligned}
&\int_{0}^{\infty}{dt \over t}\left[{\gamma^{\o}_{+}(2gt) \over 1-\e^{-t}}-{\gamma^{\o}_{-}(2gt) \over \e^{t}-1}\right](\cos{(ut)}-1) = 0\, , \\
&\int_{0}^{\infty}{dt \over t}\left[{\gamma^{\o}_{-}(2gt) \over 1-\e^{-t}}+{\gamma^{\o}_{+}(2gt) \over \e^{t}-1}\right] \sin{(ut)} =  2u\,,
\end{aligned}
\ee
that hold for $u^2 < (2g)^2$. The system of equations~(\ref{SystII}) is not equivalent to the original one~(\ref{EQN}) because it admits an infinite number of solutions. It is a sort of off-shell extension of the BES equation that turned out to be very useful in constructing its solution at strong coupling~\cite{BKK07,KSV08,BK09}. When supplemented with appropriate analytical properties for $\gamma^{\o}_{\pm}(t)$ the solution to~(\ref{SystII}) is of course unique and identical to the one obtained from~(\ref{EQN}). Here we just need to note that the equations~(\ref{SystIII}) apply to our problem.

If we aim at simplifying energy and momentum of  gauge field and bound states, another function is worth considering. It is denoted $\Gamma^{\o}(t)$ and is introduced by means of the transformation~\cite{BKK07,KSV08,BK09}
\be\label{gtoG}
\Gamma^{\o}(2gt) \equiv \left(1+i\textrm{coth}{\left(\ft{t}{2}\right)}\right)\gamma^{\o}(2gt)\, ,
\ee
with $\gamma^{\o}(t)$ defined in~(\ref{DefSmallG}). In terms of this function, the equations~(\ref{SystII}) can be combined into
\be\label{SystIII}
\int_{0}^{\infty}dt \bigg[\Gamma^{\o}(2gt)\e^{iut-i\pi/4} -\Gamma^{\o}(-2gt)\e^{-iut+i\pi/4}\bigg] = 4i\sqrt{2}\, ,
\ee
and it also assumes $u^2 < (2g)^2$. The equation~(\ref{SystIII}) is far more simpler to analyze than the previous ones~(\ref{SystII}), especially at strong coupling. The price to pay is that the function $\Gamma^{\o}(t)$ has more complicated analytical properties. It has for instance an infinite number of poles along the imaginary axis, due to the transformation~(\ref{gtoG}). The quantity of interest here is not directly $\Gamma^{\o}(t)$ but its Fourier transform, that we introduce as
\be\label{FourTr}
\Lambda(u) \equiv \int {dt \over 2\pi}\, \Gamma^{\o}(2gt)\e^{iut}\, .
\ee
It is defined for any real value fo $u$ except at $u=\pm 2g$ where it has (integrable) square-root singularities, $\Lambda(u) \sim 1/\sqrt{|u\mp 2g|}$. These two points split the analysis of $\Lambda^{\o}(u)$ into the two domains $u^2 \gtrless (2g)^2$. Here we will focus on the interval $u^2 <(2g)^2$.

Starting from the segment $u^2 < (2g)^2$, the function $\Lambda(u)$ can be analytically continued in both the upper- and lower-half $u$-plane. Indeed, after writing 
\be\label{FourTrb}
\Lambda(u) =  \int_{0}^{\infty}{dt \over 2\pi}\, \Gamma^{\o}(2gt)\e^{iut}+ \int_{0}^{\infty}{dt \over 2\pi}\, \Gamma^{\o}(-2gt)\e^{-iut}\, ,
\ee
we observe that for $u^2<(2g)^2$ we can apply the equation~(\ref{SystIII}) and get both
\be\label{FourLap}
\begin{aligned}
\Lambda(u) &= -{2\sqrt{2} \over \pi}\e^{i\pi/4} \, \, \, \, + {1\over \sqrt{2}\pi}\int_{0}^{\infty}dt\, \Gamma^{\o}(2gt)\e^{iut-i\pi/4}\, , \\
\Lambda(u) &= -{2\sqrt{2} \over \pi}\e^{-i\pi/4} + {1 \over \sqrt{2}\pi}\int_{0}^{\infty}dt\, \Gamma^{\o}(-2gt)\e^{-iut+i\pi/4} .
\end{aligned}
\ee
From the two identities in~(\ref{FourLap}), we conclude that the function $\Lambda(u)$ extends to a holomorphic function of $u$ in both the upper- and lower-half $u$-plane. We recall however that $\Lambda(u)$ has (integrable) square-root branch points at  $u^2 = (2g)^2$, such that the present analytical continuation is defined in the complex plane with a cut  along $u^2 > (2g)^2$. Notice finally that the function $\Lambda(u)$ has essential singularities at both $u=\pm i\infty$. Around these points, $\Lambda(u)$ admits expansion in inverse powers of $u$ starting as
\be\label{GoldThe}
\Lambda(u) = -{2\sqrt{2} \over \pi}\e^{\pm i\pi/4}\left[1+{\Gamma_{\textrm{cusp}}(g) \over 2u} + O(1/u^2)\right]\, , 
\ee
for $u$ in the upper- and lower-half plane, respectively. The large $u$ expansion~(\ref{GoldThe}) is merely asymptotic, however, since $\Gamma^{\o}(t)$ has poles in the complex $t$-plane.

\begin{figure}[h]%
\setlength{\unitlength}{0.14in}
\centering
\begin{picture}(14,14)%
\put(-10,0){\insertfig{4}{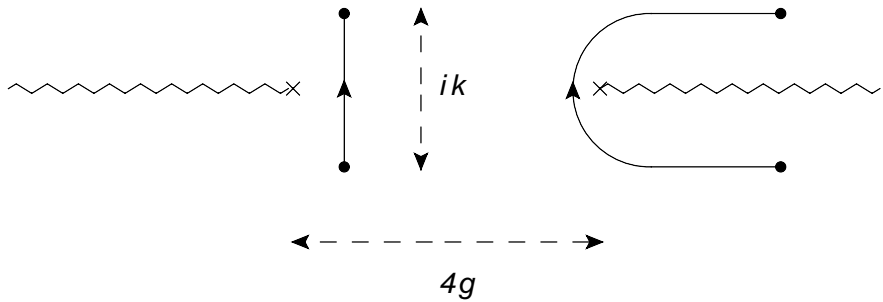}}
\end{picture}%
\caption{Two realizations for the contour of integration in the integral representation~(\ref{CIRa}), associated to a rapidity $u$ with $u^2 \gtrless (2g)^2$, respectively. Here $k = \ell$ fixes the size of the bound state, i.e., the end points $u \pm ik/2 = u\pm i\ell/2$ of the contour of integration. The latter is chosen such as to avoid the cuts in the function $\Lambda$, running from $\pm 2g$ to $\pm \infty$ and depicted by the two wavy lines.}\label{CRGF}
\end{figure}

We can now derive an alternative representation for energy and momentum of gauge field and of its bound states. Namely, starting from the general expressions~(\ref{DRGF}), we form the quantity $E_{\ell}(u) + p_{\ell}(u)$ reading
\be
\begin{aligned}
E_{\ell}(u) +p_{\ell}(u)& = \, \, 2u+2\epsilon \\
&+{1\over 2\sqrt{2}}\int_{0}^{\infty}{dt \over t}\bigg[\e^{-i\pi/4}\Gamma^{\o}(2gt)\left(\e^{i(u+i\epsilon)t}-1\right) + \e^{i\pi/4}\Gamma^{\o}(-2gt)\left(\e^{-i(u-i\epsilon)t}-1\right)\bigg]\, ,
\end{aligned}
\ee
where $\epsilon \equiv \ell/2$. It then drastically simplifies when written in terms of $\Lambda(u)$ that leads to the integral representation
\be\label{CIRa}
E_{\ell}(u) +p_{\ell}(u) = {i\pi\over 2}\int_{u-i\epsilon}^{u+i\epsilon}dv\, \Lambda(v)\, ,
\ee
with the contour of integration as depicted in Figure~\ref{CRGF}. A similar expression can be obtained for $E_{\ell}(u) -p_{\ell}(u)$, using the reflexion $u\rightarrow -u$, and reads
\be\label{CIRb}
E_{\ell}(u) -p_{\ell}(u) = {i\pi\over 2}\int_{u-i\epsilon}^{u+i\epsilon}dv\, \Lambda(-v)\, .
\ee
The two relations~(\ref{CIRa}, \ref{CIRb}) conclude this discussion. We will now come back to fermion and prove the continuation between small and large rapidity domain.

\subsubsection*{Interpolating small and large fermion}

\begin{figure}[h]%
\begin{center}%
\mbox{\begin{picture}(0,240)(150,0)%
\put(0,0){\insertfig{8}{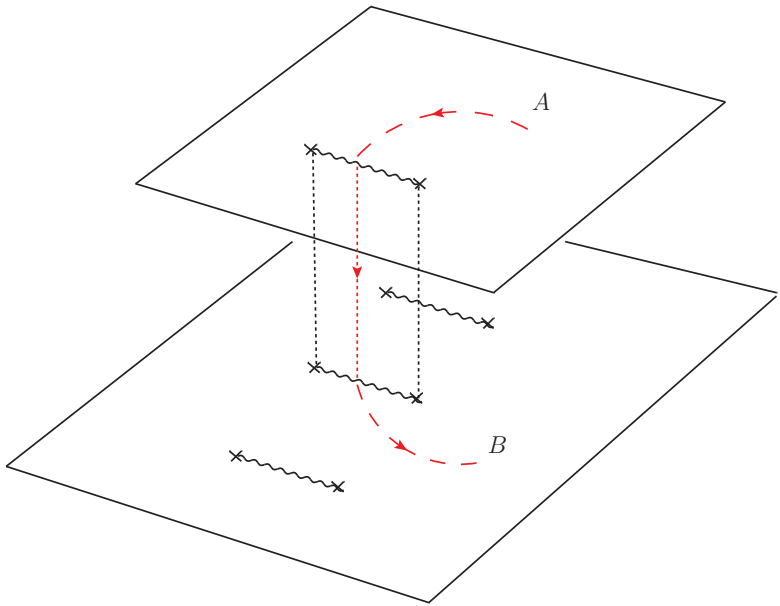}}
\end{picture}}%
\end{center}%
\caption{Sketch of the complex $u$-planes for energy and momentum of a large (lower sheet) and small (upper sheet) fermion. Crosses indicate square-root branch points. Except for the central cut, the upper sheet is free from singularities. The lower sheet contains an infinite series of cuts which extends along the imaginary axis, with a spacing between two consecutive cuts equal to $\pm i$. The dashed line indicates the path followed during the analytic continuation from the small to large fermion domain.}\label{ACpict}
\end{figure}

To prove that small and large fermion interpolate each other, we will analytically continue through the cut the energy and momentum of a small fermion and show that they become identical to those of a large fermion~(\ref{DRLF}). The path we will follow is depicted in Fig.~\ref{ACpict}. We give the proof for the energy only, which can be easily adapted to the momentum. We only perform the continuation by  going through the cut from above, but the same conclusion would be reached by analytically continuing from below.

We start with the energy in the small rapidity domain, depicted by point $A$ in Fig.~\ref{ACpict}. Hence we can write
\be\label{AnyCa}
E_{\textrm{sf}}(u) = 1-{1 \over 2}\lambda_{+}(u)\, ,
\ee
with $\lambda_{+}(u)$ as given in~(\ref{LapTrs}). We can easily move in the upper half-plane and evaluate the energy $E_{\textrm{sf}}(u+i0^+)$ right above the cut when $u^2 < (2g)^2$. We can go through the cut by writing the energy as
\be\label{AnyCc}
E_{\textrm{sf}}(u+i0^+)  = 1+{1 \over 2}\lambda_{+}(u-i0^-) -\int_{0}^{\infty}{dt\over t}\gamma^{\o}_{+}(2gt)\cos{(ut)} \equiv E_{\textrm{lf}}(u-i0^+)  \, ,
\ee
where we used the explicit form of $\lambda_{+}(u)$. Observe that in writing~(\ref{AnyCc}) we have anticipated that the function we are getting to is the energy in the large rapidity domain, thus changing the subscript `$\textrm{sf}$' into `$\textrm{lf}$' at the end of the equation. In particular, we stress that~(\ref{AnyCc}) is different from $E_{\textrm{sf}}(u-i0^+)$, which is the value of the energy evaluated below the cut in the small fermion domain.

We know that the function $\lambda_{+}(u)$ evaluated in~(\ref{AnyCc}) is analytic outside of the cut $u^2 < (2g)^2$, such that we need to consider the continuation of the integral in~(\ref{AnyCc}) to get $E_{\textrm{lf}}(u)$ away from the cut. To this end we note that the latter integral can be expressed as
\be\label{AnyCd}
-\int_{0}^{\infty}{dt\over t}\gamma^{\o}_{+}(2gt)\cos{(ut)} = \int_{0}^{\infty}{dt\over t}{\gamma^{\o}_{+}(2gt)-\gamma^{\o}_{-}(2gt) \over \e^{t}-1}\left(\cos{(ut)}-1\right)-\int_{0}^{\infty}{dt\over t}\gamma^{\o}_{+}(2gt)\, ,
\ee
by using the first equation in~(\ref{SystII}), which applies here since $u^2 <(2g)^2$. This identity is telling us that the left-hand side of~(\ref{AnyCd}) can be analytically continued into a holomorphic function of $u$ in the strip $(\Im{(u)})^2 < 1$, since this is a property of the right-hand side of~(\ref{AnyCd}). So we can easily move in this strip and get to the point $B$ in Fig.~\ref{ACpict}, which stands for a generic point in this strip. For real momentum, we need to choose a rapidity $u$ at $B$ that is real and away from the cut (i.e., $u^2 > 2g$). Then we can further simplify~(\ref{AnyCc}), since in this case $\lambda_{+}(u)$ restricts to the real function given in~(\ref{RealForm}). Combining it with~(\ref{AnyCd}), we get the expression for the energy of a large fermion quoted in~(\ref{DRLF}).

We have thus shown the interpolation between small and large fermion. The two domains, and their respective sheets, are connected through their common cut along $u^2 <(2g)^2$. This cut originates from $\lambda_{+}(u)$ in~(\ref{AnyCc}). We learnt furthermore that this is the only cut in the strip $(\Im{(u)})^2 < 1$ on the large fermion sheet. As we shall see, outside this strip, the energy of a large fermion has an infinite number of cuts, in distinction to what happens in the small fermion domain.

\subsubsection*{Singularities in the complex rapidity plane}

We now discuss the singularities of energy and momentum in the complex $u$-plane for all the excitations. A natural starting point is to inspect the weak coupling expressions quoted in~(\ref{WCH}, \ref{WCGF}, \ref{WCLF}, \ref{WCSF}). We observe that both energy and momentum have poles along the imaginary $u$-axis, whose precise locations depend on the flavor of the excitation. For scalar, for instance, we find poles at $u_{\textrm{p}} = i(2\mathbb{Z}+1)/2$, for both energy and momentum.%
\footnote{For other excitations, it is interesting to note that a few new poles shows up as we increase the loop order. At $O(g^6)$ however the pattern of poles is complete and found to be the same for both energy and momentum of a given excitation.} We notice then that the orders of the poles increase with the loop order. This is suggesting that the weak coupling expansion breaks down around these points and thus does not give a proper account of the singularities in the $u$-plane (at any given order in $g^2$).

It is then not difficult to figure out what the genuine singularities are: at any finite value of the coupling constant we only have cuts of size $4g$ connecting square-root branch points. The reason being that we can decompose energy and momentum into sums of terms of the type
\be\label{prototerm}
\lambda_{\pm}(u+im)\, ,
\ee
with $m$ belonging to some subset of $\mathbb{Z}$ or $(2\mathbb{Z}+1)/2$, depending on the flavor of the excitation.%
\footnote{In practice the sums are taken over an infinite subset of $\mathbb{Z}$ or $(2\mathbb{Z}+1)/2$, except for a small fermion where only $m=0$ appears. Then, in order to make these sums convergent, one should consider terms of the type $\lambda_{\pm}(u + im)-\lambda_{\pm}(im)$ instead of~(\ref{prototerm}). This does not affect the argument we are giving here.} The weak coupling expansion of~(\ref{prototerm}) produces poles of various orders at $u_{\textrm{p}} = -im$, but, at finite coupling, the actual singularities of~(\ref{prototerm}) are two branch points at $u_{\textrm{b-p}}= u_{\textrm{p}} \pm 2g$, respectively. Summing over $m$ leads to series of cuts in the complex $u$-plane. From this perspective, the weak coupling expansion, taken at a rapidity $u\neq u_{\textrm{p}}$, can be thought as Laurent expanding around these cuts, thus generating poles sitting at their individual centers of mass.

\begin{figure}[h]%
\setlength{\unitlength}{0.14in}
\centering
\begin{picture}(10,23)%
\put(-5.5,0){\insertfig{8}{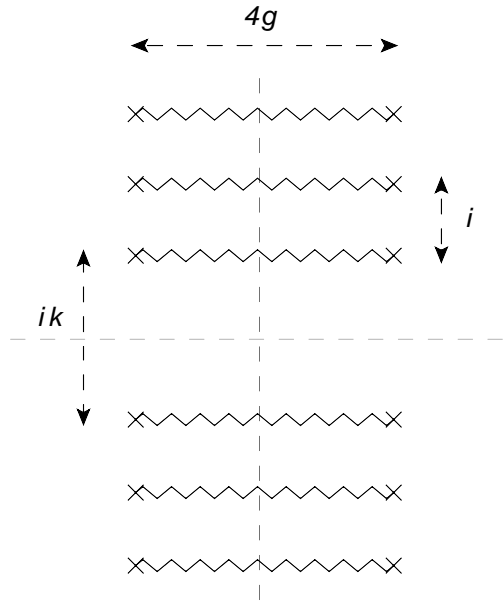}}
\end{picture}%
\caption{Complex $u$-plane for energy and momentum of scalar, gauge field, and bound states. Crosses indicate square-root branch points. Wavy lines indicate a choice of cuts that is natural from the weak coupling perspective. The dashed lines represent the real and imaginary $u$-axis, respectively. The integer $k$ depends on the excitation. It is equal to $1$ for a scalar and to $\ell$ for a bound state of $\ell$ gauge fields. For a large fermion, the picture with $k=2$ can be applied if completed with an extra cut along the real $u$-axis. For a small fermion, we would only keep the latter cut and remove the two semi-infinite towers in the upper- and lower-half plane. For all excitations, within the weak coupling expansion, the sequence of cuts degenerates into a sequence of poles along the imaginaris $u$-axis.}\label{Uplane}
\end{figure}%

Now, to complete our discussion of the singularities in the complex rapidity plane, it remains to list the center-of-mass positions $u_{\textrm{p}}$ of the cuts for all the excitations. They are the same for both energy and momentum of a given excitation. We find that $u_{\textrm{p}}= i(2\mathbb{Z}+1)/2$ for a scalar, $u_{\textrm{p}} = i\mathbb{Z}$ for a large fermion, $u_{\textrm{p}} =0$ for a small fermion, and finally $u_{\textrm{p}} = \pm i(\ell/2 + \mathbb{N})$ for gauge field ($\ell=1$) and bound states ($\ell > 1$). They are depicted in Fig.~\ref{ACpict} and~\ref{Uplane}. As we have already mentioned, some of the expressions we are manipulating are directly related to the resolvants of~\cite{KSV08}. For them the pattern of cuts is in agreement with the findings of~\cite{KSV08} and otherwise stands for a quite straighforward extension. Finally, we recall that the central cut (at $u_{\textrm{p}}=0$) of large and small fermion are glued together and that it disappears when energy and momentum are expressed in terms of $x$. It means that all the singularities in the $x$-plane lie in the large fermion domain, $|x|>g$, and more precisely at $x_{\textrm{b-p}} = x(u_{\textrm{b-p}})$ with $u_{\textrm{b-p}} = \pm 2g + i(\mathbb{Z}-\{0\})$.

\subsubsection*{Quasiclassical limit}

There is another singular point for energy and momentum in the complex rapidity plane, namely the point at infinity. There energy and momentum admit asymptotic expansion, whose leading terms turn out to be controlled by the few observables that parameterize the vacuum energy, i.e. the cusp anomalous dimension and virtual scaling function.

At large rapidity $u$, with $u$ going to infinity without crossing the cuts, energy and momentum are found to scale as
\be\label{LUA}
\begin{aligned}
E_{\star}(u) &=\Gamma_{\textrm{cusp}}(g)\left(\log{u}-\psi(1)\right) + \textrm{twist}_{\star} + {B_{2}(g) \over 2} + O(1/u)\,, \\
p_{\star}(u) &= 2u -{\pi \over 2} \, \Gamma_{\textrm{cusp}}(g) + O(1/u)\, ,
\end{aligned}
\ee
where $\textrm{twist}_{\star} =1$ for $\star =$ hole, fermion, gauge field, and $\textrm{twist}_{\star} =\ell$ for bound states ($\ell>1$). The results~(\ref{LUA}) are given here for $u\sim \infty$ with $\Re{(u)} > 0$ and are easily extended to the case $\Re{(u)} < 0$ by using the parity of energy an momentum under $u\rightarrow -u$. Note also that in the case of fermion, the asymptotics~(\ref{LUA}) holds for a large fermion, since for a small fermion the point $u=\infty$ is associated to $x=0$, that is a fermion at rest. Say differently, the asymptotics~(\ref{LUA}) is a large momentum expansion,
\be\label{LPA}
E_{\star}(p) =\Gamma_{\textrm{cusp}}(g)\left(\log{(p/2)}-\psi(1)\right) + \textrm{twist}_{\star} + {B_{2}(g) \over 2} + O(1/p)\,.
\ee

The logarithmic scaling of the energy~(\ref{LUA}) is a longstanding result~\cite{K95,BGK03,BGK06}. The fact that it translates into the scaling~(\ref{LPA}) is a more recent observation derived at strong coupling and conjectured to hold exactly in~\cite{DL10}. It is directly related to the change of asymptotics~\cite{K95,BGK03,BGK06}
\be
\Delta -S \sim A_{2}(g)\log{S}\rightarrow A_{3}(g)\log{S}\, , \qquad  A_{3}(g) = 3\Gamma_{\textrm{cusp}}(g)\, ,
\ee
accompanying the entering of a hole rapidity in the quasiclassical domain $u \sim S \gg 1$.%
\footnote{This regime is quasiclassical from the point of view of the Baxter equation~\cite{K95,BGK06}.} This large hole brings us far away from the vacuum and closer to some highly-excited trajectories in the spectrum of large spin anomalous dimensions. The string theory dual of these trajectories are the spiky strings~\cite{Kruczenski04}, that generalize the GKP string to configurations with more than two fold-points. Accordingly, at strong coupling, the asymptotics~(\ref{LUA}) is observed in the giant hole regime~\cite{DL10}, where the excitation is semiclassical and connected at large momentum to the spiky-string solutions.

One can easily verify the asymptotics~(\ref{LUA}) at weak coupling by using the formulae~(\ref{WCH}, \ref{WCGF}, \ref{WCLF}), and given the weak coupling expansion for the virtual scaling function~\cite{FRS07,FZ09}
\be
B_{2}(g) = - 24 \zeta(3)g^4 + {16\over 3}\left(\pi^2\zeta(3)+30\zeta(5)\right) g^6 +O(g^{8})\, .
\ee
It is not difficult then, following the hint of the weak coupling expansion, to rewrite appropriately the integrands in the integral representations~(\ref{DRH}, \ref{DRGF}, \ref{DRLF}) such that the result~(\ref{LUA}) becomes apparent at any value of the coupling.

\subsection{From gauge field to bound states}

We are now in position to prove the fusion identities that link energy and momentum of bound states to those of the gauge field. We recall that these relations read
\be\label{FusID}
E_{\ell}(u) = \sum_{j=0}^{\ell-1} E_{\textrm{gf}}(u+ji-i(\ell-1)/2) \, ,
\ee
with $E_{\textrm{gf}}(u) = E_{\ell = 1}(u)$, and similarly for the momenta. They are saying that the twist-$\ell$ excitation with index $\ell$ is a $\ell$-string of gauge field. These identities are actually quite manifest when looking at the contour integral representations~(\ref{CIRa}, \ref{CIRb}), such that the main point is to explain why they are obscured at weak coupling.

Let us focus on the simplest case of a bound state of two gauge fields. We have both $\ell=2$ and
\be\label{FRfbs}
E_{\textrm{first-bs}}(u) = E_{\textrm{gf}}(u+\ft{i}{2}) + E_{\textrm{gf}}(u-\ft{i}{2})\, ,
\ee
where $E_{\textrm{first-bs}}(u) = E_{\ell=2}(u)$. The equality~(\ref{FRfbs}) immediately follows  from the concatenation of the contours of integration when using~(\ref{CIRa}, \ref{CIRb}) to represent the right-hand side of~(\ref{FRfbs}). The same holds true for the momenta. On the other hand, one can easily verify that the identity~(\ref{FRfbs}) is not satisfied by the weak coupling expressions~(\ref{DRGF}). The reason for this apparent disagreement is that the fusion relation~(\ref{FRfbs}) assumes that the rapidity $u$ lies in the strip $-2g < \Re{(u)} < 2g$, while the weak coupling expansion works outside of it. Performing the shifts in~(\ref{FRfbs}) inside and outside this strip are two non-equivalent operations. This is due to the presence of cuts connecting the branch points at $u=\pm 2g\pm i(2\mathbb{Z}+1)/2$, in the complex rapidity plane for energy (and momentum) of a gauge field.

\begin{figure}[h]%
\setlength{\unitlength}{0.14in}
\centering
\begin{picture}(13,16)%
\put(-9,-1){\insertfig{5.5}{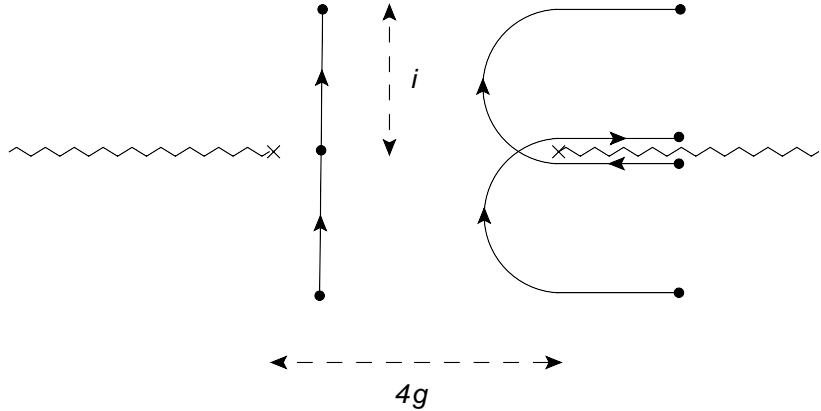}}
\end{picture}%
\caption{Fusion of two gauge-field excitations into a bound state. For a rapidity $u^2 <  (2g)^2$, depicted on the left, the two shifted contours of integration add up to form the contour of integration for the first bound state. For a rapidity $u^2 >  (2g)^2$, depicted on the right, the two shifted contours of integration hit the cut and extract an extra contribution fom the discontinuity.}\label{fusionpict}
\end{figure}

To verify the previous assertion we shall perform the shift in~(\ref{FRfbs}) for arbitrary $u$, assumed real for simplicity. We get that
\be\label{Edisc}
E_{\textrm{gf}}(u+\ft{i}{2}) + E_{\textrm{gf}}(u-\ft{i}{2}) = E_{\textrm{first-bs}}(u) + R_{\textrm{disc}}(u)\, ,
\ee
where the remainder function $R_{\textrm{disc}}(u)$ is related to the discontinuity of $\Lambda(v)$ through its cut, see Fig~\ref{fusionpict}. An explicit expression for the latter function is
\be\label{Rdisca}
R_{\textrm{disc}}(u) = \int_{0}^{\infty}{dt \over t}\left[{\gamma^{\o}_{+}(2gt) \over 1-\e^{-t}}-{\gamma^{\o}_{-}(2gt) \over \e^{t}-1}\right](\cos{(ut)}-1)\, .
\ee

By means of the representation~(\ref{Rdisca}), we verify  that $R_{\textrm{disc}}(u) = 0$ when $u^2 <(2g)^2$, in virtue of the first equation in~(\ref{SystII}) and in agreement with our previous discussion. However, the remainder function $R_{\textrm{disc}}(u)$ is non-zero for $u^2 > (2g)^2$, and this is why the identity~(\ref{FRfbs}) does not hold at weak coupling. A similar analysis carries over for the momentum, for which one can find
\be\label{pdisc}
p_{\textrm{gf}}(u+\ft{i}{2}) + p_{\textrm{gf}}(u-\ft{i}{2}) = p_{\textrm{first-bs}}(u) + R'_{\textrm{disc}}(u)\, ,
\ee
where
\be
R'_{\textrm{disc}}(u) = 2u-\int_{0}^{\infty}{dt \over t}\left[{\gamma^{\o}_{-}(2gt) \over 1-\e^{-t}}+{\gamma^{\o}_{+}(2gt) \over \e^{t}-1}\right]\sin{(ut)}\, .
\ee

To some extent, the fusion identities~(\ref{FusID}) still make sense in the domain $(\Re(u))^2 > (2g)^2$, but to interpret them correctly we need to specify that the gauge-field rapidities in~(\ref{FusID}) are not all lying on the same sheet. Only one of them can actually be on the physical sheet, that is on the sheet that contains the real-momentum line.

\section{Strong coupling regimes}\label{SCDRs}

In this section, we analyze the dispersion relations at strong coupling.  We will show that the latters admit three qualitatively distinct regimes, that parallel the ones found for a magnon over the BMN vacuum~\cite{BMN, MS06, HM}. There are the perturbative regime~\cite{FT02,AM07} with $E, p \sim g^0$, the near-flat space regime with $E, p \sim g^{1/4}$, and the giant hole regime~\cite{DL10} with $E, p \sim g$. The near-flat space regime interpolates between the perturbative and semiclassical domain. The expansion parameter is $1/g$ in the latter two cases and $1/\sqrt{g}$ in the former.

In the perturbative regime and to leading order at strong coupling, the dispersion relations are all relativistic and the pattern of masses agrees with the one obtained directly in string theory~\cite{FT02, AM07}.%
\footnote{Up to the missing mass 2 boson.} In the near-flat space regime, all excitations are found to be massless with dispersion relation $\tilde{E} = \tilde{p}$, to leading order at strong coupling and at fixed $\tilde{E} = E/g^{1/4}, \tilde{p} = p/g^{1/4}$. The leading-order dispersion relation in the giant hole regime is also independent on the flavor of the excitation. Its expression, derived below, agrees with the one obtained in~\cite{DL10} using the finite gap technology.

For scalar only, there is an extra regime of exponentially small energy and momentum, $E, p \sim \e^{-\pi g}$. The presence of this regime was predicted and elucidated in~\cite{AM07} and it originates from the non-trivial infrared dynamics of the transverse fluctuations in $\rm S^{5}$, which turns out to be controlled by the non-linear O(6) sigma model. At the perturbative level, one has 5 massless Goldstone bosons associated to the spontaneous symmetry breaking of O(6) down to O(5). But the symmetry is restored non-perturbatively and one ends up with a O(6) vector multiplet of massive particles with mass $m \sim \e^{-\pi g}$. This non-perturbative regime was previously observed to emerge from the Bethe ansatz equations in~\cite{FGR08a,BK08,FGR08b} in the study of the generalized scaling function~\cite{FRS07}.

\subsection{Scalar}

We start our analysis with the dispersion relation for a scalar in the non-perturbative regime, $E, p \sim \e^{-\pi g}$. It appears when the Bethe rapidity $u$ is kept of order $O(1)$ while taking $g\gg1$. When the rapidity gets larger, but remains in the domain $1 \ll u \ll  2g$ (for right mover), we enter the perturbative regime with $E, p \sim 1$.  We will analyze it up to $O(1/g)$ correction. For $u - 2g \sim 1$, we fall in the near-flat space regime with $E, p \sim g^{1/4}$. The perturbative expansion is reorganized in this regime and the expansion parameter becomes $1/\sqrt{g}$. We will construct the dispersion relation in the near-flat space regime up to the next-to-leading order. Eventually, for a rapidity $u\gg 2g$, we find a semiclassical regime where $E, p \sim g$. This is the giant hole regime~\cite{DL10}, that we will study to leading order at strong coupling.

\subsubsection*{Non-perturbative regime}

To deal conveniently with the non-perturbative regime, we extend and apply the representation of~\cite{BK08,BK09} providing energy and momentum in the form of hyperbolic Fourier series.%
\footnote{This representation will not be deduced here from the general formula~(\ref{DRH}) and the reader is referred to~\cite{BK08,BK09} for the methodology.}
It reads
\be\label{DRhnp}
\begin{aligned}
E &= \sum_{n\ge 0}\bigg[(-1)^{n}m_{4n+1}\cosh{((4n+1)\theta)} - (-1)^{n}m_{4n+3}\cosh{((4n+3)\theta)}\bigg]\, , \\
p &= \sum_{n\ge 0}\bigg[(-1)^{n}m_{4n+1}\sinh{((4n+1)\theta)} \, +\,   (-1)^{n} m_{4n+3}\sinh{((4n+3)\theta)}\bigg]\, ,
\end{aligned}
\ee
with $\theta$ given in terms of the Bethe rapidity as $\theta \equiv \pi u/2$.  The representation~(\ref{DRhnp}) is exact but it assumes $\theta^2 < (\pi g)^2$, i.e. $u^2 <(2g)^2$, for the sums to have a chance to converge. The coefficients $m_{4n+1}, m_{4n+3}$, are functions of the coupling constant only and they can be expressed in terms of the solution to the BES equation, see Eqs.~(\ref{GMasses}, \ref{SCGMasses}). For the time being, we only need to note that these coefficients are exponentially suppressed with the coupling as $m_{n} \sim \exp{(-n\pi g)}$.

It immediately follows that, at strong coupling and for a fixed rapidity $\theta$, the dispersion relation becomes relativistic~\cite{AM07,BK08}
\be\label{DRhnplo}
E^2-p^2 = m_{1}^2\, ,
\ee
up to subleading $\sim \exp{(-4\pi g)} \sim m_{1}^4$ corrections. The coefficient $m_{1}$, entering~(\ref{DRhnplo}), is known to admit the strong coupling expansion~\cite{FGR08a,BK08}
\be\label{mass-gap}
m_{1} = k\, g^{1/4}\e^{-\pi g}\bigg[1+{3-6\log{2} \over 32 \pi g} + O(1/g^2)\bigg]\, , \qquad k = 2^{3/4}\pi^{1/4}/\Gamma(5/4)\, ,
\ee
where the (embedding) parameter $k$ was first obtained in~\cite{AM07} from direct string theory computation~\cite{FTT06}. The relation~(\ref{DRhnplo}) allows us to identify the six scalars of the gauge theory with the vector multiplet of excitations of the non-linear O(6) sigma model, which stands for the low-energy effective theory of the GKP string~\cite{AM07}.

To go beyond the leading order expression~(\ref{DRhnplo}), it is convenient to first introduce the rescaled variables $\varepsilon = E/m_{1}, \rho = p/m_{1}$. Then the dispersion relation can be found as an expansion in powers of $m_{1}^2$ at fixed $\rho$. For the first correction to~(\ref{DRhnplo}), we need to consider
\be\label{DRnph}
\begin{aligned}
\varepsilon &= \cosh{\theta}-c\, m_{1}^2 \cosh{(3\theta)} + O(m_{1}^{4})\, , \\
\rho &= \sinh{\theta}\, +c\, m_{1}^2 \sinh{(3\theta)} \, + O(m_{1}^{4})\, ,
\end{aligned}
\ee
where $c \equiv m_{3}/m_{1}^3$. The relation between $\rho$ and $\theta$ is easily inverted as
\be\label{tr-rel}
\theta = \textrm{arcsinh}{\rho}-c\, {3\rho + 4\rho^3 \over \sqrt{1+\rho^2}}m_{1}^2 + O(m_{1}^{4})\, ,
\ee
leading to the dispersion relation
\be\label{nlDRnph}
\varepsilon = \sqrt{1+\rho^2} -c\, {1+8\rho^2+8\rho^4 \over \sqrt{1+\rho^2}} m_{1}^2 + O(m_{1}^{4})\, .
\ee
The coefficient $c$ is evaluated using~(\ref{SCGMasses}) and reads
\be\label{chnp}
c \equiv {m_{3} \over m_{1}^{3}} = {\Gamma(\ft{1}{4})^4 \over 64(12)^{1/4}\pi^{2} g} + O(1/g^2)\, .
\ee
Looking at the equality~(\ref{nlDRnph}) we note the emergence of singularities at $\rho = \pm i$. It is a fake actually, and it stands for a mass renormalization. In our units, the mass is equal to $1$ at leading order, but it receives a correction $\sim cm_{1}^2$. Taking this effect into account leads to
\be\label{DRhnpbis}
\varepsilon = \sqrt{m^2+\rho^2}\bigg[1 -8 c \rho^2 m_{1}^2 + O(m_{1}^{4})\bigg]\, ,
\ee
with the (rescaled) mass $m = 1 -cm_{1}^2 +O(m_{1}^4)$. Now the factor in square bracket includes the genuine corrections only. The dependence on the momentum $\rho$ of the first correction in~(\ref{DRhnpbis}) suggests that it originates from higher-derivative operators in the effective (low-energy) Lagrangian density for the O(6) excitations. These irrelevant interactions break the boost invariance in agreement with the fact that it is a low-energy symmetry only~\cite{AM07}.

It is not complicated, though tedious, to include subleading corrections to~(\ref{DRhnpbis}) of order $O(m_{1}^{2n})$ with $n\ge 2$. Doing so, one would observe that higher powers of $\rho$ are generated at each steps. There is a bound, however, in the sense that the large $\rho$ asymptotics of the $O(m_{1}^{2n})$ contribution cannot exceed $\rho^{2n}$. It points toward the existence of a non-trivial scaling limit when $\rho \gg 1$ with $\rho m_{1}$ kept fixed. This is, of course, the string perturbative regime, with $p = \rho m_{1}=O(1)$ and $m_{1}\rightarrow 0$, that we analyze below. As a final remark, we notice that in terms of the rapidity $\theta$ the perturbative regime requires the scaling
\be
\theta - \pi g + \ft{1}{4}\log{g} = \textrm{finite} \,,
\ee
as follows from~(\ref{tr-rel}) and~(\ref{mass-gap}), for right mover $(\theta >0)$. It means that $1 \ll \theta \ll \pi g$, or equivalently $1\ll u \ll 2g$.
 
\subsubsection*{Perturbative regime}
 
To deal with the string perturbative regime, it is convenient to first reparameterize the dispersion relation~(\ref{DRhnp}), keeping only the terms that are relevant when $m_{1} \ll p \sim 1$. For right mover, it amounts to neglecting all contributions to energy and momentum which are powers of $\e^{-\theta}$ in~(\ref{DRhnp}). Doing so and introducing a new variable, $z$, as
\be
\theta = -\log{m_{1}} + \log{z}\, , 
\ee
which is kept finite when $g\gg 1$, we get from~(\ref{DRnph})
\begin{align}\label{p-dr}
E &= \ft{1}{2}\sum_{n\ge 0}\left[(-1)^n c_{n}z^{4n+1}-(-1)^n d_{n}z^{4n+3}\right]\, , \notag \\
p &= \ft{1}{2}\sum_{n\ge 0}\left[(-1)^n c_{n}z^{4n+1}+(-1)^n d_{n}z^{4n+3}\right]\, ,
\end{align}
with the coefficients, see Eq.~(\ref{SCGMasses}),
\be\label{cdhp}
c_{n} = {m_{4n+1} \over m_{1}^{4n+1}} \sim g^{-n}\, , \qquad d_{n} = {m_{4n+3} \over m_{1}^{4n+3}} \sim g^{-n-1}\, .
\ee
Note that, by definition, $c_{0} = 1$. The relations~(\ref{p-dr}) are valid up to non-perturbative contributions $\sim m_{1}^2/p$, but they include all perturbative corrections in $1/g$. Strong coupling expansion for the coefficients $c_{n}, d_{n}$,  can be computed order by order in $1/g$ from the solution to the BES equation, and their leading expressions are given in~(\ref{SCGMasses}). Finally, we note that, at a given order in $1/g$, the infinite sums in~(\ref{p-dr}) truncate, see Eq.~(\ref{cdhp}), and the dispersion relation $E = E(p)$ is easily obtained.

Applying~(\ref{p-dr}) up to $O(1/g)$ contribution, which requires expanding~(\ref{p-dr}) up to $O(z^5)$, we find the dispersion relation%
\footnote{Incidentally, the expression~(\ref{Ehp}) could have been obtained immediately by taking the large $\rho$ asymptotics of~(\ref{DRhnpbis}). It does not mean however that there is a one-to-one correspondence between $O(1/g^n)$ correction in~(\ref{Ehp}) and $O(m_{1}^{2n})$ contribution in~(\ref{DRhnpbis}). For instance, to get the full answer at $O(1/g^2)$, from the asymptotics of~(\ref{DRhnpbis}), one would have to include corrections up to $O(m_{1}^{6})$.}
\be\label{Ehp}
E =  p\left[1 -{\Gamma(\ft{1}{4})^4 \, p^2 \over 8(12)^{5/4}\pi^{2} g}+O(1/g^2)\right]\, .
\ee
The leading term, $E=p$, is in agreement with the string theory result~\cite{FT02}. Moreover, if interpreted as a low-momentum expansion, the relation $E=p+O(p^3)$ holds true to all loops, that is up to non-pertubative corrections $\sim m^2_{1}/p$. This has to be attributed to the identification of a scalar with a Goldstone fluctuations in $\rm S^{5}$~\cite{AM07}.

For illustrative purposes, let us include the $O(1/g^2)$ correction to~(\ref{Ehp}). They are obtained by working at $O(z^9)$ in~(\ref{p-dr}). The dispersion relation then reads
\be\label{Eill}
E =  p\left[1 + {\alpha p^2 \over g} + {\beta p^2 + \gamma p^4 + \delta p^6 \over g^2} + O(1/g^3)\right]\, ,
\ee
where the explicit values of $\alpha, \ldots , \delta$, are not relevant here. We see that a new pattern appears. Namely, increasing the order in $1/g$ comes with an increment of the maximal power of $p$ by $4$ units. This is not apparent if only the $O(1/g)$ correction is considered. It suggests a non-trivial scaling when $p$ gets as large as $g^{1/4}$, with the energy given by
\be\label{Eillb}
E-p = g^{-1/4}\left[\alpha\left(p \over g^{1/4}\right)^{3} + \delta\left(p \over g^{1/4}\right)^{7} + \ldots \right] + O(1/g^{-3/4})\, .
\ee
As we shall see, the regime $p\sim g^{1/4}$ interpolates between the perturbative and semiclassical domain, and it is therefore the analogue of the near-flat space regime for magnon above the BMN vacuum~\cite{MS06}. The difference with the BMN case is that energy and momentum scale like $g^{1/4}$ and that the expansion runs in inverse powers of $1/\sqrt{g}$.

Notice that assuming that $p\sim g^{1/4}$, or equivalently $z\sim g^{1/4}$, means that $\theta -\pi g$ is kept finite for $g\gg 1$. In terms of the Bethe rapidity it means that $u-2g = O(1)$. It is important to observe that $u$ gets not rescaled when working in the near-flat space regime. Indeed, we recall that energy and momentum for a scalar have branch points at $u = 2g \pm i(2n+1)/2$, with $n\in \mathbb{N}$ (and similarly for $u<0$). Rescaling $u$ by a power of $g$ would cause the cuts to collide, leading to singular behavior at $u = 2g$~\cite{KSV08}. This is not what we want as we would like to transit between $u<2g$ and $u>2g$. Later we will perform the rescaling $u \rightarrow \bar{u} = u/(2g)$ in the giant hole regime $\bar{u}>1$.

\subsubsection*{Near-flat space regime}
 
In the near flat space limit, we define $\xi = \theta-\pi g$ that is kept fixed for $g\gg 1$. In terms of this new variable, the overlap with the large momentum limit in the perturbative regime is at $\xi = -\infty$. It is also convenient to introduce rescaled energy and momentum as $\tilde{E} = E/g^{1/4}$ and $\tilde{p} = p/g^{1/4}$. The dispersion relation can then be written parametrically as
\be\label{DRhnfs}
\tilde{E} = \chi_{0}(\xi) -  \chi_{1}(\xi)/\sqrt{g}\, , \qquad \tilde{p} = \chi_{0}(\xi) +  \chi_{1}(\xi)/\sqrt{g}\, , 
\ee
where the functions $\chi_{0, 1}(\xi)$ are given by
\be\label{Cho}
\chi_{0}(\xi) = {1 \over 2g^{1/4}}\sum_{n\ge 0}^{\infty}(-1)^n m_{4n+1}\e^{(4n+1)(\pi g+\xi)}\, , \qquad \chi_{1}(\xi) = {g^{1/4} \over 2}\sum_{n\ge 0}^{\infty}(-1)^n m_{4n+3}\e^{(4n+3)(\pi g+\xi)} \, ,
\ee
with the coefficients $m_{4n+1}, m_{4n+3},$ the same as before. Note that the dispersion relation~(\ref{DRhnfs}) is just a rewriting of~(\ref{p-dr}) and thus is also exact up to non-perturbative corrections. The main difference with the perturbative regime is that we cannot truncate the sums in~(\ref{Cho}), since all contributions are of the same order in $g$ at given $\xi$, see~(\ref{SCGMasses}). The functions $\chi_{0, 1}(\xi)$ defined in~(\ref{Cho}) are both of order $O(1)$ at strong coupling ($\xi$ fixed), and admit an expansion in $1/g$. To leading order, we find, see~(\ref{Chi}, \ref{Chiloa}, \ref{Chilob}),
\be\label{Chob}
\begin{aligned}
\chi_{0}(\xi) &= {4 \over (2\pi)^{3/4}}\int_{0}^{\infty}ds\, s^{-1/4}  \left(1+\e^{-4(\xi-s)}\right)^{-1/4} + O(1/g)\, ,\\
\chi_{1}(\xi) &= {4 \over (2\pi)^{5/4}}\int_{0}^{\infty}ds\, s^{1/4} \left(1+\e^{-4(\xi-s)}\right)^{-3/4} + O(1/g)\, .
\end{aligned}
\ee
These functions are positive and increase with $\xi$. They are exponentially small at large negative $\xi$ and scale like some fractional powers of $\xi$ at large positive $\xi$.

Since the functions $\chi_{0,1}(\xi)$ are of the same order in the coupling at given $\xi$, it immediately follows from~(\ref{DRhnfs}) that the dispersion relation in the near-flat space regime can be written as
\be
\tilde{E} = \tilde{p}  + O(1/\sqrt{g})\, . 
\ee
This result is actually universal, in the sense that its holds true for any excitation (scalar, fermion, etc) in the regime $E\sim p \sim g^{1/4}$. The subleading $O(1/\sqrt{g})$ corrections are however different in each case. For the scalar, this correction is controlled by the functions~(\ref{Chob}),
\be
\tilde{E} - \tilde{p} = - 2\chi_{1}(\xi)/\sqrt{g} + O(1/g)\, , \qquad \tilde{p} = \chi_{0}(\xi) + O(1/\sqrt{g})\, .
\ee
It does not seem possible to solve this parametric representation for generic $\xi$. But it can be done easily for the two extremal values $\xi = \pm \infty$. At large negative $\xi$, see Eq.~(\ref{lnxia}), the near-flat-space momentum $\tilde{p}$ is small, $\tilde{p} \ll 1$, and the $O(1/\sqrt{g})$ correction to the near-flat-space energy scales like $\sim \tilde{p}^3$. This is just saying that the dispersion relation takes the form~(\ref{Eillb}), which is expected as $\xi \sim -\infty$ overlaps with the large momentum limit $1 \ll p \, (\ll g^{1/4})$ taken in the perturbative regime. At large positive $\xi$, using asymptotics quoted in~(\ref{lpxia}), we find that the near-flat-space momentum is large and that the $O(1/\sqrt{g})$ correction to the energy $\tilde{E}$ reads
\be\label{lmnfsa}
\tilde{E} - \tilde{p} = \left[-{1 \over 5}\left({3\tilde{p} \over 2}\right)^{5/3} + \ldots \right]/\sqrt{g} + O(1/g)\, , \qquad \tilde{p} \gg 1.
\ee
We note that for $\xi = \theta - \pi g \gg 1$, the Bethe rapidity $u = 2\theta/\pi$ has become larger than $2g$. It means that we are entering the giant hole regime, which is the last destination of our trip. If the near-flat space regime does indeed interpolate between perturbative and semiclassical regime, the large momentum asymptotics~(\ref{lmnfsa}) should match the small momentum expansion in the giant hole regime. In the latter domain, energy and momentum are of order $O(g)$ and it is better to work with $\mathcal{E} = E/(2g) = \tilde{E}/(2g^{3/4})$ and $\mathcal{P} = p/(2g) = \tilde{p}/(2g^{3/4})$. In terms of these variables the identity~(\ref{lmnfsa}) can be rephrased as
\be\label{lmnfsb}
\mathcal{E} = \mathcal{P} -{1 \over 10}\left(3\mathcal{P}\right)^{5/3} + \ldots\, ,
\ee
and it should have the meaning of   a low momentum expansion, $\mathcal{P} \ll 1$. This is effectively the correct expression~\cite{DL10}, as proved below.

\subsubsection*{Giant hole regime}

To obtain the dispersion relation for a giant hole, it is actually simpler to start with the formula~(\ref{DRH}) and immediately take the limit $g\rightarrow \infty$ with $\bar{u} \equiv u/(2g)$ kept fixed and $\bar{u}^2 >1$. Then, to leading order, we need to evaluate
\be\label{GHexp}
\begin{aligned}
&\mathcal{E}(\bar{u}) \equiv E(u)/(2g) = \int_{0}^{\infty}{dt \over t^2}\left(\gamma^{\o}_{+}(t) - \gamma^{\o}_{-}(t)\right)\left(\cos{(\bar{u}t)}-1\right)+ O(1/g)\, , \\
&\mathcal{P}(\bar{u}) \equiv p(u)/(2g) = 2\bar{u} - \int_{0}^{\infty}{dt \over t^2}\left(\gamma^{\o}_{+}(t) + \gamma^{\o}_{-}(t)\right)\sin{(\bar{u}t)}+ O(1/g)\, ,
\end{aligned}
\ee
with $\bar{u}^2 > 1$. These expressions are actually common to any excitation carrying rapidity $\bar{u}^2>1$, meaning that the dispersion relation in the giant hole regime is flavor independent. Note that in the case of fermion, the formulae~(\ref{GHexp}) are obtained in the large rapidity domain.

The integrals~(\ref{GHexp}) are most easily taken after differentiating with respect to $\bar{u}$. The two constants of integration can be fixed afterwards with the help of the large $u$ asymptotics~(\ref{LUA}) and of the known strong coupling expression for both the cusp anomalous dimension~\cite{GKP02} and the virtual scaling function~\cite{FZ09}. Using the leading order solution to the BES equation~\cite{AABEK07},
\be
\gamma^{\o}_{+}(t) + i \gamma^{\o}_{-}(t) = {\sqrt{2}it \over \pi}\int_{-1}^{1}dv\, \left({1+v \over 1-v}\right)^{1/4}\e^{-ivt} + O\left(1/g\right)\, , 
\ee
one easily obtains that
\be
{d\over d\bar{u}}\mathcal{E} = \left({\bar{u}+1 \over \bar{u}-1}\right)^{1/4}-\left({\bar{u}-1 \over \bar{u}+1}\right)^{1/4}\, , \qquad {d\over d\bar{u}}\mathcal{P} = \left({\bar{u}+1 \over \bar{u}-1}\right)^{1/4}+\left({\bar{u}-1 \over \bar{u}+1}\right)^{1/4}\, ,
\ee
where we have dropped the $O(1/g)$ loop corrections. Now to integrate these expressions, it is helpful to switch to the Zhukowsky rapidity $\bar{x} = \bar{u} + \sqrt{\bar{u}^2-1}$. One gets
\be\label{CII}
\mathcal{E}= \int_{1}^{\bar{x}}{dz \over z}\sqrt{1-1/z^2} + C_{1}\, , \qqqquad \mathcal{P} = \int_{1}^{\bar{x}}dz\sqrt{1-1/z^2} +C_{2}\, .
\ee
The correct choice for the constants of integration will turn out to be $C_{1}=C_{2}=0$. Setting them to zero and performing the integrals, we obtain the semiclassical dispersion relation as
\be\label{DRgh}
\begin{aligned}
\mathcal{E} &= {1\over 2}\log{\left({1+\sqrt{1-1/\bar{x}^2} \over 1-\sqrt{1-1/\bar{x}^2}}\right)} - \sqrt{1-1/\bar{x}^2}\,,\\
\mathcal{P} &=  \bar{x}\sqrt{1-1/\bar{x}^2} -\textrm{arctan}{\left(\bar{x}\sqrt{1-1/\bar{x}^2}\right)} \, .
\end{aligned}
\ee
A simple reparameterization leads to the dispersion relation for a giant hole found in~\cite{DL10}.

The dispersion relation~(\ref{DRgh}) implies that both energy and momentum are small for $\bar{x} \sim 1^+$. Expanding around this point, we verify that at low-momentum the energy scales as~\cite{DL10}
\be
\mathcal{E} = \mathcal{P} -{1\over 10}\left(3\mathcal{P}\right)^{5/3} + O(\mathcal{P}^{7/3})\, ,
\ee
confirming the overlap with the large-momentum limit in the near-flat space regime, see Eqs.~(\ref{lmnfsa}, \ref{lmnfsb}). In the opposite limit, $\bar{x}\rightarrow \infty$, energy and momentum are large. They read~\cite{DL10}
\be\label{GHlu}
\mathcal{E} = \log{(4\bar{u})} -1 + O(1/\bar{u}^2)\, , \qquad \mathcal{P} = 2\bar{u} -{\pi \over 2} + O(1/\bar{u})\, ,
\ee
where we have restored the (rescaled) Bethe rapidity $\bar{u} = \bar{x}/2 + O(1/\bar{x})$, with $\bar{u} \gg 1$. Rescaling energy, momentum and rapidity, the identities~(\ref{GHlu}) become
\be
E =  2g(\log{(2u/g)}-1) + O(1/u^2)\, ,\qquad p = 2u -\pi g+ \mathcal{O}(1/u)\, .
\ee
This is in agreement with the general expression~(\ref{LUA}) for the large $u$ asymptotics of the dispersion relation, given that the cusp anomalous dimension and virtual scaling function, entering~(\ref{LUA}), are given by~\cite{GKP02,FT02,FZ09}
\be
\Gamma_{\textrm{cusp}}(g) = 2g + O(1) \, , \qquad B_{2}(g) = 4g\left(\log{(2/g)}+\psi(1)\right) - 4g + O(1)\, .
\ee
We verify, by this way, that the choice $C_{1}=C_{2}=0$ for the constants of integration in~(\ref{CII}) was the correct one.

\subsection{Fermion}

Our investigation of the dispersion relation for a fermion will be mostly restricted to the perturbative regime. This is the regime that is connected to the point $x=0$ (fermion at rest) and it is found in the small rapidity domain. It means that we shall start with the formulae~(\ref{DRSF}). The strong coupling expansion is taken at a fixed value of the rescaled rapidity $\bar{u} = u/(2g)$, with $\bar{u}^2 > 1$ for a real momentum fermion. For a small fermion it means that the physical (rescaled) rapidity $\bar{x} = x/g = \bar{u}-\sqrt{\bar{u}^2-1}$ fulfills $\bar{x}^2 < 1$. Under the same assumption on $\bar{u}$, a large fermion would have rapidity $\bar{x} = \bar{u}+\sqrt{\bar{u}^2-1}$ satisfying $\bar{x}^2 > 1$. This is the giant hole regime, with the same dispersion relation as found for a scalar, to leading order at strong coupling.%
\footnote{Indeed, starting from the representation~(\ref{DRLF}) for the dispersion relation of a large fermion, and taking the strong coupling limit at $\bar{u}$ fixed ($\bar{u}^2 >1$), one would get the same expression as the one derived in~(\ref{GHexp}) for a scalar with rapidity $\bar{u}^2 >1$.} The near-flat space regime controls the transition between $\bar{x}^2 < 1$ and $\bar{x}^2 > 1$.  Its analysis will not be done thoroughly, but we will verify that the leading-order dispersion relation $\tilde{E} = \tilde{p} + O(1/\sqrt{g})$ holds true with $\tilde{E} = E/g^{1/4}$ and $\tilde{p} = p/g^{1/4}$ kept fixed at strong coupling.
 
\subsubsection*{Perturbative regime}

For evaluating energy and momentum of a fermion in the perturbative regime, we start with the representation~(\ref{EpPsf}) and look for a strong coupling expansion, assuming $\bar{u} = u/(2g)$ fixed ($\bar{u}^2 > 1$). It is obtained by expanding at large $g$ and fixed $t$ the function $\gamma^{\o}(t)$. Using the known strong coupling expansion for the latter function~\cite{AABEK07,BKK07,KSV08,BK09} and the formula~(\ref{EpPsf}), one easily obtains the dispersion relation in the form
\be\label{DRfa}
E+p = \left(1+\bar{x} \over 1-\bar{x}\right)^{1/2}f(\bar{x})\, ,
\ee
with the function $f(\bar{x})$ given up to $O(1/g^2)$ in Appendix~\ref{AppSC}. Since $(E, p) \rightarrow (E, -p)$ under the transformation $\bar{x} \rightarrow -\bar{x}$,  energy and momentum read
\be\label{DRfc}
\begin{aligned}
E &= \left(1+\bar{x} \over 1-\bar{x}\right)^{1/2}f(\bar{x})+\left(1-\bar{x} \over 1+\bar{x}\right)^{1/2}f(-\bar{x}) \, , \\
p &= \left(1+\bar{x} \over 1-\bar{x}\right)^{1/2}f(\bar{x}) -\left(1-\bar{x} \over 1+\bar{x}\right)^{1/2}f(-\bar{x})\, ,
\end{aligned}
\ee
while the dispersion relation is most easily obtained by means of
\be\label{DRfb}
E^2-p^2 = f(\bar{x})f(-\bar{x})\, .
\ee
Notice that the representation~(\ref{DRfc}) should not be interpreted as saying that energy and momentum have branch points at $\bar{x}^2=1$. This is actually the case to any order in the strong coupling expansion, but, as we already said, energy and momentum are regular for $x$ real. What happens is that the strong coupling expansion is singular at $\bar{x}^2 = 1$. All the branch points, lying in the large fermion domain, collide at strong coupling in the rescaled rapidity $\bar{x} = x/g$. They accumulate at $\bar{x}^2=1$ generating the illusion that energy and momentum are singular at $x^2=1$. The resummation of the $1/g$ expansion would reveal that this is not the case~\cite{KSV08}.

After this brief remark, let us come back to the dispersion relation~(\ref{DRfb}). To leading order at strong coupling, one easily find that $f(\bar{x}) =1 + O(1/g)$, giving the relativistic dispersion law of a mass $1$ particle,
\be
E^2-p^2 = 1\, ,
\ee
in agreement with string theory~\cite{FT02}. The relation between momentum $p$ and rapidity $\bar{x}$ takes the form
\be\label{xpf1}
\bar{x} = {p\over \sqrt{1+p^2}}\bigg[1+O(1/g)\bigg]\, ,
\ee
meaning that $\bar{x}$ is the relativistic velocity of the fermion. It is not difficult to include higher-loop corrections. To $O(1/g^2)$ accuracy, one finds the dispersion relation (see Appendix~\ref{AppSC})
\be\label{Epfa}
E = \sqrt{1+p^2}\left[1- {p^2\over 4g} + {\left(\alpha +\beta p^{2}+\gamma p^{4} \right)p^{2} \over g^2} + O(1/g^3)\right]\, ,
\ee
with
\be\label{Epfb}
\alpha = {\textrm{K} \over 2\pi^2} - {3 \log{2} \over 16\pi}\, , \qquad  \beta = {1\over 96}+{4\textrm{K} \over \pi^2} \, , \qquad \gamma = -{1\over 12} +{4\textrm{K} \over \pi^2} \, ,
\ee
where $\textrm{K}$ is the Catalan's constant. The first correction to the $(\bar{x}, p)$ relation reads
\be\label{xpf}
\bar{x} = {p\over \sqrt{1+p^2}}\bigg[1+(a+b p^2)/g + O(1/g^2)\bigg]\, ,
\ee
with
\be\label{xpfbis}
a =  {3\log{2} \over 4\pi}\, , \qquad b =  {1\over 4} + {3\log{2} \over 2\pi}\, .
\ee

Looking at~(\ref{Epfa}), we verify the absence of correction to the mass, $E(p=0) =1$. We also observe that the identity $E(p=\pm i)=0$ is satisfied~\cite{AM07}. It is so to all orders at strong coupling. We note that we can reach the points $p=\pm i$ by sending $\bar{x}$ to infinity along the imaginary axis, $\bar{x} = \pm i \infty$, see Eq.~(\ref{xpf1}). To get to these points we can think starting at $\bar{x} = 0$, at which $p=0$ and $E=1$, and then move along the imaginary axis. At intermediate values $\bar{x}=\pm i$, we are at $p=\pm i/\sqrt{2}$ and the energy has decreased up to $E = 1/\sqrt{2}$. At these points, we are about to leave the small fermion domain, since in terms of the Bethe rapidity we are right above or below the cut at $\bar{u} = \pm 0$. If we go further we enter the large fermion domain and thus taking $\bar{x} \rightarrow \pm i\infty$ requires analytically continuing through the cuts in the upper- and lower-half plane of this domain. Here the cuts are not apparent since, as we have said before, they have collided in the current scaling. It should be clear however that the infinity we need to consider is different from the one we were discussing to establish the logarithmic scaling of the energy.

\subsubsection*{Near-flat space regime}

The near-flat space regime shows up when considering the large momentum limit of the dispersion relation~(\ref{Epfa}). Namely, keeping the leading contribution at $p\sim \infty$ order by order in the $1/g$ expansion~(\ref{Epfa}), we find that%
\footnote{One needs also to include the subleading classical contribution, $\sqrt{1+p^2} = p+1/(2p)+\ldots$, which turns out to be of the same order as the loop corrections in the near flat-space regime.}
\be\label{nfsf}
E -p = {1\over 2p}-{p^3 \over 4g} + {\gamma p^7 \over g^2} + \ldots\, ,
\ee
with $\gamma$ a numerical constant, see Eq.~(\ref{Epfb}). We immediately recognize the pattern observed previously for scalar. Namely, the suppression by a power of $1/g$ comes with an enhancement by $p^4$. The near-flat-space regime resums these contributions at fixed energy $\tilde{E} = E/g^{1/4}$ and momentum $\tilde{p} =p/g^{1/4}$. In this scaling limit, the dispersion relation~(\ref{nfsf}) reads
\be
\tilde{E}-\tilde{p} = \delta \tilde{E}/\sqrt{g}+ O(1/g)\, ,
\ee
with the correction $\delta \tilde{E}$ determined at small $\tilde{p}$ by~(\ref{nfsf}). In terms of the rapidity $\bar{x}$, the near-flat space regime $p\sim g^{1/4}$ corresponds to working around $\bar{x}^2 \sim 1$. More precisely, using~(\ref{xpf}), we get
\be
x -g = \sqrt{g}\, \delta x(\tilde{p}) + \ldots\, ,
\ee
where we have rescaled to the original rapidity, $x = g\bar{x}$, and introduced $\delta x(\tilde{p}) = -1/2\tilde{p}^2+ O(\tilde{p}^2)$. For the Bethe rapidity, $u = x+g^2/x$, it means that $u - 2g = \delta x(\tilde{p})^2 + \ldots\, $. As we vary $\tilde{p}$ from $0$ to $+\infty$, we expect that $\delta x(\tilde{p})$ will range from $-\infty$ to $+\infty$ and in particular changes sign. At some intermediate value of $\tilde{p}$ the quantity $\delta x(\tilde{p})$ would vanish meaning that we are at the boarder between small and large fermion domain. It corresponds to the case where $u-2g$ is minimal and equals to zero.

If we come back to the general representation~(\ref{EpPsf}), the near-flat space regime $u=2g +O(1)$ corresponds to expanding the function $\gamma^{\o}(t)$ at $g\gg 1$ with $t/g$ kept fixed. It is known how to construct the solution to the BES equation in this regime~\cite{KSV08,BK09} and an expression for the correction $\delta\tilde{E}$ can be derived in principle for any $\tilde{p}$, at least parametrically. It would be interesting to perform this analysis in detail and check the matching with the giant hole regime when $\tilde{p}\gg 1$.

\subsection{Gauge field and bound states}

For gauge field and bound states, the string perturbative regime, $E, p = O(1)$, corresponds to taking the strong coupling limit while keeping the rescaled Bethe rapidity $\bar{u} \equiv u/(2g)$ fixed and such that $\bar{u}^2 <1$. We will study this regime up to $O(1/g^2)$, but before going into detail it is interesting to see why this regime requires $\bar{u}^2 < 1$. To this end, let us rewrite the relation~(\ref{CIRa}), giving energy and momentum of gauge field ($\ell=1$) and bound states ($\ell>1$), as
\be\label{DRgfa}
E +p = i\pi g\int_{\bar{u}-i\bar{\epsilon}}^{\bar{u}+i\bar{\epsilon}}d\bar{v}\, \Lambda(2g\bar{v}) \, ,
\ee
where $\bar{\epsilon} = \ell/(4g)$. When the coupling constant gets large, we have $\bar{\epsilon} \rightarrow 0$ and the contour of integration in~(\ref{DRgfa}) becomes small, leading to
\be
E +p = -{\pi \ell \over 2}\Lambda(2g\bar{u})\, ,
\ee
in a first approximation. So the smallness of $\bar{\epsilon}$ has compensated the large factor of $g$, and energy and momentum are of the order of the function $\Lambda(2g\bar{u})$, which is of order $O(1)$ as we shall see. We stress that this is correct for $\bar{u}^2 <1$. Otherwise, when $\bar{u}^2 >1$, since the integrand $\Lambda(2g\bar{v})$ in~(\ref{DRgfa}) has cuts running from $\bar{v}= \pm 1$ to $\pm \infty$, respectively, the contour of integration does not become small: it wraps around one of the cuts, and energy and momentum are of order $O(g)$. This is the giant hole regime~\cite{DL10}, whose analysis will not be given below because to leading order the dispersion relation is the same as for scalar and fermion, as said before. The transition through the points $\bar{u} = \pm 1$ requires a resummation of the perturbative expansion that can be done by working with $u\pm 2g = O(1)$. This is the near-flat space regime that we will consider mostly to leading order. Finally, we will consider another semiclassical limit reached when $\ell \sim g \gg 1$. In this limit, we have $\bar{\epsilon} = O(1)$, meaning that the coutour of integration in~(\ref{DRgfa}) does not shrink at strong coupling, and both energy and momentum are then of order $O(g)$. Physically, this limit amounts to giving  to the GKP string a (macroscopic) transverse spin $S_{\perp} = \ell \sim g$. We will derive the leading order expression for the energy at arbitrary $\bar{\epsilon}$, assuming an excitation at rest, $p =0$. 

\subsubsection*{Perturbative regime}

As said before, in the perturbative regime, $g\gg 1$, $\bar{u} = u/(2g)$ fixed and $\bar{u}^2 < 1$, energy and momentum are $O(1)$. Starting from~(\ref{DRgfa}) and performing a bit of algebra, as explained in Appendix~\ref{AppSC}, one observes that the dispersion relation can be written as
\be\label{DRgfb}
E+p = \sqrt{2}\, \ell\left({1+\bar{u} \over 1-\bar{u}}\right)^{1/4}f(\bar{u})\, ,
\ee
where $f(\bar{u})$ admits a $1/g$ expansion at fixed $\bar{u}$.%
\footnote{Note that, in spite of the notation, the function $f(\bar{u})$ in~(\ref{DRgfb}) is different from the one entering the relation~(\ref{DRfa}) for the fermion. Their strong coupling expansions are however very similar when both expressed in terms of $\bar{u}$.}
From the knowledge of $f(\bar{u})$, one can easily reconstruct both energy and momentum by separating odd and even parts in $\bar{u}$,
\be
\begin{aligned}
E &= {\ell \over \sqrt{2}}\bigg[\left({1+\bar{u} \over 1-\bar{u}}\right)^{1/4}f(\bar{u}) + \left({1-\bar{u} \over 1+\bar{u}}\right)^{1/4}f(-\bar{u})\bigg]\, , \\
p &= {\ell \over \sqrt{2}} \bigg[\left({1+\bar{u} \over 1-\bar{u}}\right)^{1/4}f(\bar{u}) - \left({1-\bar{u} \over 1+\bar{u}}\right)^{1/4}f(-\bar{u})\bigg]\, .
\end{aligned}
\ee
For the dispersion relation, it is more convenient to observe that
\be\label{DRgfsc}
E^2-p^2 = 2\ell^2f(\bar{u})f(-\bar{u})\, .
\ee
For instance, using the leading order solution to the BES equation~\cite{AABEK07}, one immediately finds that (see Appendix~(\ref{AppSC}) for details)
\be
f(\bar{u}) = 1 + O(1/g)\, ,
\ee
giving
\be\label{DRgflo}
E^2 - p^2 = 2\ell^2\, .
\ee
This is a relativistic dispersion law for a particle with mass $\sqrt{2}\, \ell$. In particular, it is saying that the gauge field $(\ell=1)$ has mass $\sqrt{2}$ at strong coupling. There are exactly two bosons with precisely this mass in the spectrum of string fluctuations~\cite{FT02}, that we verify to correspond to the two twist-one excitations $F_{+, 1\pm i2}$ of the gauge theory. The linearity in $\ell$ of the mass formula indicates that a bound state is no more than a pile of $\ell$ gauge fields at strong coupling. This is presumably a consequence of the fact that there is no interaction between them at this order. As we shall see the masses still satisfy a linear law at one loop and the gauge fields get bounded starting at $O(1/g^2)$.

Including radiative corrections to the dispersion relation~(\ref{DRgflo}) can be done order by order in the $1/g$ expansion with the help of the results of~\cite{BKK07,KSV08,BK09} for the solution to the BES equation. The details of this analysis for corrections up to $O(1/g^2)$ are given in Appendix~\ref{AppSC}. The dispersion relation so obtained is simplified when expressed in terms of \textit{rescaled} energy and momentum, $E_{r} = E/\ell, p_{r} = p/\ell$. For further simplification, we will simply omit to make use of the subscript $r$ in the following, keeping in mind that the correct expressions require the substitutions $E \rightarrow E/\ell, p \rightarrow p/\ell$. After these brief remarks, using the expression quoted in~(\ref{Appfgf}) for the the function $f(\bar{u})$, the dispersion relation~(\ref{DRgfsc}) is found to be
\be\label{Epgfa}
E = \sqrt{2+p^2}\left[1-{(1+p^2)^2 \over 8g (2+p^2)} + {(\alpha + \beta p^2 +\gamma p^4 +\delta p^6)(1+p^2)^2 \over g^2(2+p^2)^2} + O(1/g^3)\right]\, ,
\ee
with
\be\label{Epgfb}
\begin{aligned}
\alpha &= -{7\over 384}-{\ell^2 \over 96} + {\textrm{K} \over 2\pi^2} - {3 \log{2} \over 16\pi}\, , \qquad  \beta = -{5\over 48}-{5\ell^2\over 192}+{9\textrm{K} \over 4\pi^2}-{3\log{2} \over 32\pi} \, , \\
\gamma &= -{47 \over 384}-{\ell^2 \over 48}+{2\textrm{K} \over \pi^2}\, , \qqqquad \qquad \, \, \delta = -{7\over 192} -{\ell^2 \over 192} + {\textrm{K} \over 2\pi^2}\, .
\end{aligned}
\ee
The $(\bar{u}, p)$ relation reads, up to $O(1/g)$,
\be\label{xpgf}
\bar{u} = {p\sqrt{2+p^2} \over 1+p^2}\left[1+{a+b p^2 \over g (2+p^2)} + O(1/g^2)\right]\, ,
\ee
with
\be
a = {1\over 8}-{3\log{2} \over 2\pi}\, , \qquad b = {1\over 8}-{3\log{2} \over 4\pi}\, .
\ee
As a particular application of~(\ref{Epgfa}, \ref{Epgfb}), we find the (rescaled) mass
\be\label{renMgf}
m = E(p=0) = \sqrt{2}\left[1 - {1 \over 16g} + {\alpha \over 4g^2} + O(1/g^3) \right]\, .
\ee
We note that the first correction to the mass is negative, indicating that it decreases with the coupling. This is in qualitative agreement with the expectation that the mass increases monotonically between its weak and strong coupling values, $1$ and $\sqrt{2}$, respectively. Next, recalling that the mass $m$ in~(\ref{renMgf}) stands for a rescaled one, we observe that the breakdown of the linearity in $\ell$ appears at $O(1/g^2)$ and is controlled by the term $\sim -\ell^2$ in $\alpha$, see Eq.~(\ref{Epgfb}). Due to the negative sign of the latter contribution, we conclude that the gauge fields are effectively bounded by the interaction. It is curious nevertheless that the latter phenomenon is delayed to $O(1/g^2)$, and it would be interesting to look at the scattering between gauge fields at strong coupling to shed light on it.

Examining futher~(\ref{Epgfa}) we find singularities at $p^2 = -2$. They are of the same sort as the ones encountered for scalar in the non-perturbative regime. They account for renormalization of the mass and one can get ride of them easily by rewriting~(\ref{Epgfa}) as
\be\label{Epgfaren}
E = \sqrt{m^2+p^2}\left[1-{p^2 \over 8g} + {p^2(\beta' +\gamma' p^2 +\delta p^4) \over g^2} + O(1/g^3)\right]\, ,
\ee
with $m$ given in~(\ref{renMgf}) and the coefficients $\beta', \gamma'$, found as 
\be\label{Epgfbren}
\beta' = -{1\over 24}-{\ell^2 \over 96} + {3\textrm{K} \over 4\pi^2} - {3 \log{2} \over 32\pi}\, , \qquad  \gamma' = -{19\over 384}-{\ell^2\over 96}+{\textrm{K} \over \pi^2} \, .
\ee
The expression~(\ref{Epgfaren}) is obviously simpler than the previous one~(\ref{Epgfa}). However, it obscures that the dispersion relation satisfies the identity~\cite{AM07}
\be\label{Epi}
E(p=\pm i) = 1\, ,
\ee
which is apparent in~(\ref{Epgfa}). This identity can be proved to be exact. The proof will not be given here and we simply note, using~(\ref{xpgf}), that the points $p=\pm i$ corresponds to $\bar{u} = \pm i\infty$. Notice also that after rescaling to original units, the relation~(\ref{Epi}) reads $E(p=\pm i\ell) = \ell$, with $E$ the energy of a bound state of $\ell$ gauge fields.

\subsubsection*{Near-flat space regime}

The strong coupling expansion of the function $f(\bar{u})$ that gives energy and momentum~(\ref{DRgfb}) is singular around the points $\bar{u}^2=1$, see Appendix~\ref{AppSC}. Around $\bar{u}=1$, say, it turns into an expansion in inverse powers of $2g(1-\bar{u}) = 2g-u$ indicating that a resummation is needed. Using the expression~(\ref{xpgf}), we note that the point $\bar{u}=1$ requires taking the momentum $p$ to be large.  More precisely, the asymptotics at large $p$ reads
\be
\bar{u} = 1-{1 \over 2p^4} + O(1/p^6)\, ,
\ee
to leading order at strong coupling. Thus working around the points $u\sim 2g$ is equivalent to considering momentum $p\sim g^{1/4}$. This is the near-flat space regime.  From the result~(\ref{Epgfa}) we find that the energy $\tilde{E} = E/g^{1/4}$ is given by
\be
\tilde{E} = \tilde{p} + \delta \tilde{E}(\tilde{p})/\sqrt{g} + O(1/g)\, ,
\ee
where $\tilde{p} = p/g^{1/4}$ is kept fixed and with $\delta \tilde{E}(\tilde{p})$ determined at small $\tilde{p}$ by~(\ref{Epgfa}) as
\be\label{SubNFSgf}
\delta \tilde{E}(\tilde{p}) = {1\over \tilde{p}} -{1 \over 8} \tilde{p}^3 +\delta \tilde{p}^7 + O(\tilde{p}^{11})\, ,
\ee
with $\delta$ given in~(\ref{Epgfb}). We note that to leading order $\tilde{E} = \tilde{p} + O(1/\sqrt{g})$, even after the rescaling to the natural units $(E, p) \rightarrow (E/\ell,p/\ell)$. We verify thus the flavor independence of the leading dispersion relation in the near-flat space limit. The subleading correction $\delta \tilde{E}(\tilde{p})$ does depend on $\ell$ however, via the coefficient $\delta$ in~(\ref{SubNFSgf}), see Eq.~(\ref{Epgfb}). 

Note finally that it is possible to derive a parametric representation for the subleading dispersion relation valid for any $\tilde{p}$. It is however difficult to deal with it and the matching with the giant hole domain is not obviously observed. It would be interesting to work it out in detail.

\subsubsection*{Semiclassical bound state}

As mentioned before, it is possible for bound states to reach another semiclassical regime without taking the momentum to be large. We can assume $p=0$ and choose $\ell \sim g$. It corresponds to giving a macroscopic transverse spin $S_{\perp} = \ell \sim g$ to the GKP string. For $p=0$ the energy is minimal at given spin $S_{\perp}$. This regime is easy to study to leading order at strong coupling, the mass of a bound state reading
\be\label{HSexp}
E(0) = 2\sqrt{2} g\int_{-\bar{\epsilon}}^{\bar{\epsilon}}d\bar{v}\, \left({1+i\bar{v} \over 1-i\bar{v}}\right)^{1/4}\, + O(g^0),
\ee
with $\bar{\epsilon} = \ell/4g = S_{\perp}/4g$. The relation~(\ref{HSexp}) is a direct consequence of the general formula~(\ref{DRgfa}) given the strong coupling expression for $\Lambda(2g\bar{v})$, see Appendix~\ref{AppSC}. Note also that energy in~(\ref{HSexp}) has not been rescaled by $\ell$, in distinction with the convention of the previous subsection. The integral in~(\ref{HSexp}) can be taken exactly yielding
\be\label{HeavyMass}
E(0) = 4 g\left(\sqrt{\eta(\eta-1)}+\log{\left[\eta^{1/2}+\sqrt{\eta-1}\right]}\right) + O(g^0)\, ,
\ee
where $\eta \equiv \sqrt{1+\bar{\epsilon}^2}$. The mass is thus a smooth, monotonic increasing function of $\bar{\epsilon}$, interpolating between
\be\label{hsse}
E(0) = 4\sqrt{2}g\left[\bar{\epsilon} -{\bar{\epsilon}^{\, 3} \over 24} + O\left(\bar{\epsilon}^{\, 5}\right)\right] + O(g^0) = \sqrt{2}S_{\perp} + \ldots\, ,
\ee
at $\bar{\epsilon} \sim 0$, and
\be\label{hsle}
E(0) = 2g\bigg[2\bar{\epsilon} + \left(\log{(4\bar{\epsilon})}-1\right) + O(1/\bar{\epsilon})\bigg] + O(g^0) = S_{\perp} + 2g\left(\log{\left({S_{\perp}\over g}\right)}-1\right) + \ldots\, ,
\ee
at $\bar{\epsilon} \sim \infty$. It is interesting to note that the quantity $E(0)/(\sqrt{2}S_{\perp})$ is always smaller than $1$ and decreases when $S_{\perp}$ increases. We verify again that the mass of the bound state is smaller than the sum of the masses of its gauge-field constituents.

Soliton above the GKP vacuum carrying spin $S_{\perp} \sim g$ can be found directly in string theory, and, remarkably enough, has energy at rest exactly given by~(\ref{HeavyMass}).%
\footnote{I am grateful to J.~Maldacena for sharing with me his notes on this solution.} This soliton should also be some special case of the two $\rm AdS_{5}$-spins solutions constructed in~\cite{TT09}. In the limit $S_{\perp} = \ell \gg g$, i.e. $\bar{\epsilon} \gg 1$, the asymptotics behavior~(\ref{hsle}) is also found to be in agreement with the result of~\cite{FRZ09}.%
\footnote{In the notations of~\cite{FRZ09}, the limit $\bar{\epsilon} \gg 1$ should correspond to $\alpha \rightarrow \infty$ (and $L=3$), with $\alpha = S/\ell$ and with $S$ the total spin. Note also that our result gives the GKP energy of the semiclassical bound state measured above the vacuum. So one should substract the vacuum contribution in the result of~\cite{FRZ09} to make the comparison.}

\section{Conclusion and outlook}\label{Concl}

In this paper we have classified excitations propagating on top of the long GKP string and studied their dispersion relations both at weak and strong coupling. Developing ideas of seminal analyses~\cite{K95,BGK06,ES06,BES06,FRS07} we have derived large-spin integral equations from the all-loop asymptotic Bethe ansatz equations for the spectrum of scaling dimensions~\cite{BS05}. The equations devised in the current consideration are of the type encountered previously in the literature~\cite{ES06,BES06,FRS07}. There exists an extensive methodology for their solution~\cite{Benna06,AABEK07,BKK07,KSV08,BK08,BK09}. In particular, owing to their close relationship with the BES equation, which controls the vacuum distribution of roots, we were able to construct an exact parametric representation for the dispersion relations, which, though not totally explicit, only requires the knowledge of the solution to the BES equation. This reduction in complexity allowed us to analyze the dispersion relations in various regimes both at weak and strong coupling. As particular applications, we have derived the first few terms in their weak coupling expansion and computed corrections to the spectrum of string fluctuations at strong coupling. A natural next step is a thorough analysis of the complex momentum plane for all excitations, and, in particular, of the Wick rotation symmetry proposed in~\cite{AGMSV10}. 

At a more dynamical level, we have identified the mechanism that leads to the restoration of the $SU(2)$ symmetry accompanying the shift from the BMN to the GKP vacuum. We have also observed the enhancement of the residual symmetry group of the GKP vacuum in the weak coupling limit to $SL(2|4)$. This peculiar dynamical aspect of fermions was related at the level of their dispersion relation to the necessity of working with two Bethe rapidity planes to cover their full kinematics.

Among the dynamical issues that would be worth pursuing is the fate of the missing mass 2 boson. There are possible decay channels for it, in two fermions, for instance, but it would be enlightening to give a more quantitative estimate of their amplitudes. One might want to understand, for instance, whether the mass 2 boson is stable to any order at strong coupling and gets destabilized by non-perturbative corrections.%
\footnote{This seems a quite reasonnable expectation since its presence in the spectrum of string fluctuations is tied to the breakdown of the $SO(6)$ symmetry down to $SO(5)$, which is only restored at the non-perturbative level.} The question of stability of the mass 2 boson is also related to the completeness of the basis of excitations that have been identified so far. Another dynamical issue concerns the size of the wrapping effects. They were assumed to be associated with $O(1/S)$ corrections based on physical expectations and direct computations in the $\mathfrak{sl}(2)$ subsector~\cite{BJL08}. The latter result only covers the case of scalar excitations and it would be interesting to check whether the estimate holds true for other flavors as well, by means of the Y-system proposed in~\cite{GKV}.

There are possible applications of the analysis performed in this paper to some related problems. One of the most interesting one concerns the computation of scattering amplitudes along the lines of~\cite{AGMSV10}. Another possible application is to the construction of the spectrum of excitations around the GKP string embedded into the string dual of the Aharony-Bergman-Jafferis-Maldacena theory~\cite{ABJM}. The all-loop asymptotic Bethe ansatz equations are known for this theory as well~\cite{MZ08,GV08} and are closely related to the ones for the planar $\mathcal{N}=4$ SYM theory. From their structure we expect that the spectrum of large spin excitations can be analyzed along the lines presented here. Moreover, up to the replacement of the coupling constant by the interpolating function~\cite{Gaiotto,GV08} of the Gromov-Vieira equations, the dispersion relations should be directly expressible in terms of the ones constructed here.  

\section*{Acknowledgments}

I am especially indebt to Juan Maldacena, Pedro Vieira, and Amit Sever, for illuminating discussions and for sharing with me their views and results on the issue adressed in this paper. I would like to thank Andre\"i Belitsky for his comments on the manuscript and for valuable discussions on the matter it contains. I am also grateful to Igor Klebanov, Dmitry Krotov, Dmytro Volin, and Alexander Zhiboedov, for discussions. 

\appendix

\section{Basic integrals}\label{AALE}

In this appendix, we provide basic integrals, mostly borrowed from~\cite{ES06}, that are useful to performing the all-loop analysis given in the Section~\ref{ALA}.

\subsubsection*{Basic one-loop integrals}

Expressions of the type
\be
I_{r}(u) = -i\log{\left(r+2iu \over r-2iu\right)} = 2\, \textrm{arctan}{\left({2u\over r}\right)}\, ,
\ee
with asymptotics $\pm \pi$ at $u=\pm \infty$, respectively, make their appearance below, after separating the one-loop scattering phases from their radiative corrections. A convenient representation can be found as
\be\label{OLEIrep}
I_{r}(u) = 2\int_{0}^{\infty}{dt\over t}\sin{(ut)}\e^{-rt/2}\, .
\ee

\subsubsection*{Basic all-loop integrals}

The building blocks for the formulae given below are the following integral representations for inverse powers of the shifted all-loop spectral parameter $x$. They read
\be\label{EIrep}
{1\over n}\left({g \over x^{[\pm p]}}\right)^{n} = (\mp i)^n \int_{0}^{\infty}{dt \over t} J_{n}(2gt) \, \e^{\pm iut-pt/2}\, ,
\ee
where $x^{[\pm p]}(u) \equiv x(u\pm ip/2)$ and $x(u) = (u+\sqrt{u^2-(2g)^2})/2$. Here, the index $n$ belongs to $\mathbb{N}^*$ and the function $J_{n}(t)$ is the $n^{\textrm{th}}$ Bessel's function. When $p=0$, one can also find the representations (for $u^2 > (2g)^2$)
\be
\begin{aligned}\label{Elrep0}
&\bigg({g \over x}\bigg)^{2n} \, \, \, \,  \, = \bigg({x' \over g}\bigg)^{2n} \, \, \, \, \, \, = (-1)^n (2n)\int_{0}^{\infty}{dt \over t} J_{2n}(2gt) \, \cos{(ut)}\, ,  \\
&\bigg({g \over x}\bigg)^{2n-1} = \bigg({x' \over g}\bigg)^{2n-1} =  (-1)^{n-1} (2n-1)\int_{0}^{\infty}{dt \over t} J_{2n-1}(2gt) \, \sin{(ut)}\, ,
\end{aligned}
\ee
where $x'$ is the second solution to $u=x+g^2/x$, reading $x'(u) = (u-\sqrt{u^2-(2g)^2})/2$. The identities~(\ref{Elrep0}) are consequences of~(\ref{EIrep}) when taking into account that
\be\label{Beq}
\int_{0}^{\infty}{dt \over t} J_{2n}(2gt) \sin{(ut)} = 0\, , \qquad \int_{0}^{\infty}{dt \over t} J_{2n-1}(2gt) \cos{(ut)} = 0\, .
\ee
for $u^2 > (2g)^2$ and $n \ge 1$.

\subsubsection*{Spin-chain higher conserved charges}

With the help of~(\ref{EIrep}), the spin-chain higher conserved charges
\be\label{HCE}
Q_{r+1} = {i \over r}\left({1 \over x^{+r}}-{1 \over x^{-r}}\right)\, ,
\ee
where $x^{\pm r} = (x^{\pm}(u))^r$, can be found as
\be\label{HCch}
Q_{2n}  = {2(-1)^{n-1}\over g^{2n-1}}\int_{0}^{\infty}{dt \over t} J_{2n-1}(2gt) \cos{(ut)} \e^{-t/2}\, , \qquad Q_{2n+1}  = {2(-1)^{n-1}\over g^{2n}}\int_{0}^{\infty}{dt \over t} J_{2n}(2gt) \sin{(ut)} \e^{-t/2}\, ,
\ee
with $n\geqslant 1$.

\subsubsection*{Integrals for inhomogeneous terms}

The all-loop large spin equations considered in Section~\ref{ALA} contain various inhomogeneous terms. Useful integral representations for them can be obtained with the help of few equalities.

The first identity holds for
\be
I(x) \equiv -i\log{\left(-{x^{-} \over x^{+}}\right)} = -i\log{\left({1 + 2iu \over 1-2iu}\right)} -i\log{\left({1+g^2/x^{+\, 2} \over 1+g^2/x^{-\, 2}}\right)}\,,
\ee
that involves the shifted spectral parameter $x^{\pm} = x(u\pm i/2)$, and where $x^{\pm\, 2} = (x^{\pm})^2$. By Taylor expanding the right-hand side of~(\ref{Kexpa}) around $g=0$, and making use of the integral representations~(\ref{OLEIrep}, \ref{EIrep}), one obtains
\be\label{Kexpa}
I(x) = 2\int_{0}^{\infty}{dt \over t} J_{0}(2gt) \sin{(ut)} \e^{-t/2}\, .
\ee

The second equality holds for
\be\label{Mexpa}
I_{p}(x,y) \equiv -i\log{\left({1-g^2/x^{+}y^{[-p]} \over 1-g^2/x^{-}y^{[+p]}}\right)}\, ,
\ee
that involves the two shifted spectral parameters $x^{\pm} = x(u\pm i/2)$ and  $y^{[\pm p]} = y(v\pm ip/2)$. Going along the same lines before, one gets
\be
\begin{aligned}\label{Mexpb}
I_{p}(x,y) &= (2g)^2\int_{0}^{\infty}dt \cos{(ut)}\e^{-t/2}\int_{0}^{\infty}ds K(2gt, 2gs) \sin{(vs)}\e^{-ps/2} \\
 &\, -(2g)^2\int_{0}^{\infty}dt \sin{(ut)}\e^{-t/2}\int_{0}^{\infty}ds K(2gt, 2gs) \cos{(vs)}\e^{-ps/2}\, ,
\end{aligned}
\ee
where the symmetric kernel $K(t,s) = K(s,t)$ is given in~(\ref{Kern}). In the particular case $p=0$, we have
\be\label{Mexpc}
I_{0}(x,y) = -i\log{\left({1-g^2/x^{+}y \over 1-g^2/x^{-}y}\right)} =  -i\log{\left({1-y'/x^{+} \over 1-y'/x^{-}}\right)} \, ,
\ee
with $y = (v+\sqrt{v^2-(2g)^2})/2$ and $y'=(v-\sqrt{v^2-(2g)^2})/2$. It can be evaluated as
\be
\begin{aligned}\label{Mexpcb}
I_{0}(x, y) &= (2g)^2\int_{0}^{\infty}dt \cos{(ut)}\e^{-t/2}\int_{0}^{\infty}ds K_{+}(2gt, 2gs) \sin{(vs)} \\
 &\, -(2g)^2\int_{0}^{\infty}dt \sin{(ut)}\e^{-t/2}\int_{0}^{\infty}ds K_{-}(2gt, 2gs) \cos{(vs)}\, .
\end{aligned}
\ee
for $v^2>(2g)^2$ and with the kernels $K_{\pm}(t,s)$ given in~(\ref{Kernpm}). The last two relations are easily deduced from~(\ref{Mexpb}, \ref{Kern}, \ref{Kernpm}), by applying~(\ref{Beq}).

Finally, to obtain the inhomogeneous terms for the gauge field and its bound states, we need to consider
\begin{align}
\hat{I}_{p}(x, y) \equiv -i\log{\left({1-g^2/x^{+}y^{[+p]} \over 1-g^2/x^{-}y^{[+p]}}\right)}-i\log{\left({1-g^2/x^{+}y^{[-p]} \over 1-g^2/x^{-}y^{[-p]}}\right)}\, .
\end{align}
We find that
\be
\begin{aligned}\label{Mexpbs}
\hat{I}_{p}(x,y) &= 2(2g)^2\int_{0}^{\infty}dt \cos{(ut)}\e^{-t/2}\int_{0}^{\infty}ds K_{+}(2gt, 2gs) \sin{(vs)}\e^{-ps/2} \\
 &\, -2(2g)^2\int_{0}^{\infty}dt \sin{(ut)}\e^{-t/2}\int_{0}^{\infty}ds K_{-}(2gt, 2gs) \cos{(vs)}\e^{-ps/2}\, .
\end{aligned}
\ee

\section{Formulae for all-loop analysis}\label{AALS}

In this appendix, we explain how to derive the expressions for $\mathcal{I}_{\textrm{dressing}}(u), \tilde{\mathcal{I}}_{\textrm{dressing}}(u)$, starting from the representation given in~\cite{BES06} for the dressing phase. Then, for all of the momentum-carrying excitations considered in this paper, we construct the inhomogeneous contributions $\mathcal{I}_{\star}(u, u_{\star}), \tilde{\mathcal{I}}_{\star}(u, u_{\star})$. We give also the defining relations for the auxiliary functions that permit to cast the integral equations into the form of infinite systems of equations. Finally, we provide expressions for the infinite series of source terms entering the latter set of equations. 

\subsubsection*{Dressing phase}

According to~\cite{BS05,BHL06,BES06}, the dressing phase $\sigma(u,v) \equiv \exp{(i\theta(u,v))}$ admits a bilinear expansion over the individual charges $q_{r}(u), q_{s}(v)$, see Eq.~(\ref{HCE}). An explicit expression for it was proposed in~\cite{BES06} and reads, after separating even/odd parts in $u$,
\be
\begin{aligned}\label{DPexp}
\theta(u,v)|_{\textrm{odd in $u$}} &= 2\sum_{n\ge 1}(2n)\int_{0}^{\infty}{dt \over t}J_{2n}(2gt)\sin{(ut)}\e^{-t/2} \\
&\times \sum_{m\ge 1}(-1)^m 2(2m-1)g^{2m-1}\int_{0}^{\infty}{ds \over s}{J_{2n}(2gs)J_{2m-1}(2gs) \over \e^{s}-1} q_{2m}(v)\, , \\
\theta(u,v)|_{\textrm{even in $u$}} &= -2\sum_{n\ge 1}(2n-1)\int_{0}^{\infty}{dt \over t}J_{2n-1}(2gt)\cos{(ut)}\e^{-t/2} \\
&\times \sum_{m\ge 1}(-1)^m 2(2m)g^{2m}\int_{0}^{\infty}{ds \over s}{J_{2n-1}(2gs)J_{2m}(2gs) \over \e^{s}-1}q_{2m+1}(v)\, ,
\end{aligned}
\ee
with $J_{n}(t)$ the Bessel's functions. To get the expressions for $\mathcal{I}_{\textrm{dressing}}(u), \tilde{\mathcal{I}}_{\textrm{dressing}}(u)$, we perfom the replacement $v\rightarrow u_{4, j}$ in~(\ref{DPexp}) and then sum over all the main roots $u_{4, j}$. This has the effect of substituting $q_{r}(v)$ in~(\ref{DPexp}) by the total charges $Q_{r}$, since the roots $u_{4}$ are the only spin-chain momentum carriers. For our purposes, we only need to consider total charges measured above the vacuum, $q_{r} = Q_{r} - q^{\textrm{vacuum}}_{r}$. Then, introducing the generating functions $q_{-}(s), q_{+}(s)$, as in~(\ref{HCC}), it is not difficult to evaluate
\be
\mathcal{I}_{\textrm{dressing}}(u) +\tilde{\mathcal{I}}_{\textrm{dressing}}(u) \equiv 2\partial_{u}\sum_{j=1}^{K_{4}}\theta(u, u_{4, j})\, ,
\ee
and to cast it into the form~(\ref{Idress}).

\subsubsection*{Scalar}

By definition~(\ref{ITE}, \ref{ITEbis}), the inhomogeneous terms $\mathcal{I}_{\textrm{h}}(u, u_{\textrm{h}}),\tilde{\mathcal{I}}_{\textrm{h}}(u, u_{\textrm{h}})$, associated with a hole carrying a rapidity $u_{\textrm{h}}$, immediately follow from the knowledge of the undressed scattering phase between main roots. The latter quantity is given in~(\ref{Zmb}). Hence, combining~(\ref{ITE}, \ref{ITEbis}, \ref{Zmb}) and using the formulae~(\ref{OLEIrep}, \ref{Kexpa}, \ref{Mexpb}) of Appendix~\ref{AALE}, one easily finds that
\be
\begin{aligned}
\mathcal{I}_{\textrm{h}}(u, u_{\textrm{h}}) &= -2\int_{0}^{\infty}dt\cos{(ut)}\e^{-t/2}\pi_{\textrm{h}}(t, u_{\textrm{h}})\, , \\
\tilde{\mathcal{I}}_{\textrm{h}}(u, u_{\textrm{h}}) &= -2\int_{0}^{\infty}dt\sin{(ut)}\e^{-t/2}\tilde{\pi}_{\textrm{h}}(t, u_{\textrm{h}})\, ,
\end{aligned}
\ee
with
\be
\begin{aligned}\label{PIh}
\pi_{\textrm{h}}(t, u_{\textrm{h}}) &= \cos{(u_{\textrm{h}}t)}\e^{-t/2}-J_{0}(2gt)-(2g)^2\int_{0}^{\infty}ds\, t K(2gt, 2gs)\cos{(u_{\textrm{h}}s)}\e^{-s/2}\, , \\
\tilde{\pi}_{\textrm{h}}(t, u_{\textrm{h}}) &= \sin{(u_{\textrm{h}}t)}\e^{-t/2}-(2g)^2 \int_{0}^{\infty}ds\, t K(2gt, 2gs)\sin{(u_{\textrm{h}}s)}\e^{-s/2}\, ,
\end{aligned}
\ee
with the kernel density $K(t,s)$ as in~(\ref{Kern}, \ref{Kernpm}).

The defining relations for $\gamma^{\textrm{h}}_{\pm}(t), \tilde{\gamma}^{\textrm{h}}_{\pm}(t)$, that permit to cast the integral equations~(\ref{EqTF}) into the form of the systems of equations~(\ref{EQN}), are
\be
\begin{aligned}\label{SGH}
\gamma^{\textrm{h}}_{+}(2gt)+\gamma^{\textrm{h}}_{-}(2gt) &= (\e^{t}-1)\Omega_{\textrm{h}}(t) - \cos{(u_{\textrm{h}}t)}\e^{-t/2} + J_{0}(2gt)\, , \\
\tilde{\gamma}^{\textrm{h}}_{+}(2gt)+\tilde{\gamma}^{\textrm{h}}_{-}(2gt) &= (\e^{t}-1)\tilde{\Omega}_{\textrm{h}}(t)  - \sin{(u_{\textrm{h}}t)}\e^{-t/2}\, .
\end{aligned}
\ee
The infinite series of source terms for the system~(\ref{EQN}) read
\be
\begin{aligned}\label{KH}
\kappa^{\textrm{h}}_{n}(u_{\textrm{h}}) &= -\int_{0}^{\infty}{dt \over t}{J_{n}(2gt)\over \e^{t}-1}\left(\cos{(u_{\textrm{h}}t)}\e^{t/2}-J_{0}(2gt)\right)\, , \\
\tilde{\kappa}^{\textrm{h}}_{n}(u_{\textrm{h}}) &= -\int_{0}^{\infty}{dt \over t}{J_{n}(2gt)\over \e^{t}-1}\sin{(u_{\textrm{h}}t)}\e^{t/2}\, .
\end{aligned}
\ee
Finally, evaluating~(\ref{SGH}) at small $t$, we get the relations
\be\label{MRH}
\Omega_{\textrm{h}}(0) =-{1 \over 2} + 2g\gamma^{\textrm{h}}_{1}(g)\, , \qquad \tilde{\Omega}_{\textrm{h}}(0) = u_{\textrm{h}} + 2g\tilde{\gamma}^{\textrm{h}}_{1}(g)\, ,
\ee
where the coefficients $\gamma^{\textrm{h}}_{1}, \tilde{\gamma}^{\textrm{h}}_{1}$, originate from the small-$t$ asymptotics
\be
\gamma^{\textrm{h}}_{-}(2gt) =2g\gamma^{\textrm{h}}_{1} t +O(t^3)\, , \qquad \tilde{\gamma}^{\textrm{h}}_{-}(2gt) =2g\tilde{\gamma}^{\textrm{h}}_{1} t +O(t^3)\, .
\ee
Note that we used the fact that $\gamma_{+}^{\textrm{h}}(t) \sim \tilde{\gamma}^{\textrm{h}}_{+}(t) \sim t^2$ at small $t$.

\subsubsection*{Gauge field and bound states}

To compute the inhomogeneous terms~(\ref{ITE}) associated to a gauge field ($\ell=2$) or its bound states ($\ell>1$), carrying a rapidity $u_{\ell}$, we need the expression for their scattering phases with a main root $u$. These excitations are embedded into the asymptotic Bethe ansatz equations~(\ref{ABAE}) in the form of stack of strings~\cite{FRZ09}. For instance, considering gauge field and bound states with positive $\mathfrak{so}(2)$ charges, they are found as a stack of one $(\ell+1)$-string of roots $u_{3}$, one $\ell$-string of roots $u_{2}$ and one $(\ell-1)$-string of roots $u_{1}$, with a (real) center-of-mass rapidity $u_{\ell}$. Then the scattering phase for a main root $u$ going through the stack is obtained by taking the product
\be\label{Sstack}
S_{4\ell}(u, u_{\ell}) = \prod_{j =1}^{\ell+1}S_{43}(u, u_{3, j})\prod_{j = 1}^{\ell}S_{42}(u, u_{2, j})\prod_{j =1}^{\ell-1}S_{41}(u, u_{1, j})\, ,
\ee
where
\be\label{stack}
u_{3, j}-u_{\ell} = {i\over 2}( 2j-2-\ell)\, , \qquad u_{2, j}-u_{\ell} = {i \over 2}(2j-1-\ell)\, , \qquad u_{1, j}-u_{\ell} = {i \over 2}(2j -\ell)\, .
\ee
The notations $S_{43}, S_{42}, S_{41}$, in~(\ref{Sstack}) stand for the elementary scattering phases encoded in the middle-node equations of the asymptotic Bethe ansatz equations~(\ref{ABAE}). They read explicitely as
\be\label{SPstack}
S_{43}(u, u_{3, j}) = {x^{+}-x_{3, j} \over x^{-}-x_{3, j}}\, , \qquad S_{42}(u, u_{2, j}) = 0\, , \qquad S_{41}(u, u_{1, j}) = {1-g^2/x^{+}x_{1, j} \over 1-g^2/x^{-}x_{1, j}}\, .
\ee
Plugging~(\ref{stack}) into~(\ref{Sstack}) and making use of~(\ref{SPstack}), one finds after a bit of algebra
\be\label{SPgf}
S_{4\ell}(u, u_{\ell}) =  {u-u_{\ell}+\ft{i(\ell+1)}{2} \over u-u_{\ell}-\ft{i(\ell+1)}{2}} {1-g^2/x^{-}x^{[+\ell]}_{\ell} \over 1-g^2/x^{+}x^{[+\ell]}_{m}} {1-g^2/x^{-}x^{[-\ell]}_{\ell} \over 1-g^2/x^{+}x^{[-\ell]}_{\ell}}\, ,
\ee
where $x_{\ell}^{[\pm \ell]} = x(u_{\ell}+i\ell/2)$. For later convenience, we have factored out the one-loop scattering phase in~(\ref{SPgf}), which is the one expressed in terms of the difference of the Bethe rapidities $u -u_{\ell}$. That was done by using algebraic identities between the deformed spectral parameters $x, x_{\ell}$, and the Bethe rapidities $u, u_{\ell}$ (see~\cite{BS05}). The result~(\ref{SPgf}) is a straighforward extension to $u_{\ell}\neq 0$ of the findings of~\cite{FRZ09}.

Equipped with the expression~(\ref{SPgf}), we can now evaluate the inhomogeneous terms $\mathcal{I}_{\ell}(u, u_{\ell})$, \, $\tilde{\mathcal{I}}_{\ell}(u, u_{\ell})$, as defined in~(\ref{ITE}). Using the formulae~(\ref{OLEIrep}, \ref{Kexpa}, \ref{Mexpbs}), one can show that
\be
\begin{aligned}
\mathcal{I}_{\ell}(u, u_{\ell}) &= -2\int_{0}^{\infty}dt\cos{(ut)}\e^{-t/2}\pi_{\ell}(t, u_{\ell})\, , \\
\tilde{\mathcal{I}}_{\ell}(u, u_{\ell}) &= -2\int_{0}^{\infty}dt\sin{(ut)}\e^{-t/2}\tilde{\pi}_{\ell}(t, u_{\ell})\, ,
\end{aligned}
\ee
where
\be
\begin{aligned}\label{PIgf}
\pi_{\ell}(t, u_{\ell}) &= \cos{(u_{\ell}t)}\e^{-\ell t/2}-J_{0}(2gt)-(2g)^2\int_{0}^{\infty}ds\, t K_{-}(2gt, 2gs)\cos{(u_{\ell}s)}\e^{-\ell s/2}\, , \\
\tilde{\pi}_{\ell}(t, u_{\ell}) &= \sin{(u_{\ell}t)}\e^{-\ell t/2}-(2g)^2 \int_{0}^{\infty}ds\, t K_{+}(2gt, 2gs)\sin{(u_{\ell}s)}\e^{-\ell s/2}\, .
\end{aligned}
\ee
with the even/odd kernel densities $K_{\pm}(t,s)$ as in~(\ref{Kernpm}).

The defining relations for $\gamma^{\ell}_{\pm}(t), \tilde{\gamma}^{\ell}_{\pm}(t),$ are
\be
\begin{aligned}\label{SGGF}
\gamma^{\ell}_{+}(2gt)+\gamma^{\ell}_{-}(2gt) &= (\e^{t}-1)\Omega_{\ell}(t) - \cos{(ut)}\e^{-\ell t/2} + J_{0}(2gt)\, , \\
\tilde{\gamma}^{\ell}_{+}(2gt)+\tilde{\gamma}^{\ell}_{-}(2gt) &= (\e^{t}-1)\tilde{\Omega}_{\ell}(t)  - \sin{(ut)}\e^{-\ell t/2}\, .
\end{aligned}
\ee
For the two infinite series of source terms in~(\ref{EQN}), we found
\be
\begin{aligned}\label{KGFa}
\kappa^{\ell}_{2n}(u_{\ell}) &= -\int_{0}^{\infty}{dt \over t}{J_{2n}(2gt)\over 1-\e^{-t}}\left(\cos{(u_{\ell}t)}\e^{-\ell t/2}-J_{0}(2gt)\right)\, , \\
\kappa_{2n-1}^{\ell}(u_{\ell}) &=-\int_{0}^{\infty}{dt \over t}{J_{2n-1}(2gt)\over \e^{t}-1}\left(\cos{(u_{\ell}t)}\e^{-\ell t/2}-J_{0}(2gt)\right)\, ,
\end{aligned}
\ee
and
\be
\begin{aligned}\label{KGFb}
\tilde{\kappa}^{\ell}_{2n}(u_{\ell}) &= -\int_{0}^{\infty}{dt \over t}{J_{2n}(2gt)\over \e^{t}-1}\sin{(u_{\ell}t)}\e^{-\ell t/2}\, , \\
\tilde{\kappa}^{\ell}_{2n-1}(u_{\ell}) &= -\int_{0}^{\infty}{dt \over t}{J_{2n-1}(2gt)\over 1-\e^{-t}}\sin{(u_{\ell}t)}\e^{-\ell t/2} \, .
\end{aligned}
\ee
Finally, we note that
\be\label{MRGF}
\Omega_{\ell}(0) =-{\ell \over 2} + 2g\gamma^{\ell}_{1}(g)\, , \qquad \tilde{\Omega}_{\ell}(0) = u_{\ell} + 2g\tilde{\gamma}^{\ell}_{1}(g)\, ,
\ee
where the coefficients $\gamma^{\ell}_{1}, \tilde{\gamma}^{\ell}_{1}$, originate from the small-$t$ asymptotics
\be
\gamma^{\ell}_{-}(2gt) =2g\gamma^{\ell}_{1} t +O(t^3)\, , \qquad \tilde{\gamma}^{\ell}_{-}(2gt) =2g\tilde{\gamma}^{\ell}_{1} t +O(t^3)\, .
\ee

\subsubsection*{Large fermion}

For a large fermion, the physical rapidity $x_{\textrm{f}}$ has modulus bigger than $g$ and is given in terms of the Bethe rapidity $u_{\textrm{lf}}$ as
\be\label{Xlf}
x_{\textrm{f}} = {1 \over 2}\left(u_{\textrm{lf}}+\sqrt{u_{\textrm{lf}}^2-(2g)^2}\right)\, .
\ee
We use the subscript `$\textrm{lf}$' (for `large fermion') to keep track of this information when working with the Bethe rapidity. Note that for real rapidity $x_{\textrm{f}}$ we need $u^2_{\textrm{lf}} >(2g)^2$.

The phase for the scattering of a main root $u$ with a fermion carrying a large rapidity $x_{\textrm{f}}$ is conveniently written as 
\be\label{SPf}
S_{4\textrm{f}}(u, x_{\textrm{f}}) = {u-u_{\textrm{lf}}+\ft{i}{2} \over u-u_{\textrm{lf}}-\ft{i}{2}} {1-g^2/x^{-}x_{\textrm{f}} \over 1-g^2/x^{+}x_{\textrm{f}}}\, .
\ee
Given the expression~(\ref{SPf}), we can evaluate the source terms $\mathcal{I}_{\textrm{lf}}(u, u_{\textrm{lf}}), \tilde{\mathcal{I}}_{\textrm{lf}}(u, u_{\textrm{lf}})$, as defined in~(\ref{ITE}). Using the formulae~(\ref{OLEIrep}, \ref{Kexpa}, \ref{Mexpc}, \ref{Mexpcb}), we find
\be
\begin{aligned}
\mathcal{I}_{\textrm{lf}}(u, u_{\textrm{lf}}) = & -2\int_{0}^{\infty}dt\cos{(ut)}\e^{-t/2}\pi_{\textrm{lf}}(t, u_{\textrm{lf}})\, , \\
\tilde{\mathcal{I}}_{\textrm{lf}}(u, u_{\textrm{lf}}) = & -2\int_{0}^{\infty}dt\sin{(ut)}\e^{-t/2}\pi_{\textrm{lf}}(t, u_{\textrm{lf}})\, ,
\end{aligned}
\ee
where
\be
\begin{aligned}\label{PIlf}
\pi_{\textrm{lf}}(t, u_{\textrm{lf}})=& \cos{(u_{\textrm{lf}}t)}-J_{0}(2gt) - 2g^2 \int_{0}^{\infty}ds\, t K_{-}(2gt, 2gs)\cos{(u_{\textrm{lf}}s)}\, , \\
\tilde{\pi}_{\textrm{lf}}(t, u_{\textrm{lf}}) =& \sin{(u_{\textrm{lf}}t)}- 2g^2\int_{0}^{\infty}ds\, t K_{+}(2gt, 2gs)\sin{(u_{\textrm{lf}}s)}\, .
\end{aligned}
\ee
with $K_{\pm}(t, s)$ given in~(\ref{Kernpm}).

The defining relations for $\gamma^{\textrm{lf}}_{\pm}(t), \tilde{\gamma}^{\textrm{lf}}_{\pm}(t),$ are
\be
\begin{aligned}\label{SGLF}
\gamma^{\textrm{lf}}_{+}(2gt)+\gamma^{\textrm{lf}}_{-}(2gt) &= (\e^{t}-1)\Omega_{\textrm{lf}}(t) - \cos{(u_{\textrm{lf}}t)}-J_{0}(2gt)\, , \\
\tilde{\gamma}^{\textrm{lf}}_{+}(2gt)+\tilde{\gamma}^{\textrm{lf}}_{-}(2gt) &= (\e^{t}-1)\tilde{\Omega}_{\textrm{lf}}(t) - \sin{(u_{\textrm{lf}}t)}\, .
\end{aligned}
\ee
The source terms in~(\ref{EQN}) are found to be
\be
\begin{aligned}\label{KLFa}
\kappa^{\textrm{lf}}_{2n}(u_{\textrm{lf}}) &= -\int_{0}^{\infty}{dt \over t}{J_{2n}(2gt)\over \e^{t}-1}\left(\cos{(u_{\textrm{lf}}t)}-J_{0}(2gt)\right)-{1\over 2}\int_{0}^{\infty}{dt \over t}J_{2n}(2gt)\cos{(u_{\textrm{lf}}t)}\, ,  \\
\kappa^{\textrm{lf}}_{2n-1}(u_{\textrm{lf}}) &= -\int_{0}^{\infty}{dt \over t}{J_{2n-1}(2gt)\over \e^{t}-1}\left(\cos{(u_{\textrm{lf}}t)}-J_{0}(2gt)\right)\, ,
\end{aligned}
\ee
and
\be
\begin{aligned}\label{KLFb}
\tilde{\kappa}^{\textrm{lf}}_{2n}(u_{\textrm{lf}}) &= -\int_{0}^{\infty}{dt \over t}{J_{2n}(2gt)\over \e^{t}-1}\sin{(u_{\textrm{lf}}t)}\, , \\
\tilde{\kappa}^{\textrm{lf}}_{2n-1}(u_{\textrm{lf}}) &= -\int_{0}^{\infty}{dt \over t}{J_{2n-1}(2gt)\over \e^{t}-1}\sin{(u_{\textrm{lf}}t)}-{1\over 2}\int_{0}^{\infty}{dt \over t}J_{2n-1}(2gt)\sin{(u_{\textrm{lf}}t)}\, .
\end{aligned}
\ee
We note also that
\be\label{MRLF}
\Omega_{\textrm{lf}}(0) = 2g\gamma^{\textrm{lf}}_{1}\, , \qquad \tilde{\Omega}_{\textrm{lf}}(0) = u_{\textrm{lf}} + 2g\tilde{\gamma}^{\textrm{lf}}_{1}\, ,
\ee
where the coefficients $\gamma^{\textrm{lf}}_{1}, \tilde{\gamma}^{\textrm{lf}}_{1}$, come from
\be
\gamma^{\textrm{lf}}_{-}(2gt) =2g\gamma^{\textrm{lf}}_{1} t +O(t^3)\, , \qquad \tilde{\gamma}^{\textrm{lf}}_{-}(2gt) =2g\tilde{\gamma}^{\textrm{lf}}_{1} t +O(t^3)\, .
\ee

\subsubsection*{Small fermion}

For a small fermion, the physical rapidity $x_{\textrm{f}}$ has modulus smaller than $g$ and is given in terms of the Bethe rapidity $u_{\textrm{sf}}$ by
\be\label{Xsf}
x_{\textrm{f}} = {1 \over 2}\left(u_{\textrm{sf}}-\sqrt{u_{\textrm{sf}}^2-(2g)^2}\right)\, .
\ee
Same remarks as for large fermion (see below Eq.~(\ref{Xlf})).

The phase for the scattering of a main root $u$ with a fermion carrying a small rapidity $x_{\textrm{f}}$ is conveniently written as 
\be
S_{4\textrm{f}}(u, x_{\textrm{f}}) = {x^{+} \over x^{-}}\left({1-x_{\textrm{f}}/x^{+}\over 1-x_{\textrm{f}}/x^{-}}\right)\, .
\ee
Plugging this expression into~(\ref{ITE}), we find that the inhomogeneous terms, which source corrections to the twist-two solution, are given by
\be\label{ITsf}
\mathcal{I}_{\textrm{sf}}(u, u_{\textrm{sf}}) + \tilde{\mathcal{I}}_{\textrm{sf}}(u, u_{\textrm{sf}}) = -i\partial_{u}\log{\left({1-x_{\textrm{f}}/x^{+}(u)\over 1-x_{\textrm{f}}/x^{-}(u)}\right)} \, ,
\ee
where $\mathcal{I}_{\textrm{sf}}(u, u_{\textrm{sf}}), \tilde{\mathcal{I}}_{\textrm{sf}}(u, u_{\textrm{sf}})$, are even, odd, in $u$, respectively. Using the formulae~(\ref{Mexpc}, \ref{Mexpcb}), the latter quantities can be found to admit the representation
\be
\begin{aligned}
\mathcal{I}_{\textrm{sf}}(u, u_{\textrm{sf}}) = & -2\int_{0}^{\infty}dt\cos{(ut)}\e^{-t/2}\pi_{\textrm{sf}}(t, u_{\textrm{sf}})\, , \\
\tilde{\mathcal{I}}_{\textrm{sf}}(u, u_{\textrm{sf}}) = & -2\int_{0}^{\infty}dt\sin{(ut)}\e^{-t/2}\pi_{\textrm{sf}}(t, u_{\textrm{sf}})\, ,
\end{aligned}
\ee
where
\be
\begin{aligned}\label{PIsf}
\pi_{\textrm{sf}}(t, u_{\textrm{sf}})=&  \, 2g^2 \int_{0}^{\infty}ds\, t K_{-}(2gt, 2gs)\cos{(u_{\textrm{sf}}s)}\, , \\
\tilde{\pi}_{\textrm{sf}}(t, u_{\textrm{sf}}) =& \, 2g^2\int_{0}^{\infty}ds\, t K_{+}(2gt, 2gs)\sin{(u_{\textrm{sf}}s)}\, .
\end{aligned}
\ee
The kernels $K_{\pm}(t, s)$ are given in~(\ref{Kernpm}).

The defining relations for $\gamma^{\textrm{sf}}_{\pm}(t), \tilde{\gamma}^{\textrm{sf}}_{\pm}(t),$ are especially simple. They read
\be
\begin{aligned}\label{SGSF}
\gamma^{\textrm{sf}}_{+}(2gt)+\gamma^{\textrm{sf}}_{-}(2gt) &= (\e^{t}-1)\Omega_{\textrm{sf}}(t)\, , \\
\tilde{\gamma}^{\textrm{sf}}_{+}(2gt)+\tilde{\gamma}^{\textrm{sf}}_{-}(2gt) &= (\e^{t}-1)\tilde{\Omega}_{\textrm{sf}}(t)\, .
\end{aligned}
\ee
The \textit{non-zero} source terms in~(\ref{EQN}) are
\be
\begin{aligned}\label{KSF}
\kappa^{\textrm{sf}}_{2n}(u_{\textrm{sf}}) &= {1 \over 2}\int_{0}^{\infty}{dt \over t}J_{2n}(2gt)\cos{(u_{\textrm{sf}}t)} \, , \\
\tilde{\kappa}^{\textrm{sf}}_{2n-1}(u_{\textrm{sf}}) &= {1 \over 2}\int_{0}^{\infty}{dt \over t}J_{2n-1}(2gt)\sin{(u_{\textrm{sf}}t)}\, .
\end{aligned}
\ee
We have furthermore the relations
\be\label{MRSF}
\Omega_{\textrm{sf}}(0) = 2g\gamma^{\textrm{sf}}_{1}\, , \qquad \tilde{\Omega}_{\textrm{sf}}(0) = 2g\tilde{\gamma}^{\textrm{sf}}_{1}\, ,
\ee
where the coefficients $\gamma^{\textrm{sf}}_{1}, \tilde{\gamma}^{\textrm{sf}}_{1}$, come from
\be
\gamma^{\textrm{sf}}_{-}(2gt) =2g\gamma^{\textrm{sf}}_{1} t +O(t^3)\, , \qquad \tilde{\gamma}^{\textrm{sf}}_{-}(2gt) =2g\tilde{\gamma}^{\textrm{sf}}_{1} t +O(t^3)\, .
\ee

\subsubsection*{Vacuum distribution}
 
The vacuum distribution density can be written in the large spin limit as
\be
\begin{aligned}\label{RVAC}
\rho_{\textrm{vacuum}}(u) =& \, \, {2\over \pi}\left(\log{S}-\psi(1)\right)+{2\over \pi}\int_{0}^{\infty}dt\, {\cos{(ut)}\e^{t/2}J_{0}(2gt)-1 \over \e^{t}-1} \\
&-\left(\log{S}-\psi(1)\right)\sigma^{\o}(u) -\delta \sigma^{\o}(u)\, ,
\end{aligned}
\ee
where $\sigma^{\o}(u) = \sigma^{\o}(-u)$ can be found as solution to the BES equation~\cite{BES06}. The function $\delta \sigma^{\o}(u)=\delta \sigma^{\o}(-u)$ captures the subleading, $O(\log^{0}{S})$, correction to the vacuum (twist-two) distribution density, considered in~\cite{FZ09}.

One can introduce the functions $\Omega^{\o}(t), \delta\Omega^{\o}(t)$, as the Fourier-like transforms of  $\sigma^{\o}(u)$, $\delta \sigma^{\o}(u)$, respectively, following~(\ref{FLtra}, \ref{FLtrb}). Then, defining $\gamma^{\o}_{+}(t), \delta\gamma^{\o}_{+}(t)$, and $\gamma^{\o}_{-}(t), \delta\gamma^{\o}_{-}(t)$, respectively even and odd in $t$, by
\be
\begin{aligned}\label{OGV}
\gamma^{\o}_{+}(2gt)+\gamma^{\o}_{-}(2gt) &= (\e^{t}-1)\Omega^{\o}(t)\, , \\
\delta\gamma^{\o}_{+}(2gt)+\delta\gamma^{\o}_{-}(2gt) &= (\e^{t}-1)\delta\Omega^{\o}(t)\, ,
\end{aligned}
\ee
one finds that $\gamma^{\o}_{\pm}(t), \delta\gamma^{\o}_{\pm}(t),$ solve the system of equations~(\ref{EQN}) with source terms
\be
\begin{aligned}
\kappa_{n}^{\o} &= 2g\, \delta_{n, 1}\, , \\
\delta\kappa_{n}^{\o} &= 2\int_{0}^{\infty}{dt\over t}{J_{n}(2gt)J_{0}(2gt)-gt\delta_{n, 1} \over \e^{t}-1}\, ,
\end{aligned}
\ee
respectively. Note that our convention for introducing $\gamma^{\o}_{\pm}(t)$ differs from the one used in~\cite{BK09} by the overall factor `$4g$'. Namely, we have $\gamma^{\o}_{\pm}(t) = 4g\gamma_{\pm}(t)$, where $\gamma_{\pm}(t)$ are the functions analyzed in~\cite{BK09}. The difference between the coefficients $\delta\kappa_{n}^{\o}$ given here and the ones used in~\cite{FZ09} originates from the decomposition~(\ref{RVAC}), which is different in the two cases. A redefinition of the type $\delta\gamma^{\o}_{+}(t) \rightarrow \delta\gamma^{\o}_{+}(t) + \alpha(1-J_{0}(t))$, for some constant $\alpha$, would lead to the coefficients of~\cite{FZ09}. The reason for decomposing as in~(\ref{RVAC}) is that it improves the large $n$ behavior of the coefficients $\delta\kappa_{n}^{\o}$.

\section{Isotopic $SU(2)$ root}\label{AISR}
 
In this appendix, we perform the all-loop analysis of the isotopic root $u_{\textrm{b}}$ which implements the restoration of the $SU(2)\subset SU(4)$ symmetry. We recall that this root is a stack formed out of one 2-string of main roots, one root $u_{3}$ and one root $u_{5}$, as
\be\label{StackIso}
u_{4}^{\pm} = u_{\textrm{b}}\pm \ft{i}{2}\, , \qquad u_{3} = u_{5} = u_{\textrm{b}}\, ,
\ee
where $u_{\textrm{b}}$ is taken to be real.%
\footnote{This assumption can probably be relaxed to permit complex roots $u_{\textrm{b}, j}$.}
Our analysis also assumes that $u_{\textrm{b}}^2 > (2g)^2$ to avoid complication with the cuts structure of the all-loop asymptotic Bethe ansatz equations~\cite{BS05}. At the end the results show no dependence at all on this bound and can be easily analytically continued. It would be interesting nevertheless to understand how we should deal with the cuts structure to get explicitely the results when $u_{\textrm{b}}^2 < (2g)^2$.

The appendix is organized as follows. We first prove that the stack~(\ref{StackIso}) does not carry energy and momentum. The analysis also shows that they do not carry any of the higher spin-chain conserved charges.%
\footnote{They contribute however to the total spin-chain momentum $\e^{iP}$ by a factor $(-1)^{K_{\textrm{b}}}$ where $K_{\textrm{b}}$ is the total number of root $u_{\textrm{b}}$.}
To find the energy and momentum of the stack~(\ref{StackIso}), we need to compute the correction to the distribution of main roots sourced by~(\ref{StackIso}). This is done by contructing the scattering phase between main root and root $u_{\textrm{b}}$, from which an inhomogeneous term for the large spin integral equation is obtained. Then we show that the solution to this equation is exactly  equal to its one-loop expression. This is enough to prove the statement that the stack is isotopic~(\ref{StackIso}). Then, to establish that the stack~(\ref{StackIso}) implements the restored $SU(2)$ symmetry, we show that the effective Bethe ansatz equations for the roots $u_{\textrm{b}, j}$ are identical to the ones of an inhomogeneous $SU(2)$ Heisenberg spin chain. The length of this spin chain is fixed by the total number of scalars (holes) and the inhomogeneities are given by the holes rapidities.
 
\subsection{Energy and momentum}
 
The energy of the stack~(\ref{StackIso}) is the sum of the bare energy carried by the 2-string of main roots $u_{\textrm{b}}^{\pm}$ plus the energy induced by the deformation of the background. To get the latter contribution, we need the expression $\rho_{\textrm{b}}(u)$ for the correction to the distribution density sourced by~(\ref{StackIso}). This function is given as the solution to the large spin integral equation. The main ingredient of this equation is the inhomogeneous term which follows from the undressed scattering phase between the probe main root $u$ and the stack~(\ref{StackIso}). The kernel of the equation is the same as the one already considered in Section~\ref{ALA}. The main simplicity here, as compared to the momentum carriers studied in Section~\ref{ALA}, is that the integral equation can be solved exactly for a stack. We recall finally that in our approach we consider separately the even and odd part in $u$ of the correction density
\be
\rho_{\textrm{b}}(u) = -\sigma_{\textrm{b}}(u)-\tilde{\sigma}_{\textrm{b}}(u)\, ,
\ee
with the notations of Section~\ref{ALA}.

We start by constructing the inhomogeneous term. The phase for the scattering of a main root $u$ through an isotopic root $u_{\textrm{b}}$ is given by
\be\label{SisoApp}
S_{4\textrm{b}}(u, u_{\textrm{b}}) = S_{44}\left(u, u_{\textrm{b}}^{+}\right)S_{44}\left(u, u_{\textrm{b}}^{-}\right)S_{43}\left(u, u_{\textrm{b}}\right)S_{45}\left(u, u_{\textrm{b}}\right)\, ,
\ee
where $S_{44}, S_{43}, S_{45}$, stand for the elementary scattering phases between a main root $u$ and a root $u_{4}, u_{3}, u_{5}$, respectively. Their expressions are given on the right-hand side of the middle-node equations in~(\ref{ABAE}). After isolating the contribution coming from the dressing phase, we find that the undressed scattering amplitude, denoted $S^{\circ}_{4\textrm{b}}$, reads
\be\label{SP4a}
S^{\circ}_{4\textrm{b}}(u, u_{\textrm{b}}) = {u-u_{\textrm{b}}-\ft{3i}{2} \over u-u_{\textrm{b}}+\ft{3i}{2}}{u-u_{\textrm{b}}+\ft{i}{2} \over u-u_{\textrm{b}}-\ft{i}{2}}\left({1-g^2/x^{+}x_{\textrm{b}}^{[-2]} \over 1-g^2/x^{-}x_{\textrm{b}}^{[+2]}}\right)^2\, ,
\ee
where $x^{[\pm2]}_{\textrm{b}} = x(u_{a}\pm i)$. The phase~(\ref{SP4a}) generates the inhomogeneous terms $\mathcal{I}_{\textrm{b}}(u, u_{\textrm{b}}), \tilde{\mathcal{I}}_{\textrm{b}}(u, u_{\textrm{b}})$,  that are given by
\be
-i\partial_{u}\log{S^{\circ}_{4\textrm{b}}}(u, u_{\textrm{b}}) \equiv \mathcal{I}_{\textrm{b}}(u, u_{\textrm{b}}) + \tilde{\mathcal{I}}_{\textrm{b}}(u, u_{\textrm{b}})\, ,
\ee
where $\mathcal{I}_{\textrm{b}}(u, u_{\textrm{b}}), \tilde{\mathcal{I}}_{\textrm{b}}(u, u_{\textrm{b}})$, are even, odd, in $u$ respectively. Evaluating $\mathcal{I}_{\textrm{b}}(u, u_{\textrm{b}}), \tilde{\mathcal{I}}_{\textrm{b}}(u, u_{\textrm{b}})$, with the help of formulae given in Appendix~\ref{AALE}, we find
\be
\begin{aligned}
\mathcal{I}_{\textrm{b}}(u, u_{\textrm{b}}) =&  -2\int_{0}^{\infty}dt\, \cos{(ut)}\e^{-t/2}\pi_{a}(t, u_{\textrm{b}})\, , \\
\tilde{\mathcal{I}}_{\textrm{b}}(u, u_{\textrm{b}}) =&  -2\int_{0}^{\infty}dt\, \sin{(ut)}\e^{-t/2}\tilde{\pi}_{a}(t, u_{\textrm{b}})\, ,
\end{aligned}
\ee
where
\be\label{InhApp}
\begin{aligned}
\pi_{\textrm{b}}(t, u_{\textrm{b}}) &= \left(1-\e^{-t}\right)\cos{(u_{\textrm{b}}t)} + (2g)^2 \int_{0}^{\infty}ds\, t K(2gt, 2gs)\cos{(u_{\textrm{b}}s)}\e^{-s}\, , \\
\tilde{\pi}_{\textrm{b}}(t, u_{\textrm{b}}) &=\left(1-\e^{-t}\right)\sin{(u_{\textrm{b}}t)} + (2g)^2 \int_{0}^{\infty}ds\, t K(2gt, 2gs)\sin{(u_{\textrm{b}}s)}\e^{-s}\, . 
\end{aligned}
\ee
The kernel $K(t, s)$ is given in~(\ref{Kern}).

Now the claim is that the solutions $\Omega_{\textrm{b}}(t), \tilde{\Omega}_{\textrm{b}}(t)$, to the equations~(\ref{EqTF}) sourced by the terms~(\ref{InhApp}) are
\be\label{IrD}
\Omega_{\textrm{b}}(t) = \cos{(u_{\textrm{b}}t)}\e^{-t}\, , \qquad \tilde{\Omega}_{\textrm{b}}(t) = \sin{(u_{\textrm{b}}t)}\e^{-t}\, ,
\ee
where we recall that $\Omega_{\textrm{b}}(t), \tilde{\Omega}_{\textrm{b}}(t)$, are the Fourier-like transforms~(\ref{FLtra}, \ref{Dec}) of $\sigma_{\textrm{b}}(u), \tilde{\sigma}_{\textrm{b}}(u)$, respectively. The fact that the two functions~(\ref{IrD}) solve the equations would be straightforward if it were not for the dressing-phase contributions in~(\ref{EqTF}). So let us see if the two functions~(\ref{IrD}) annihilate these terms separately. For the dressing-phase contributions to vanish, we need all the conserved charges sourced by a stack to be zero. The latter charges can be found as sum of a bare and induced contributions. Evaluating the induced contribution associated to~(\ref{IrD}) we find
\begin{align}
q^{\textrm{induced}}_{2n} &= {2(-1)^{n} \over g^{2n-1}}\int_{0}^{\infty}{dt \over t}J_{2n-1}(2gt)\Omega_{a}(t) = -{i \over 2n}\left[\left({1 \over x^{[+2]}_{\textrm{b}}}\right)^{2n}-\left({1 \over x^{[-2]}_{\textrm{b}}}\right)^{2n}\right]\, , \notag \\ 
q^{\textrm{induced}}_{2n+1} &= {2(-1)^{n} \over g^{2n}}\int_{0}^{\infty}{dt \over t}J_{2n}(2gt)\tilde{\Omega}_{a}(t) = -{i \over 2n-1}\left[\left({1 \over x^{[+2]}_{\textrm{b}}}\right)^{2n-1}-\left({1 \over x^{[-2]}_{\textrm{b}}}\right)^{2n-1}\right]\, .
\end{align}
But these are exactly \textit{minus} the charges for a $2$-string of main roots. It means that induced and bare conserved charges cancel one another. The dressing-phase contributions vanish for~(\ref{IrD}) and the functions~(\ref{IrD}) are exact solutions.

We have shown that a stack with rapidity $u_{\textrm{b}}$ do not carry any of the spin-chain charges. Therefore it does contribute to the anomalous dimension and cannot carry GKP energy. Moreover the correction to the density sourced by $u_{\textrm{b}}$ is one-loop exact,
\be\label{DisoApp}
\rho_{\textrm{b}}(u) \equiv -\sigma_{\textrm{b}}(u)-\tilde{\sigma}_{\textrm{b}}(u) = -{1\over \pi}\int_{0}^{\infty}dt\, \cos{((u-u_{\textrm{b}})t)}\e^{-t/2}\, .
\ee
From the large $u$ asymptotics of $\rho_{\textrm{b}}(u)$, we get the GKP momentum by looking at the coefficient in front of the leading $\sim 1/u$ contribution, see Eq.~(\ref{SPLU}). But, the density~(\ref{DisoApp}) scales like $\sim 1/u^2$ at large $u$, meaning that the stack is exactly momentum-less, $p_{\textrm{b}} = 0$. This concludes the proof that the stack~(\ref{StackIso}) is isotopic. 

\subsection{Effective equations}

To get the effective equations for the stack rapidities $u_{\textrm{b}, j}$, we pick up a probe stack, with a rapidity $u_{\textrm{b}, k}$, that we transport once around the spin chain. Gathering the phases for the scattering with all the possible roots, composing a generic state, we get the equations by imposing the condition that the total phase equals $1$. The last identity correponds to imposing periodic boundary condition to the Bethe wave function.

The total phase can be decomposed into a product of a bare and an induced scattering factor, giving the equations as
\be\label{EffEqApp}
1 = S^{\textrm{induced}}_{\textrm{b}}S^{\textrm{bare}}_{\textrm{b}}\, .
\ee
The induced phase takes into account  the scattering with the main roots,
\be\label{IndScatt}
S^{\textrm{induced}}_{\textrm{b}} = \prod_{j=1}^{K_{4}}S_{\textrm{b}4}(u_{\textrm{b}, k}, u_{4, j})\, .
\ee
Note that we take product over all the main roots, real plus complex, thus including the ones that form part of the stacks $u_{\textrm{b}, j}$, with $j\neq k$.%
\footnote{We should not include scattering of a stack with itself, hence the condition that any product over the set of stacks should be restricted to $j\neq k$.} Though it may seem more natural to define the induced scattering phase by taking product over the \emph{real} main roots only, we found simpler to gather the full set of main roots into a single factor. The immediate consequence of this prescription is that the bare scattering of the stack $u_{\textrm{b}, j}$ with its congeners boils down to a product of scattering phases with a set of fermionic roots carrying rapidities $u_{3, j} = u_{5, j} = u_{\textrm{b}, j}$. Since there is no bare interaction between fermionic roots, the only phases come from the scattering of the 2-string $u_{4, k}^{\pm} = u_{\textrm{b}, k} \pm i/2$ with the roots $u_{3, j} = u_{5, j} = u_{\textrm{b}, j}$, with $j\neq k$. Using~(\ref{ABAE}) we get
\be
S^{\textrm{bare}}_{\textrm{b}\textrm{b}} = \prod_{j\neq k}^{K_{\textrm{b}}} \left({x^{[+2]}_{\textrm{b}, k}-x_{\textrm{b}, j} \over x^{[-2]}_{\textrm{b}, k}-x_{\textrm{b}, j}}\right)^2\, ,
\ee
with $x^{[\pm2]}_{\textrm{b}, k} = x(u_{\textrm{b}, k}\pm i)$. The equations are thus given by~(\ref{EffEqApp}) with the induced scattering phase~(\ref{IndScatt}) and the bare scattering phase
\be\label{BareScatt}
S^{\textrm{bare}}_{\textrm{b}} =  \left({x^{[-2]}_{\textrm{b}} \over x^{[+2]}_{\textrm{b}}}\right)^{2+M} S^{\textrm{bare}}_{\textrm{b}\textrm{b}}\prod_{\star =1}^{M} S^{\textrm{bare}}_{\textrm{b}\star}\, ,
\ee
where the first term on the right-hand of~(\ref{BareScatt}) comes from the propagation of the 2-string around the spin chain with length $2+M$,%
\footnote{See Eq.~(\ref{EffLength}) and comments following it for the origin of this quantity.} while the last term is for the bare scattering with a generic set of $M$ momentum-carrying excitations. With the help of the expressions computed below, we find that the equations~(\ref{EffEqApp}) read
\be\label{EffEqAppf}
\prod_{j=1}^{K_{\textrm{h}}}{u_{\textrm{b}, k}-u_{\textrm{h}, j}+\ft{i}{2} \over u_{\textrm{b}, k}-u_{\textrm{h}, j}-\ft{i}{2}} = \prod_{j\neq k}^{K_{\textrm{b}}}{u_{\textrm{b}, k}-u_{\textrm{b}, j}+i \over u_{\textrm{b}, k}-u_{\textrm{b}, j}-i}\, .
\ee
They are identical to the equations of a $\mathfrak{su}(2)$ inhomogeneous spin chain of lenth $K_{\textrm{h}}$, with inhomogeneities given by the set of holes rapidities $u_{\textrm{h}, j}$. We recall that $K_{\textrm{h}}$ is the total number of scalars, whatever $\mathfrak{su}(4)$ polarizations they carry. Including scattering with the $\mathfrak{su}(2)\times \mathfrak{su}(2) \subset \mathfrak{su}(4)$ isotopic roots $u_{2}, u_{6}$, the equations~(\ref{EffEqAppf}) become the middle-node equations for an inhomogeneous $\mathfrak{so}(6)$ spin chain, as follows from multiplying the right-hand side of~(\ref{EffEqAppf}) by the factor  
\be\label{EffEqAppfb}
\prod_{j=1}^{K'_{2}}{u_{\textrm{b}, k}-u_{2, j}-\ft{i}{2} \over u_{\textrm{b}, k}-u_{2, j}+\ft{i}{2}}\prod_{j=1}^{K'_{6}}{u_{\textrm{b}, k}-u_{6, j}-\ft{i}{2} \over u_{\textrm{b}, k}-u_{6, j}+\ft{i}{2}}\, ,
\ee
where $K'_{2}, K'_{6},$ are the numbers of solitary roots $u_{2}, u_{6}$ (i.e., those roots which are not part of some stacks). The simplicity of the expression~(\ref{EffEqAppfb}) comes from the absence of higher-loops corrections to the second and sixth node asymptotic Bethe ansatz equations~(\ref{ABAE}), combined with the fact that there is no induced scattering phase since there is no coupling between roots of type $2, 6$, and main roots. The simplicity of~(\ref{EffEqAppf}), however, is due to a wealth of cancellations between the induced and bare scattering phase. To prove~(\ref{EffEqAppf}), we will now compute $S^{\textrm{bare}}_{\textrm{b}\star}$ and $S^{\textrm{induced}}_{\textrm{b}}$, starting with the former.

\subsubsection*{Bare scattering}

We list below the various bare sattering phases $S^{\textrm{bare}}_{\textrm{b}\star}$ for $\star = \textrm{h}, \textrm{f}, \bar{\textrm{f}}, \ell, \bar{\ell}$. A simplification comes from the observation that the isotopic stack is invariant under the replacement $(u_{1}, u_{2}, u_{3}) \rightarrow (u_{7}, u_{6}, u_{5})$ changing the sign of the $\mathfrak{so}(2)$ spin. Therefore, it is not necessary to distinguish between $\textrm{f}, \ell$, and $\bar{\textrm{f}}, \bar{\ell}$.
\begin{itemize}
\item There is no bare scattering with a hole,
\be
S^{\textrm{bare}}_{\textrm{b}\textrm{h}} = 1\, .
\ee
\item The bare scattering of the stack with a fermion reads
\be
S^{\textrm{bare}}_{\textrm{b}\textrm{f}} = {x^{[+2]}_{\textrm{b}, k}-x_{\textrm{f}} \over x^{[-2]}_{\textrm{b}, k}-x_{\textrm{f}}}\, .
\ee
\item The bare scattering of the stack with a gauge field ($\ell=1$) or bound states ($\ell > 1$) is given by
\be
S^{\textrm{bare}}_{\textrm{b}\ell} = {u_{\textrm{b}, k}-u_{\ell} +i(1+\ell/2) \over u_{\textrm{b}, k}-u_{\ell} - i(1+\ell/2)}\left({1-g^2/x^{[-2]}_{\textrm{b}, k}x^{[+\ell]}_{\ell} \over 1-g^2/x^{[+2]}_{\textrm{b}, k}x^{[+\ell]}_{\ell}}{1-g^2/x^{[-2]}_{\textrm{b}, k}x^{[-\ell]}_{\ell} \over 1-g^2/x^{[+2]}_{\textrm{b}, k}x^{[-\ell]}_{\ell}}\right)\, .
\ee
It is obtained by taking product of elementary scattering phases between the isotopic stack and the $\ell$-stack.

\end{itemize}
This is the complete list of bare scattering phases for excitations considered in this paper.

\subsubsection*{Induced scattering}
 
The induced scattering phase~(\ref{IndScatt}) is more difficult to estimate, since it requires integrating over the continuum of real main roots. The starting point is the phase for the scattering of the root $u_{\textrm{b}, k}$ with a main root $u$,
\be\label{SelApp}
S_{\textrm{b}4}(u_{\textrm{b}, k}, u) = S^{\circ}_{\textrm{b}4}(u_{\textrm{b}, k}, u)S^{\textrm{dressing}}_{\textrm{b}4}(u_{\textrm{b}, k}, u) \, ,
\ee
where for convenience we have factored out the the dressing phase contribution that reads
\be
S^{\textrm{dressing}}_{\textrm{b}4}(u_{\textrm{b}, k}, u) = \sigma^2(u_{\textrm{b}, k}^{+}, u)\sigma^2(u_{\textrm{b}, k}^{-}, u)\, .
\ee
The undressed scattering phase $S^{\circ}_{\textrm{b}4}(u_{\textrm{b}, k}, u)$ is given by $S^{\circ}_{\textrm{b}4}(u_{\textrm{b}, k}, u) = S^{\circ\, -1}_{4\textrm{b}}(u, u_{\textrm{b}, k})$, which, after using~(\ref{SP4a}), yields
\be\label{SelAppb}
S_{\textrm{b}4}^{\circ}(u_{\textrm{b}, k}, u) = {u_{\textrm{b}, k}-u-\ft{3i}{2} \over u_{\textrm{b}, k}-u+\ft{3i}{2}}{u_{\textrm{b}, k}-u+\ft{i}{2} \over u_{\textrm{b}, k}-u-\ft{i}{2}}\left({1-g^2/x_{\textrm{b}, k}^{[+2]}x^{-}\over 1-g^2/x_{\textrm{b}, k}^{[-2]}x^{+}}\right)^{2}\, .
\ee
To proceed further it is useful to take the $\log$ of~(\ref{SelApp}). We find that the latter quantity can be written as
\be\label{PhiApp}
-i\log{S_{\textrm{b}4}(u_{\textrm{b}, k}, u)} = \int_{0}^{\infty}{dt \over t}\sin{(u_{\textrm{b}, k}t)}\e^{-t}\phi(t, u) - \int_{0}^{\infty}{dt \over t}\cos{(u_{\textrm{b}, k}t)}\e^{-t}\tilde{\phi}(t, u) \, ,
\ee
where
\be
\phi(t, u) = \phi^{\circ}(t, u) +\phi^{\textrm{dressing}}(t, u) \, , \qquad \tilde{\phi}(t, u) =  \tilde{\phi}^{\circ}(t, u) + \tilde{\phi}^{\textrm{dressing}}(t, u)\, ,
\ee
reflecting the decomposition~(\ref{SelApp}). After a bit of algebra, using identities given in Appendix~\ref{AALE}, we get
\be\label{PhiAppa}
\begin{aligned}
\phi^{\circ}(t, u) =& \, -4\sinh{\left(t/2\right)}\cos{(ut)} - 2(2g)^2\int_{0}^{\infty}ds\, t K(2gt,2gs)\cos{(us)}\e^{-s/2}\, , \\
\tilde{\phi}^{\circ}(t, u) =& \,  -4\sinh{\left(t/2\right)}\sin{(ut)} - 2(2g)^2\int_{0}^{\infty}ds\, t K(2gt,2gs)\sin{(us)}\e^{-s/2}\, ,
\end{aligned}
\ee
and
\be\label{PhiAppb}
\begin{aligned}
\phi^{\textrm{dressing}}(t, u) &= 2(2g)^2\int_{0}^{\infty}ds\, tK_{-}(2gt, 2gs) {q_{-}(2gs, u)\over \e^{t}-1}\, , \\
\tilde{\phi}^{\textrm{dressing}}(t, u) &= 2(2g)^2\int_{0}^{\infty}ds\, tK_{+}(2gt, 2gs) {\tilde{q}_{+}(2gs, u)\over \e^{t}-1}\, .
\end{aligned}
\ee
The kernels $K(t,s), K_{+}(t, s), K_{-}(t, s),$ are given in~(\ref{Kern}, \ref{Kernpm}). In analogy with the analysis in Section~\ref{ALA}, we have introduced generating functions $q_{-}(t, u), \tilde{q}_{+}(t, u)$, for even, odd, charges $q_{2n}(u), q_{2n-1}(u)$, respectively, as
\be\label{InCCApp}
\begin{aligned}
q_{-}(t, u) &= 2\sum_{n\ge 1}(2n-1)(-1)^{n}g^{2n-1}q_{2n}(u)J_{2n-1}(t)\, ,\\
\tilde{q}_{+}(t, u) &= 2\sum_{n\ge 1}(2n)(-1)^{n}g^{2n}q_{2n+1}(u)J_{2n}(t)\, .
\end{aligned}
\ee

The expressions above hold for a generic main root $u$. To obtain the complete induced scattering phase, we substitute $u\rightarrow u_{4, j}$ in~(\ref{PhiApp}), sum over the set of main roots and take the large spin limit. Performing these steps leads to the decomposition
\be\label{IndScattb}
\sum_{j=1}^{K_{4}}\phi^{\circ}(t, u_{4, j}) = \int dv\, \rho(v)\phi^{\circ}(t, v) - \sum_{j=1}^{K_{\textrm{h}}}\phi^{\circ}(t, u_{\textrm{h}, j}) + \sum_{j\neq k}^{K_{\textrm{b}}}\phi^{\circ}(t, u_{\textrm{b}, j}+\ft{i}{2}) + \sum_{j\neq k}^{K_{\textrm{b}}}\phi^{\circ}(t, u_{\textrm{b}, j}-\ft{i}{2})\, ,
\ee
and similarly for  $\sum_{j=1}^{K_{4}}\tilde{\phi}^{\circ}(t, u_{4, j})$. Note that we took care of isolating the contribution from the set of 2-strings of main roots, $u_{\textrm{b}, j}\pm i/2$, since they are not part of the continuum of real main roots. We also substracted the interaction of the stack $u_{\textrm{b}, k}$ with itself by restricting the sums to $j\neq k$ in the right-hand side of~(\ref{IndScattb}). We could perform the same steps for the dressing phase contributions, but it is more convenient to note that since $\phi^{\textrm{dressing}}, \tilde{\phi}^{\textrm{dressing}}$, are given in terms of the generating functions for the individual charges, summing over the main roots amounts to replacing the latters by the generating functions for the total charges. Then we can write
\be
\sum_{j=1}^{K_{4}}q_{-}(t, u_{4, j}) = q_{-}^{\textrm{vacuum}}(t) + q_{-}(t)\, , \qquad q_{-}(t) = \sum_{\star=1}^{M} q_{-}^{\star}(t)\, ,
\ee 
where $q_{-}(t)$ is the generating function of the even charges measured above the vacuum. A similar identity holds for the generating function of odd charges, with $\tilde{q}_{+}^{\textrm{vacuum}}(t) = 0$, and we recall that there is no contribution to $q_{-}(t), \tilde{q}_{+}(t)$, from a stack $u_{\textrm{b}, j}$.

It remains to take the integral over the density $\rho(v)$ in~(\ref{IndScattb}). In principle we should consider separately the contribution from the vacuum and the ones from the excitations. But there is actually a trick to unit the two. It turns out to be possible to trade the vacuum for a pair of two holes carrying the rapidities $u_{\textrm{h}, \pm} = \pm S$.%
\footnote{This is reminiscent of the two large holes, with rapidities $\sim \pm S/\sqrt{2}$, that appear in the Baxter approach~\cite{BGK06}.  We stress however that here we need to choose the rapidities as $u_{\textrm{h}, \pm} = \pm S$ to get the right result up to subleading $\sim \log^{0}S$ contributions.} More precisely, the identity
\be\label{Trick}
\rho_{\textrm{vacuum}}(u) = \rho_{\textrm{h}}(u; u_{\textrm{h}, +}) + \rho_{\textrm{h}}(u; u_{\textrm{h}, -})
\ee
is valid for both the $\sim \log{S}$ and $\sim \log^0{S}$ parts of the vacuum distribution density, where $\rho_{\textrm{h}}(u; u_{\textrm{h}, \pm})$ is the correction to the density sourced by a hole carrying rapidity $u_{\textrm{h}, \pm}$.%
\footnote{This claim may look suspicious given that we said before that an important difference between the vacuum density and the correction sourced by an excitation $\rho_{\star}(u)$ is that the former scales like $\sim \log{(S/u)}$ at large $u$ while the latter is suppressed as $\sim 1/u$. There is no contradiction however but simply an order of limit phenomenon. Namely, the large $u$ behavior $\rho_{\star}(u) \sim 1/u$ is correct only if the rapidity $u_{\star}$ of the excitation is kept fixed in the process (it is actually an asymptotics for $u \gg u_{\star}$). However, if the excitation is sent to infinity $u_{\star} \sim S\gg 1$ and if only the leading $\sim \log{u_{\star}}$ and subleading $\sim\log^0{u_{\star}}$ contributions are retained, then the large $u$ asymptotics is transmuted into $\sim \log{(u_{\star}/u)} \sim \log{(S/u)}$, in agreement with the scaling of the vacuum density. Note finally that we need to send a momentum-carrying excitation to infinity for this to apply and since the vacuum has twist two we need two large twist-one excitations, that can be choosen to be two holes with zero total momentum.}
We will not prove this relation here but we note that it generalizes the observation that the energy of a hole with a large rapidity $u_{\textrm{h}}$ is exactly half the vacuum energy if $u_{\textrm{h}} = \pm S$, see Eq.~(\ref{LUA}). Accordingly, to simplify the analysis, we will set the vacuum density to zero and work with an extended set of excitations. The two extra holes will be given the large rapidities $u_{\textrm{h}, \pm} = \pm S$ at the end. This trick can be apply to the charges in~(\ref{Trick}) as well, such that we can completely ignore the vacuum contribution.

We come to the most remarkable simplification of the computation of the induced scattering. Namely, we notice that after integrating with the density $\rho(v) = -\sigma(v)-\tilde{\sigma}(v)$ in~(\ref{IndScattb}) and adding the dressing phase contribution, we get precisely the left-hand side of the large spin integral equation, see Eqs.~(\ref{EqTF}, \ref{FLtra}). Using this equation then leads to
\be\label{Phif}
\sum_{j=1}^{K_{4}}\phi(t, u_{4, j}) = 2\sum_{\star=1}^{M+2}\pi_{\star}(t, u_{\star}) - \sum_{j=1}^{K_{\textrm{h}}}\phi^{\circ}(t, u_{\textrm{h}, j})+2\sum_{j=1}^{K_{\textrm{b}}}\pi_{\textrm{b}}(t, u_{\textrm{b}, j}) + \sum_{j\neq k}^{K_{\textrm{b}}}\phi^{\circ}(t, u_{\textrm{b}, j}+\ft{i}{2}) + \sum_{j\neq k}^{K_{\textrm{b}}}\phi^{\circ}(t, u_{\textrm{b}, j}-\ft{i}{2})\, ,
\ee
and similarly for  $\sum_{j=1}^{K_{4}}\tilde{\phi}(t, u_{4, j})$. The inhomogeneous terms $\pi_{\star}(t, u_{\star}), \pi_{\textrm{b}}(t, u_{\textrm{b}})$, are given in Appendix~\ref{AALS} for the momentum-carrying excitations and in~(\ref{InhApp}) for the isotopic stack. It is now straightforward to compute the induced scattering phase since all terms in~(\ref{Phif}) are known explicitely. The result is
\be\label{Sindf}
S^{\textrm{induced}}_{\textrm{b}} = \prod_{j\neq k}^{K_{\textrm{b}}} {u_{\textrm{b}, k}-u_{\textrm{b}, j}-i \over u_{\textrm{b}, k}-u_{\textrm{b}, j}+i}\left({1-g^2/x^{[+2]}_{\textrm{b}, k}x_{\textrm{b}, j} \over 1-g^2/x^{[-2]}_{\textrm{b}, k}x_{\textrm{b}, j}}\right)^2\prod_{\star =1}^{M+2}S^{\textrm{induced}}_{\textrm{b}\star}\, ,
\ee
where
\begin{itemize}
\item for a hole (scalar) with rapidity $u_{\textrm{h}}$, 
\be
S^{\textrm{induced}}_{\textrm{b}\textrm{h}} = \left({x^{[+2]}_{\textrm{b}, k} \over x^{[-2]}_{\textrm{b}, k}}\right){u_{\textrm{b}, k}-u_{\textrm{h}} -\ft{i}{2} \over u_{\textrm{b}, k}-u_{\textrm{h}} + \ft{i}{2}}\, ,
\ee
\item for a fermion with rapidity $x_{\textrm{f}}$, either in the small or large rapidity domain,
\be
S^{\textrm{induced}}_{\textrm{b}\textrm{f}} = \left({x^{[+2]}_{\textrm{b}, k} \over x^{[-2]}_{\textrm{b}, k}}\right){x^{[-2]}_{\textrm{b}, k}-x_{\textrm{f}} \over x^{[+2]}_{\textrm{b}, k}-x_{\textrm{f}}}\, ,
\ee
\item for a gauge field ($\ell=1$), or bound states ($\ell>1$), with rapidity $u_{\ell}$,
\be
S^{\textrm{induced}}_{\textrm{b}\ell} = \left({x^{[+2]}_{\textrm{b}, k} \over x^{[-2]}_{\textrm{b}, k}}\right){u_{\textrm{b}, k}-u_{\ell} -i(1+\ell/2) \over u_{\textrm{b}, k}-u_{\ell} + i(1+\ell/2)}\left({1-g^2/x^{[+2]}_{\textrm{b}, k}x^{[+\ell]}_{\ell} \over 1-g^2/x^{[-2]}_{\textrm{b}, k}x^{[+\ell]}_{\ell}}{1-g^2/x^{[+2]}_{\textrm{b}, k}x^{[-\ell]}_{\ell} \over 1-g^2/x^{[-2]}_{\textrm{b}, k}x^{[-\ell]}_{\ell}}\right)\, .
\ee

\end{itemize}
We observe that a large part of the induced scattering phase~(\ref{Sindf}) is cancelled by the bare scattering one. After simplification, we find
\be\label{EffEqfApp}
1 = S^{\textrm{bare}}_{\textrm{b}}S^{\textrm{induced}}_{\textrm{b}} = \prod_{j=1}^{K_{\textrm{h}}}{u_{\textrm{b}, k}-u_{\textrm{h}, j}-\ft{i}{2} \over u_{\textrm{b}, k}-u_{\textrm{h}, j}+\ft{i}{2}} \prod_{j\neq k}^{K_{\textrm{b}}}{u_{\textrm{b}, k}-u_{\textrm{b}, j}+i \over u_{\textrm{b}, k}-u_{\textrm{b}, j}-i}\, .
\ee
in agreement with~(\ref{EffEqAppf}). Note that in~(\ref{EffEqfApp}) we have already sent to infinity the two vacuum-building holes, with rapidities $u_{\textrm{h}, \pm} = \pm S$.

\section{Strong coupling solution to the BES equation}\label{AppSC}

In this appendix we recall few results about the strong coupling expansion of the functions $\gamma^{\o}_{\pm}(t)$ and $\Gamma^{\o}(t)$, at fixed $t$. They are relevant to the computation of the dispersion relations for fermion, gauge field and bound states, in the (string theory) perturbative regime. We also provide expressions for the analysis of the dispersion relation for a scalar. All the results given below can be obtained from~\cite{BKK07,KSV08,BK08,BK09}, but we will mainly apply the representations used in~\cite{BK09}.

\subsection{Formulae for fermion}

The (even/odd) functions $\gamma^{\o}_{\pm}(t)$ can be found to admit the representation%
\footnote{We recall that one needs to perform a rescaling to derive expressions used in this paper from the ones given in~\cite{BK09}. Namely, $\gamma^{\o}_{\pm}(t) = 4g\gamma_{\pm}(t)$, where $\gamma_{\pm}(t)$ are the functions constructed in~\cite{BK09}.}
\be\label{g-sola}
\gamma^{\o}(it) \equiv \gamma^{\o}_{+}(it)+i\gamma^{\o}_{-}(it) = h^{\o}_{0}(t) I_{0}(t)+h^{\o}_{1}(t) I_{1}(t)\, .
\ee
The special functions $I_{0, 1}(t)$ are defined as
\be\label{g-solb}
I_{0}(t) = {\sqrt{2} \over \pi}\int_{-1}^{1}dv\, \left({1+v \over 1-v}\right)^{1/4}\e^{vt}\, , \qquad I_{1}(t) = {\sqrt{2} \over \pi}\int_{-1}^{1}dv\, \left({1+v \over 1-v}\right)^{1/4}{\e^{vt} \over 1+v}\, ,
\ee
and normalized as $I_{n}(0) = 1+n$ with $n=0, 1$. The factors $h^{\o}_{0, 1}(t)$ are given at strong coupling and fixed $t$ as
\be\label{g-solc}
\begin{aligned}
h^{\o}_{0}(-it) &= it - a(g) t^2/g - ic(g) t^3/g^2+ O(t^4/g^3)\, , \\
h^{\o}_{1}(-it) &= -ib(g) t/g + d(g)t^2/g^2 + O(t^3/g^3)\, .
\end{aligned}
\ee
The expansion coefficients read
\be\label{g-sold}
\begin{aligned}
a(g) &= {1\over 8}+{3\log{2} \over 4\pi} +\left({\textrm{K} \over 16\pi^2}-{9\log^2{2}\over 64\pi^2}\right)/g + O(1/g^2)\, , \,\,\,\,\,\,\, b(g) =  {3\log{2} \over 8\pi} + {\textrm{K} \over 32\pi^2}/g + O(1/g^2)\, ,\\
c(g) &= {13\over 384}+{\textrm{K} \over 4\pi^2} +{3\log{2} \over 32\pi}+{9\log^2{2} \over 32\pi^2}+O(1/g)\, ,  \qqqquad d(g) = {\textrm{K} \over 8\pi^2}+{3\log{2} \over 64\pi} +O(1/g)\, ,
\end{aligned}
\ee
where $\textrm{K}$ is the Catalan's constant.

For computing the fermion dispersion relation in the perturbative regime, one needs to evaluate
\be\label{g-sole}
E+p = 1-{1\over 2}\lambda(u)\, , \qquad \lambda(u) = \int_{0}^{\infty}{dt\over t}\gamma^{\o}(2gt)\e^{iut} = \int_{0}^{\infty}{dt\over t}\gamma^{\o}(t)\e^{i\bar{u}t}\, ,
\ee
with $\bar{u} = u/(2g)$ and $\bar{u}^2 > (2g)^2$ for a real-momentum fermion. The computation of~(\ref{g-sole}) is easily done, order by order in the $1/g$ expansion, by using~(\ref{g-sola}, \ref{g-solb}, \ref{g-solc}, \ref{g-sold}), and after noting the basic integrals
\be\label{Inpf}
\begin{aligned}
&i\int_{0}^{\infty}dt\, I_{0}(-it)\e^{i\bar{u}t} = 2-2\left({\bar{u}+1 \over \bar{u}-1}\right)^{1/4}\, , \\
&i\int_{0}^{\infty}dt\, I_{1}(-it)\e^{i\bar{u}t} = -{2\over 1+\bar{u}}\left({\bar{u}+1 \over \bar{u}-1}\right)^{1/4}\, . \\
\end{aligned}
\ee
When analytically continued to complex $\bar{u}$, outside the cut $\bar{u}^2<1$, the expressions in the right-hand side of~(\ref{Inpf}) can be used to evaluate the dispersion relation, in the perturbative regime, for physical (rescaled) rapidity $\bar{x} = x/g = \bar{u}-\sqrt{\bar{u}^2-1}$ with modulus smaller than $1$. Note that the zero momentum limit corresponds to $\bar{x}=0$ and thus to $\bar{u}=\infty$. For illustration, to leading order at strong coupling one gets that
\be
\lambda(2g\bar{u}) = 2-2\left({\bar{u}+1 \over \bar{u}-1}\right)^{1/4} = 2-2\left({1+\bar{x} \over 1-\bar{x}}\right)^{1/2} + O(1/g)\, ,
\ee
leading to
\be\label{Repp}
E+p = \left({1+\bar{x} \over 1-\bar{x}}\right)^{1/2} + O(1/g)\, .
\ee
As $E, p$, are respectively even, odd, in $\bar{x}$, or equivalently $\bar{u}$, one can extract the dispersion relation, $E=E(p)$, from representation like~(\ref{Repp}). To include higher-loop corrections, one can take derivatives with respect to $\bar{u}$ on both sides of~(\ref{Inpf}) in order to generate new identities, and then apply~(\ref{g-sola}, \ref{g-solb}, \ref{g-solc}, \ref{g-sold}) to compute~(\ref{g-sole}). Doing so, one finds that the dispersion relation can be written as
\be
E+p = \left({1+\bar{x} \over 1-\bar{x}}\right)^{1/2} f(\bar{x})\, ,
\ee
where $f(\bar{x})$ is given to two-loops by
\be\label{Appff}
f(\bar{x}) = 1-{2\bar{x}(a\bar{x}+b(1-\bar{x})^2) \over g(1-\bar{x}^2)^2} + {4\bar{x}^2(c\bar{x}(2+\bar{x}+2\bar{x}^2)+d(1-\bar{x})^2(1-\bar{x}+\bar{x}^2)) \over g^2(1-\bar{x}^2)^4} + O(1/g^3)\, ,
\ee
with the \textit{coupling dependent} coefficients $a, \ldots , d$, as in~(\ref{g-sold}). 

\subsection{Formulae for gauge field and bound states}

To evaluate the dispersion relations for gauge field and bound states, in the perturbative regime, we need the Fourier transform of $\Gamma^{\o}(2gt)$,
\be
\Lambda(u) = \int {dt \over 2\pi}\, \Gamma^{\o}(2gt) \e^{iut}\, ,
\ee
for $\bar{u}^2 < 1$, where $\bar{u} = u/(2g)$. It can be obtained from the analysis of~\cite{BK09}. For $\bar{u}^2 < 1$, it is found in the form of a $1/g$ expansion taken at fixed $\bar{u}$, and it reads
\be\label{cg-solb}
\Lambda(2g\bar{u}) = -{2\sqrt{2} \over \pi}\left({1+\bar{u} \over 1-\bar{u}}\right)^{1/4}\tau(\bar{u})\, ,
\ee
with
\be\label{cg-solc}
\tau(\bar{u}) = 1+ {a(g) \over g(1-\bar{u})} + {b(g) \over g(1+\bar{u})} + {c(g) \over g^2(1-\bar{u})^2} + {d(g) \over g^2(1+\bar{u})^2} + O(1/g^3)\, .
\ee
The coefficients $a(g), \ldots$, admit a $1/g$ expansion starting as
\be\label{cg-sold}
\begin{aligned}
a(g) &= -{1\over 32} + {3\log{2} \over 16\pi}+ \left(-{7 \over 6144} +{\textrm{K} \over 64 \pi^2} -{9\log^2{2} \over 512 \pi^2}\right)/g + O(1/g^2)\, , \\
b(g) &= -{1\over 32} - {3\log{2} \over 16\pi} + \left(-{7 \over 6144} -{\textrm{K} \over 64 \pi^2} -{9\log^2{2} \over 512 \pi^2}\right)/g + O(1/g^2) \, , \\
c(g) &= -{35 \over 6144} +{5\textrm{K} \over 64 \pi^2} -{15\log{2} \over 512 \pi} +{45\log^2{2} \over 512\pi^2}+ O(1/g)\, , \\
d(g) &= \, \, \,  \, {7 \over 2048} + {3\textrm{K} \over 64 \pi^2} - {9\log{2} \over 512 \pi} - {27\log^2{2} \over 512\pi^2}+O(1/g)\, .
\end{aligned}
\ee
Now, to evaluate the dispersion relation, we need to integrate the function $\Lambda(2g\bar{u})$,
\be\label{AppDRgfa}
E +p = i\pi g\int_{\bar{u}-i\bar{\epsilon}}^{\bar{u}+i\bar{\epsilon}}d\bar{v}\, \Lambda(2g\bar{v}) \, ,
\ee
where $\bar{\epsilon} = \ell/(4g)$. At strong coupling, we have $\bar{\epsilon}\rightarrow 0$ and for $\bar{u}^2 < 1$ the contour of integration in~(\ref{AppDRgfa}) becomes small. Hence, one can Taylor expand in $\bar{\epsilon}$ to get
\be\label{AppDRgfb}
E +p = -{\pi\ell\over 2}\left[\Lambda(2g\bar{u}) -{\ell^2 \over 96 g^2}{d^2 \over d\bar{u}^2}\Lambda(2g\bar{u}) + O(1/g^4)\right] \, .
\ee
Making use of~(\ref{cg-solb}) one concludes that Eq.~(\ref{AppDRgfb}) can be casted into the form
\be\label{AppDRgfc}
E+p = \sqrt{2}\ell\left({1+\bar{u} \over 1-\bar{u}}\right)^{1/4}f(\bar{u})\, ,
\ee
with
\be\label{Appfgf}
f(\bar{u}) = \tau(\bar{u}) -{\ell^2 \over 96 g^2}{1+4\bar{u} \over 4(1-\bar{u}^2)^2} + O(1/g^3)\, .
\ee
Applying the relations~(\ref{Appfgf}, \ref{cg-solc}, \ref{cg-sold}) to~(\ref{AppDRgfc}) we find the dispersion relations given in~(\ref{Epgfa}, \ref{Epgfb}).

\subsection{Formulae for scalar}

The dispersion relation for a hole in the non-perturbative regime is given in~(\ref{DRhnp}) in terms of an infinite set of coefficients which depend only on the coupling. They admit the representation
\be\label{GMasses}
\begin{aligned}
m_{4n+1} &= {8\sqrt{2} \over (4n+1)\pi}\e^{-(4n+1)\pi g}\left[{1\over (4n+1)\pi}- {1\over 4\sqrt{2}}\textrm{Re}\left\{\int_{0}^{\infty}{dt \e^{i(t-\pi/4)} \over t+i(4n+1)\pi g} \Gamma^{\o}(t)\right\}\right]\, , \\
m_{4n+3} &= {8\sqrt{2} \over (4n+3)\pi}\e^{-(4n+3)\pi g}\left[{1\over (4n+3)\pi}+{1\over 4\sqrt{2}}\textrm{Re}\left\{\int_{0}^{\infty}{dt \e^{i(t+\pi/4)} \over t+i(4n+3)\pi g} \bar{\Gamma}^{\o}(t)\right\}\right] \, ,
\end{aligned}
\ee
with $\bar{\Gamma}^{\o}(t) \equiv \Gamma^{\o}(-t)$. Note that despite their appearance, the relations~(\ref{GMasses}) do not assume that $g$ is large, but only $g\geqslant 0$. At $g=0$ for instance, the integrals in~(\ref{GMasses}) are subleading, giving
\be\label{GMassesZero}
m_{n}(g=0) = {8\sqrt{2} \over n^2\pi^2}\, .
\ee
Here we are mainly interested in their expression at strong coupling. Going along the lines of~\cite{BK08,BK09} we get the coefficients~(\ref{GMasses}) in this limit as
\be\label{SCGMasses}
\begin{aligned}
m_{4n+1} &= (8g)\e^{-(4n+1)\pi g}\Gamma(\ft{3}{4})(2(4n+1)\pi g)^{-3/4}{\Gamma(n+\ft{1}{4}) \over \Gamma(n+1)\Gamma(\ft{1}{4})} \\
& \qqqquad \qquad \, \, \,   \times \bigg[1 + \left({3\over 32(4n+1)\pi}-{3\log{2}\over 16\pi}\right)/g + O(1/g^2)\bigg]\, , \\
m_{4n+3} &= (8g)\e^{-(4n+3)\pi g}\Gamma(\ft{5}{4})(2(4n+3)\pi g)^{-5/4}{\Gamma(n+\ft{3}{4}) \over \Gamma(n+1)\Gamma(\ft{3}{4})}\\
&\qqqquad \qquad \, \, \,  \times \bigg[1 - \left({5\over 32(4n+3)\pi}-{3\log{2}\over 16\pi}\right)/g+ O(1/g^2)\bigg]\, .
\end{aligned}
\ee
We recover for $m_{1}$ the expression of the mass gap of the O(6) sigma model obtained in~\cite{AM07,FGR08a,BK08}. 

For the dispersion relation of a scalar in the near flat space limit, we need to evaluate the two sums
\be\label{Chi}
\begin{aligned}
\chi_{0}(\xi) &= {1 \over 2g^{1/4}}\sum_{n\ge 0}^{\infty}(-1)^n m_{4n+1}\e^{(4n+1)(\pi g+\xi)}\, , \\
\chi_{1}(\xi) &= {g^{1/4} \over 2}\sum_{n\ge 0}^{\infty}(-1)^n m_{4n+3}\e^{(4n+3)(\pi g+\xi)} \, .
\end{aligned}
\ee
Keeping the leading contribution to each of the coefficients $m_{4n+1}, m_{4n+3}$, we find from~(\ref{SCGMasses})
\be\label{Chiloa}
\begin{aligned}
\chi^{\textrm{LO}}_{0}(\xi) &={4 \over (2\pi)^{3/4}}\sum_{n\ge 0}^{\infty}(-1)^n \Gamma(\ft{3}{4})(4n+1)^{-3/4}{\Gamma(n+\ft{1}{4}) \over \Gamma(n+1)\Gamma(\ft{1}{4})}\e^{(4n+1)\xi}\, , \\
\chi^{\textrm{LO}}_{1}(\xi) &= {4 \over (2\pi)^{5/4}}\sum_{n\ge 0}^{\infty}(-1)^n \Gamma(\ft{5}{4})(4n+3)^{-5/4}{\Gamma(n+\ft{3}{4}) \over \Gamma(n+1)\Gamma(\ft{3}{4})}\e^{(4n+3)\xi}\, .
\end{aligned}
\ee
To perform the two sums above, we introduce the integral representations
\be
\Gamma(\ft{3}{4})(4n+1)^{-3/4} = \int_{0}^{\infty}ds\, s^{-1/4} \e^{-(4n+1)s}\, , \qquad \Gamma(\ft{5}{4})(4n+3)^{-5/4} = \int_{0}^{\infty}ds\, s^{1/4} \e^{-(4n+3)s}\, , 
\ee
and, after summing under the integrals, we obtain
\begin{align}\label{Chilob}
\chi^{\textrm{LO}}_{0}(\xi) &= {4 \over (2\pi)^{3/4}}\int_{0}^{\infty}ds\, s^{-1/4} \big(1+\e^{-4(\xi-s)}\big)^{-1/4}\, , \notag \\
\chi^{\textrm{LO}}_{1}(\xi) &= {4 \over (2\pi)^{5/4}}\int_{0}^{\infty}ds\, s^{1/4} \big(1+\e^{-4(\xi-s)}\big)^{-3/4} \, .
\end{align}
The asymptotics at $\xi \sim -\infty$ are given by
\be\label{lnxia}
\begin{aligned}
\chi^{\textrm{LO}}_{0}(\xi) &= {4\Gamma(\ft{3}{4}) \over (2\pi)^{3/4}}\e^{\xi} \, \, +\,  O(\e^{5\xi})\, , \\
\chi^{\textrm{LO}}_{1}(\xi) &= {\Gamma(\ft{1}{4}) \over (6\pi)^{5/4}}\e^{3\xi} + \, O(\e^{7\xi})\, .
\end{aligned}
\ee
For $\xi \sim \infty$, the integrals~(\ref{Chilob}) are dominated by $s < \xi$, leading to
\be\label{lpxia}
\begin{aligned}
\chi^{\textrm{LO}}_{0}(\xi) &= {4 \over (2\pi)^{3/4}}\int_{0}^{\xi}ds\, s^{-1/4} + \ldots = {16 \over 3(2\pi)^{3/4}}\xi^{3/4} + O(\xi^{-1/4}) \, ,\\
\chi^{\textrm{LO}}_{1}(\xi) &=  {4 \over (2\pi)^{5/4}}\int_{0}^{\xi}ds\, s^{1/4} + \ldots \, \, \, \, = {16 \over 5(2\pi)^{5/4}}\xi^{5/4} + O(\xi^{1/4}) \, .
\end{aligned}
\ee

\end{document}